\documentclass[a4paper,twocolumn,10pt,unpublished]{quantumarticle}
\pdfoutput=1
\usepackage[utf8]{inputenc}
\usepackage[english]{babel}
\usepackage[T1]{fontenc}
\usepackage{amsmath,amssymb,amsthm}
\usepackage{txfonts}
\usepackage{dsfont}
\usepackage{graphicx}
\usepackage{xcolor}
\usepackage{ragged2e}
\usepackage{caption}
\usepackage[numbers,sort&compress]{natbib}
\usepackage[colorlinks=true,linkcolor=blue,citecolor=blue,urlcolor=blue]{hyperref}
\usepackage{xurl}

\newcommand{\R}{\mathbb{R}}
\newcommand{\iden}{\mathds{1}}
\newcommand{\diag}[1]{\mathrm{diag}\left( #1 \right)}
\newcommand{\bra}[1]{\langle {#1} |}
\newcommand{\ket}[1]{| {#1} \rangle}
\newcommand{\braket}[1]{\langle #1\rangle}
\newcommand{\ketbra}[2]{\ensuremath{\left|#1\right\rangle\!\!\left\langle#2\right|}}
\DeclareMathOperator{\tr}{tr}
\newcommand{\bvec}[1]{\boldsymbol{#1}}

\newtheorem{theorem}{Theorem}
\newtheorem{corollary}{Corollary}
\newtheorem{definition}{Definition}
\newtheorem{proposition}{Proposition}
\newtheorem{lemma}{Lemma}
\begin{document}

\title{Exact monogamy of steering and remote Wigner negativity for multimode Gaussian nodes}

\author{Sudip Chakrabarty}
\email{sudip27042000@gmail.com}
\affiliation{S. N. Bose National Centre for Basic Sciences, Block JD, Sector III, Salt Lake, Kolkata 700106, India}

\maketitle

\begin{abstract}
Optical quantum networks route correlations through nodes that hold several modes and are linked to many neighbors. Steering is the correlation behind one-sided device-independent tasks and behind the remote preparation of non-Gaussian states. It is monogamous for a single mode, which can be steered by only one party, but how many neighbors can steer a node with $k$ modes was not known. We show that for Gaussian states the answer is exactly $2k-1$ when the neighbors use Gaussian measurements, with any number of measurement settings. The limit follows from a simple bound. To every neighbor we assign a share, computed from the covariance matrix, and we prove that the shares of all neighbors add up to at most $k$, irrespective of the amount of squeezing, amplification or noise the network contains. A neighbor steers the node by Gaussian measurements exactly when its share exceeds one half, and a network built from a harmonic tight frame reaches the limit. The same number governs remote Wigner negativity. A heralded non-Gaussian operation at the node can make at most $2k-1$ neighbors Wigner negative, and in our network a single heralded projection makes all of them negative in the same run. With $n$ settings of arbitrary measurements, at most $nk-1$ neighbors can steer the node or violate a Bell inequality with it. More than $2k-1$ neighbors can therefore steer the node or violate a Bell inequality with it only if they use non-Gaussian measurements with at least three settings, whereas entanglement can reach any number of neighbors. These results determine exactly how many neighbors one multimode node can serve with steering and with remote Wigner negativity.
\end{abstract}

%\tableofcontents

%=====================================================================
\section{Introduction}\label{sec:intro}
%=====================================================================

\subsection{Background and motivation}\label{sec:background}

A quantum network is useful because of the correlations it creates between distant parties. These correlations secure keys against any eavesdropper~\cite{Pirandola2020}, improve the precision of measurements spread over several locations~\cite{Zhuang2018,Guo2020}, and join small processors into a larger quantum computer~\cite{Kimble2008,Wehner2018,AghaeeRad2025}. Light is a natural carrier for such networks, and continuous-variable optics has made remarkable progress towards them. Squeezed light and beam splitters produce entangled Gaussian states on demand, cluster states of thousands of modes have been generated by multiplexing in time and frequency~\cite{Chen2014,Roslund2014,Larsen2019,Asavanant2019}, and measurement-based gates and teleportation networks have been built from the same ingredients~\cite{Menicucci2006,Larsen2021,vanLoock2000,Yonezawa2004,Braunstein2005,Weedbrook2012}. In these architectures a node rarely holds a single mode. It holds several modes, and through one multimode Gaussian state it is linked to many neighbors at the same time.

Among the correlations that such a node can offer its neighbors, Einstein-Podolsky-Rosen steering plays a special role~\cite{EPR1935,Schrodinger1935,Wiseman2007,Reid2009,Cavalcanti2017,Uola2020}. A neighbor steers the node when its measurements prepare conditional states of the node that no local-hidden-state model can explain. Steering is the resource of one-sided device-independent key distribution, in which the devices of one party need not be trusted~\cite{Branciard2012}, and it is required for continuous-variable teleportation beyond the no-cloning fidelity~\cite{He2015}. It also has a second, less obvious role, which concerns the scarcest resource of optical quantum technologies.

That resource is Wigner negativity. Gaussian states have nonnegative Wigner functions~\cite{Wigner1932,Hudson1974}, and every process in which states, operations and measurements all have nonnegative Wigner functions can be simulated efficiently on a classical computer~\cite{Mari2012,Veitch2013}. Gaussian operations cannot distill entanglement from Gaussian states~\cite{Eisert2002,Giedke2002,Fiurasek2002}, and universal computation as well as fault-tolerant encodings such as the Gottesman-Kitaev-Preskill code require a non-Gaussian element~\cite{Lloyd1999,Gottesman2001,Bourassa2021,Konno2024}. Even Gaussian boson sampling draws its hardness from photon counting at the output~\cite{Hamilton2017,Madsen2022}. Wigner negativity is therefore a key resource for any quantum advantage with light~\cite{Kenfack2004,Albarelli2018,Takagi2018,Walschaers2021}. In optics it is created by photon subtraction, photon counting and related operations, which are heralded and succeed only in a small fraction of the runs~\cite{Dakna1997,Wenger2004,Ourjoumtsev2006,NeergaardNielsen2006,Ra2020}. Every heralded event is precious, and steering lets its effect travel. A heralded operation at one node can make a distant neighbor Wigner negative, although the neighbor itself performs no non-Gaussian operation~\cite{Walschaers2020,WPT2020,Liu2022,Xiang2022,Walschaers2023,Chakrabarty2026Remote}. Steering of the node by the neighbor is necessary for this, and for a single neighbor it is also sufficient~\cite{Walschaers2020,WPT2020}. The number of neighbors that can steer a node is therefore also the number of neighbors that one rare non-Gaussian event at the node can reach.

Steering sits in a hierarchy of correlations. Entanglement~\cite{Horodecki2009,Guhne2009} is weaker, and it suffices for teleportation and key distribution with trusted devices. Bell nonlocality~\cite{Bell1964,Brunner2014} is stronger, it removes the need to trust any device, and it underlies device-independent key distribution and certified randomness~\cite{Acin2007,Pironio2010}. For Gaussian states, Bell nonlocality can only be revealed by non-Gaussian measurements~\cite{Banaszek1998,Banaszek1999,Jeong2003}.

Strong correlations cannot be handed out freely. They are monogamous~\cite{Coffman2000}, a property tied to the security of quantum cryptography and to the impossibility of cloning~\cite{Wootters1982,Barnum1996}. For steering by Gaussian measurements the rule is strict when the steered party is a single mode. Only one party can steer it, even when different parties use different observables~\cite{Reid2013,Ji2015}. When the steered party holds two or more modes this strict monogamy fails, as shown with Schur-complement inequalities and steering monogamy relations~\cite{Ji2015,Lami2016,Xiang2017} and observed in Gaussian cluster states~\cite{Deng2017}. These works establish strict monogamy for one steered mode and monogamy inequalities for steering quantifiers, but to our knowledge none of them bounds the number of single-mode parties that can steer a party with two or more modes. Quantitative monogamy relations for steering and entanglement quantifiers~\cite{Kogias2015,Xiang2017,Lami2016,Hiroshima2007,Adesso2007} bound sums of nonlinear measures, which can be small for every party, so they do not count parties either. Gaussian extendibility limits how many identical copies of one bipartite correlation can coexist~\cite{Terhal2003,Doherty2004,Lami2019}, whereas the neighbors of a node generally hold different correlations with it. For Bell nonlocality, no-signalling forbids two parties to violate the Clauser-Horne-Shimony-Holt (CHSH) inequality~\cite{CHSH1969} with a common partner that uses the same measurements in both tests~\cite{Toner2009,Pawlowski2009}, while in multipartite scenarios nonlocality can even be polygamous~\cite{Cieslinski2024}. For remote Wigner negativity, it is known that one photon subtraction cannot make two remote modes negative~\cite{Xiang2022}.

The question that matters for a multimode node is therefore still open. How many neighbors can steer a node of $k$ modes, and how many can one heralded event at the node make Wigner negative? Intuition offers two answers, and they contradict each other. Strict monogamy of a single mode suggests that each mode serves one neighbor, so that $k$ modes serve $k$ neighbors. Amplification suggests that there is no limit at all. An amplifier of gain $G$ followed by a balanced splitter into $N$ outputs gives every output the fraction $G/N$ of the signal, which exceeds one half for any $N$ once the gain is large enough, and a loss channel that transmits more than half of the signal already allows steering~\cite{Caruso2006}. This paper shows that neither answer is right, and it gives the exact one.

\subsection{Main results}\label{sec:mainresults}

The central result is a number. A node with $k$ modes can be steered by Gaussian measurements by at most $2k-1$ single-mode neighbors, and this number is reached. Here a Gaussian measurement means any homodyne, heterodyne or general-dyne measurement, and a neighbor may use as many of them as it wishes. The number $2k-1$ is therefore not tied to two settings. It is the limit for everything a neighbor can achieve with Gaussian measurements. Each mode of the node thus serves almost two neighbors, more than strict monogamy suggests, and no squeezing, amplification or noise can raise the number further.

Behind the number lies a simple accounting principle. To every neighbor $Y_j$ we attach a share $\zeta_j$, computed from the covariance matrix. The node and its neighbors can always be viewed as $k$ entangled pairs whose partner halves reach the neighbors through a Gaussian channel, and the share is the gain of this channel minus the noise that comes with the gain. We prove that the shares of all neighbors add up to at most $k$. Every steering neighbor needs a share above one half, so fewer than $2k$ neighbors can steer the node. The share is also the critical parameter of a known family of matrix inequalities that contains the Gaussian criteria for separability~\cite{Simon2000,Werner2001}, steering~\cite{Wiseman2007} and extendibility~\cite{Lami2019}, and for a single pair it coincides with the fractional extendibility parameter introduced recently for Gaussian repeaters~\cite{Ahmed2026}. These single-pair facts are known. What is new is that the shares of many neighbors with different reduced states constrain one another, and what follows from this constraint. Our results, ordered by their strength, are the following.
\begin{itemize}
\item \emph{Exact count for steering} (Theorem~\ref{thm:gsteer}). At most $2k-1$ neighbors can steer a node of $k$ modes by Gaussian measurements, with any number of settings, and a linear-optical network built from a harmonic tight frame, which we call the tight-frame network, reaches this number for every squeezing. The count holds for neighbors with any number of modes (Proposition~\ref{prop:multimode}).
\item \emph{Exact count for remote Wigner negativity} (Corollary~\ref{cor:negativity} and Theorem~\ref{thm:saturation}). If only the node applies non-Gaussian operations, one heralded operation makes at most $2k-1$ neighbors Wigner negative. In the tight-frame network a single heralded projection makes all $2k-1$ neighbors Wigner negative in the same run. A general criterion decides when one operation can make several neighbors negative at once (Theorem~\ref{thm:criterion}).
\item \emph{The share bound} (Theorem~\ref{thm:budget}). The shares of all neighbors add up to at most $k$. Every set of shares in $[0,1]$ that respects this bound is realized by $k$ two-mode squeezed vacua and passive optics (Proposition~\ref{prop:realize}), and the share does not increase under local Gaussian channels (Proposition~\ref{prop:mono}).
\item \emph{Arbitrary measurements} (Theorems~\ref{thm:count} and \ref{thm:tight}). With $n$ measurement settings at the neighbors, at most $nk-1$ neighbors can steer the node or violate a Bell inequality with it, whatever the measurements are. For two settings this is again $2k-1$, so two settings of any measurement do no better than Gaussian measurements with any number of settings, and only non-Gaussian measurements with at least three settings can serve more neighbors. For two settings the tight-frame network reaches this count, with a CHSH violation for every neighbor when idealized photon-number observables are used. For three settings, semidefinite programming with certified steering inequalities shows that a frame network reaches the steering count $3k-1$ for every $k$ we tested, up to $k=50$.
\item \emph{Teleportation and entanglement} (Theorem~\ref{thm:tel} and Proposition~\ref{prop:ent}). In unit-gain teleportation of coherent states into the node, the fidelity of a neighbor is at most $1/(2-\zeta_j)$, so at most $2k-1$ neighbors beat the Gaussian no-cloning fidelity $2/3$, and this count is reached. Entanglement, in contrast, can reach any number of neighbors.
\end{itemize}

For $k=1$ the counts reproduce strict single-mode monogamy~\cite{Reid2013,Ji2015}. For $k\geq2$ they tell exactly how far a multimode node escapes it. For remote Wigner negativity this is a hard limit on how far one heralded event can spread through Gaussian links. The optimal neighbors steer overlapping collective modes of the node, and for $k\geq3$ none of them steers any single physical mode, so only an operation on several modes of the node can make them negative. Because the shares follow from the covariance matrix, homodyne data alone tell which correlations each neighbor of a Gaussian node can at most hold, before any non-Gaussian experiment is attempted.

Two remarks fix the scope of these statements. The counts for steering, Bell nonlocality, teleportation and entanglement concern pairwise properties of one global state, each tested in its own runs, as in a network in which one fixed source prepares the state of the node and all its neighbors (Sec.~\ref{sec:setting}). The counts tell the designer of such a source how many neighbors it can serve with each correlation, whatever the source is. The result on remote Wigner negativity is stronger in kind, because there one heralded operation acts once and all neighbors are examined in the same run. The tightness results also differ in strength. The counts for Gaussian steering, remote negativity and teleportation are reached exactly and analytically. The two-setting Bell count is reached with idealized observables and a violation of order $1/(8(2k-1)^2)$. The three-setting steering count is reached, with certified numerics, for every $k$ we tested, and the three-setting Bell count only for $k=1$.

\subsection{Structure of the paper}\label{sec:structure}

Section~\ref{sec:pre} fixes the notation and the setting. Section~\ref{sec:share} introduces the share and explains what it measures. Section~\ref{sec:budget} proves that the shares add up to at most $k$ and constructs the tight-frame network, which reaches this bound. Section~\ref{sec:gsteer} derives the exact count for steering. Section~\ref{sec:wigner} turns to remote Wigner negativity and shows that one heralded projection reaches the same count in a single run. Section~\ref{sec:count} extends the counts to arbitrary measurements and Bell nonlocality, and Sec.~\ref{sec:tel} treats teleportation and entanglement. Section~\ref{sec:disc} compares the results with earlier work, collects all counts and discusses an experimental test, and Sec.~\ref{sec:conc} concludes. The appendices contain the technical proofs, the certificates for the three-setting numerics and a description of the numerical checks.

\begin{figure*}[t]
\centering
\includegraphics[width=\textwidth]{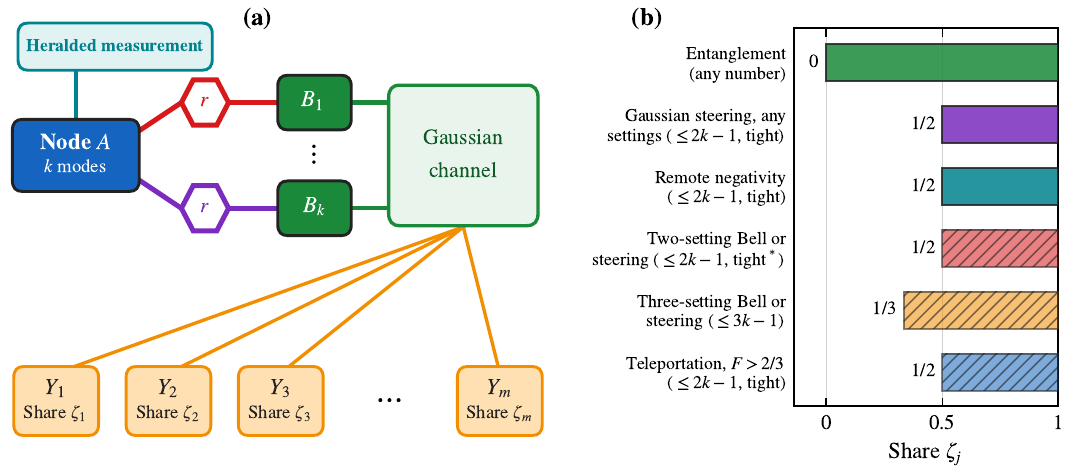}
\caption{\justifying\small (a) Setting. A node $A$ with $k$ modes (blue box) holds two-mode squeezed vacua (red and purple hexagons, squeezing $r$) with partner modes $B_1,\dots,B_k$ (green boxes). A Gaussian channel (light green box) distributes the partner modes, together with vacuum modes, among single-mode neighbors $Y_1,\dots,Y_m$ (light orange boxes). By the Gaussian Schmidt decomposition every Gaussian state of a node and its neighbors arises in this way, possibly with different squeezings for different pairs (Sec.~\ref{sec:purif}). The node may apply a heralded measurement (teal box) to create Wigner negativity at the neighbors (Sec.~\ref{sec:wignerq}). Each neighbor $Y_j$ carries a share $\zeta_j$, and the shares add up to at most $k$ (Theorem~\ref{thm:budget}). (b) Conditions on the share of one neighbor and the resulting counts. The left end of each bar is the threshold. Solid bars mark conditions that are necessary and sufficient, for entanglement (green), for steering by Gaussian measurements with any number of settings (purple) and for the existence of a heralded operation that makes the neighbor Wigner negative (teal). Hatched bars mark necessary conditions, for steering or Bell nonlocality with two settings (red) or three settings (orange) and arbitrary measurements, and for unit-gain teleportation into the node beyond the Gaussian no-cloning value $2/3$ (blue). Each label gives the largest number of neighbors that can hold the correlation and, where this number is attained, says so. The asterisk marks that the two-setting Bell count is attained only with the idealized observables of Eq.~\eqref{eq:obs}. For three settings the count $3k-1$ is attained numerically for steering (Sec.~\ref{sec:tight3}).}
\label{fig:scheme}
\end{figure*}

%=====================================================================
\section{Setting and tools}\label{sec:pre}
%=====================================================================

Before we can count neighbors, we need a precise language for the node, its neighbors and the correlations between them. This section provides it. We recall how bosonic modes and Gaussian states are described in phase space, define the node and its neighbors and explain what it means for a node to serve several of them, collect two elementary matrix facts that carry most of the proofs, and finally describe a purified picture of the global state, in which the correlations of every neighbor acquire a simple physical form.

\subsection{Phase space}\label{sec:phase}

We work with bosonic modes with annihilation operators $\hat a$ and quadratures $\hat x=\hat a+\hat a^\dagger$ and $\hat p=-i(\hat a-\hat a^\dagger)$. For $N$ modes we collect the quadratures in the vector $\hat{\bvec r}=(\hat x_1,\hat p_1,\dots,\hat x_N,\hat p_N)^T$, and the canonical commutation relations take the form
\begin{align}
[\hat r_k,\hat r_l]=2i\,\Omega_{kl},\quad
\Omega=\bigoplus_{j=1}^{N}J,\quad
J=\begin{pmatrix}0&1\\-1&0\end{pmatrix}.\label{eq:ccr}
\end{align}
The second moments of a state $\rho$ are collected in its covariance matrix, the real symmetric matrix
\begin{align}
V_{kl}=\tfrac12\tr\!\left(\rho\left\{\hat r_k-\langle\hat r_k\rangle,\,\hat r_l-\langle\hat r_l\rangle\right\}\right),\label{eq:cov}
\end{align}
which is normalized so that the vacuum has $V=\iden$. The commutation relations impose on every covariance matrix the uncertainty relation~\cite{Simon2000,Serafini2017}
\begin{align}
V+i\Omega\geq0.\label{eq:unc}
\end{align}
A Gaussian state is fixed by its first moments and its covariance matrix. Displacements play no role in this paper, so we set the first moments to zero. A Gaussian unitary acts on the quadratures as $\hat{\bvec r}\mapsto S\hat{\bvec r}+\bvec d$, where $\bvec d$ is a displacement and $S$ is a symplectic matrix, $S\Omega S^T=\Omega$, and it maps $V\mapsto SVS^T$. For a single mode the symplectic matrices are exactly the real $2\times2$ matrices of determinant one, which satisfy $SJS^T=J$. The basic entangled resource of continuous-variable optics is the two-mode squeezed vacuum with squeezing $r$, whose covariance matrix is
\begin{align}
V_{\rm TMSV}=\begin{pmatrix} v\,\iden_2 & s\,Z\\ s\,Z & v\,\iden_2\end{pmatrix},\label{eq:tmsv}
\end{align}
with $v=\cosh 2r$, $s=\sinh 2r$ and $Z=\diag{1,-1}$. In the Fock basis it reads $\sqrt{1-\lambda^2}\sum_n\lambda^n\ket{n,n}$ with $\lambda=\tanh r$, and we quote squeezing in decibels as $10\log_{10}e^{2r}$.

\subsection{The node, its neighbors and what it means to serve them}\label{sec:setting}

With these conventions we can describe the setting. A node $A$ holds $k$ modes. Its single-mode neighbors are denoted $Y_1,\dots,Y_m$, and all of them hold one Gaussian state together with the node. For the node and neighbor $Y_j$ we write the covariance matrix in block form,
\begin{align}
V^{(AY_j)}=\begin{pmatrix}V_A&V_{AY_j}\\ V_{Y_jA}&V_{Y_j}\end{pmatrix},\label{eq:block}
\end{align}
where $V_A$ is $2k\times2k$, $V_{Y_j}$ is $2\times2$, and $V_{Y_jA}=V_{AY_j}^T$.

Entanglement, steering, Bell nonlocality and the resource for teleportation are all properties of the reduced state $\rho_{AY_j}$ of the node and one neighbor. The question of the introduction is therefore about the compatibility of several such pairwise properties within one global state. In an experiment, the node and neighbor $Y_j$ perform their test in their own runs on copies of the global state, and the node may use different operations for different neighbors. This is what we mean when we say that a node serves several neighbors. It does not involve the simultaneous use of the node by several neighbors in one run, and it should not be confused with network nonlocality, where independent sources connect different parties, or with genuinely multipartite correlations. Remote Wigner negativity, studied in Sec.~\ref{sec:wigner}, is the exception, because there one heralded operation acts once on the node and all neighbors are examined in the same run.

A mode of the node that is in a pure state in some symplectic basis is in a product state with everything else, and so it is uncorrelated with every neighbor. Removing it changes no reduced state $\rho_{AY_j}$ apart from a factor that plays no role, and it lowers the number of modes of the node. We therefore assume throughout, without loss of generality, that all symplectic eigenvalues of $V_A$ exceed one, which is equivalent to
\begin{align}
V_A+i\Omega_A>0,\label{eq:posdef}
\end{align}
that is, $V_A+i\Omega_A$ is positive definite and in particular invertible. For a node with pure modes, every statement below holds for the reduced node with $k'<k$ modes, and bounds in terms of $k'$ imply those in terms of $k$. In other words, $k$ always counts the modes of the node that are not in a pure state. Table~\ref{tab:symbols} collects the main symbols used in this paper.

\begin{table}[!t]
\caption{\justifying\small Main symbols used in this paper.}
\label{tab:symbols}
\centering\small\setlength{\tabcolsep}{2pt}
\begin{tabular}{lp{0.70\columnwidth}}
\hline\hline
Symbol & Meaning\\
\hline
$A$, $k$ & Node and its number of modes\\
$Y_j$, $m$ & Single-mode neighbor $j$ and number of neighbors\\
$V$, $V_A$, $V_{AY_j}$ & Covariance matrix and its blocks, Eq.~\eqref{eq:block}\\
$\Omega$, $J$ & Symplectic forms, Eq.~\eqref{eq:ccr}\\
$K_j$, $N_j$ & Signal and noise blocks of the channel to $Y_j$, Eq.~\eqref{eq:channel}\\
$Q_j$, $M_j$, $\kappa_j$, $\mu_j$ & Schur complement and its parts, Eqs.~\eqref{eq:Q} and \eqref{eq:Mk}\\
$\zeta_j$, $\zeta_j^+$ & Share of $Y_j$ and its positive part, Eq.~\eqref{eq:share}\\
$\bvec u_j$, $\hat g_j$, $\eta$ & Frame vectors, collective node modes and transmission of the tight-frame network, Sec.~\ref{sec:network}\\
$\Gamma_j$, $\Sigma_j$ & Conditional mean map and covariance of the node given $Y_j$, Eq.~\eqref{eq:conditional}\\
$\hat E$, $\tau(\bvec\mu,\Sigma)$ & Effect of a heralded operation and quasi-state, Eq.~\eqref{eq:effect} and Def.~\ref{def:quasi}\\
$\ket{\Phi_k}$ & Heralding state of the node, Eq.~\eqref{eq:Phi}\\
$\mathcal B_j$ & CHSH value of $Y_j$, Eq.~\eqref{eq:chshk}\\
$\Xi_j$, $F_j$ & Teleportation noise matrix and optimal fidelity, Sec.~\ref{sec:telprot}\\
\hline\hline
\end{tabular}
\end{table}

\subsection{Two matrix tools}\label{sec:tools}

Almost every statement of this paper reduces to the question whether some Hermitian matrix is positive semidefinite. Two elementary facts make such questions short. The first concerns Hermitian $2\times2$ matrices, and every condition on a single neighbor will eventually be reduced to it.

\begin{lemma}\label{lem:2x2}
Let $M$ be a real symmetric $2\times2$ matrix and $b$ a real number. Then $M+ibJ\geq0$ if and only if $\tr M\geq0$ and $\det M\geq b^2$.
\end{lemma}

\begin{proof}
Write $M=\begin{pmatrix}m_1&m_3\\m_3&m_2\end{pmatrix}$. The matrix $M+ibJ$ has diagonal entries $m_1$ and $m_2$ and off-diagonal entries $m_3\pm ib$. Its trace is $\tr M$ and its determinant is $m_1m_2-(m_3+ib)(m_3-ib)=\det M-b^2$. A Hermitian $2\times2$ matrix is positive semidefinite exactly when its trace and its determinant are nonnegative, so $M+ibJ\geq0$ holds exactly when $\tr M\geq0$ and $\det M\geq b^2$. This completes the proof.
\end{proof}

The second fact is the Schur complement. It reduces a positivity condition on a large matrix to a condition on a smaller one, and it will turn conditions on the node and a neighbor together into conditions on the neighbor alone. Let
\begin{align}
G=\begin{pmatrix}G_{AA}&G_{AY}\\ G_{YA}&G_{YY}\end{pmatrix}\label{eq:G}
\end{align}
be Hermitian with $G_{AA}$ positive definite, and define the Schur complement $G/G_{AA}=G_{YY}-G_{YA}G_{AA}^{-1}G_{AY}$. Then
\begin{align}
G\geq0\iff G/G_{AA}\geq0,\label{eq:schur1}
\end{align}
and for every vector $z$,
\begin{align}
\min_{a}\,(a\oplus z)^\dagger G\,(a\oplus z)=z^\dagger\left(G/G_{AA}\right)z,\label{eq:schur2}
\end{align}
with the minimum attained at $a=-G_{AA}^{-1}G_{AY}z$. Both statements follow from the factorization $G=L^\dagger(G_{AA}\oplus G/G_{AA})L$ with the invertible matrix~\cite{Bhatia1997}
\[L=\begin{pmatrix}\iden&G_{AA}^{-1}G_{AY}\\0&\iden\end{pmatrix}.\]

\subsection{The purified picture}\label{sec:purif}

The covariance matrix fixes the Gaussian state, but it hides how the neighbors obtain their correlations with the node. A purification makes this visible. It will give the share its physical meaning in Sec.~\ref{sec:share}, and it is the starting point of the proof that the shares add up to at most $k$.

Any Gaussian state of the node and the neighbors can be completed to a pure Gaussian state by adding environment modes. By the Gaussian Schmidt decomposition~\cite{Holevo2001,Botero2003}, a Gaussian unitary on the node and a Gaussian unitary on everything else bring this pure state to a product of $k$ two-mode squeezed vacua, each pairing one node mode $\hat a_l$ with a partner mode $\hat b_l$, times the vacuum of all remaining modes. All $k$ squeezings are positive because of Eq.~\eqref{eq:posdef}. Undoing the unitary on the neighbors' side, the neighbors and the environment become the outputs of one symplectic transformation $S$, whose inputs are the $k$ partner modes and some vacuum modes. This is the picture of Fig.~\ref{fig:scheme}(a). It is a mathematical representation of the global state, used to interpret and to prove statements about quantities defined from the covariance matrix alone, and it does not require that the neighbors are physically connected to the node through separately accessible channels.

From the point of view of one neighbor $Y_j$, the transformation $S$ acts as a Gaussian channel from the partner modes to $Y_j$. Let $K_j$ be the $2\times2k$ block of the two rows of $S$ for $Y_j$ that acts on the partner modes, and $N_j$ the block that acts on the vacuum inputs. The quadratures of $Y_j$ are then
\begin{align}
\hat{\bvec r}_{Y_j}=K_j\,\hat{\bvec r}_B+N_j\,\hat{\bvec r}_{\rm vac},\label{eq:channel}
\end{align}
so $K_j$ carries the signal from the partners and $N_j$ adds noise from the vacuum inputs. Because $S$ is symplectic, the commutator of the neighbor's two quadratures is split between the two terms,
\begin{align}
K_j\Omega_BK_j^T+N_j\Omega_{\rm vac}N_j^T=J.\label{eq:commsplit}
\end{align}
Both terms are antisymmetric $2\times2$ matrices and hence multiples of $J$. How the unit commutator of a neighbor is divided between the partners and the environment is the physical content of the share, as we now explain.

%=====================================================================
\section{The share of a neighbor}\label{sec:share}
%=====================================================================

To count the neighbors that can steer the node, we need a number for each neighbor that satisfies two demands. It must decide whether the neighbor can steer the node, and it must be possible to add these numbers over all neighbors and bound the total. The usual quantifiers of steering fail the second demand, because they are nonlinear and can be small for every neighbor even when many neighbors steer. In this section we construct a number that meets both demands, the share. We define it from the covariance matrix, show that it is the gain of the channel from the node's partner modes to the neighbor minus the noise that comes with this gain, record its basic properties, write it as an optimum in two ways that will allow us to add it over neighbors, compute it for common channels, and finally show that it decides not only steering but a whole family of correlations.

\subsection{Definition and physical meaning}\label{sec:sharedef}

The natural starting point is the condition under which a neighbor cannot steer the node by Gaussian measurements, $V^{(AY_j)}+i(\Omega_A\oplus0)\geq0$~\cite{Wiseman2007}. By the Schur complement~\eqref{eq:schur1}, this condition on the whole pair is equivalent to a condition on a $2\times2$ matrix that belongs to the neighbor alone. This matrix is the Schur complement of $V^{(AY_j)}+i(\Omega_A\oplus0)$ with respect to its upper left block,
\begin{align}
Q_j=V_{Y_j}-V_{Y_jA}\left(V_A+i\Omega_A\right)^{-1}V_{AY_j}.\label{eq:Q}
\end{align}
Every Hermitian $2\times2$ matrix is a real symmetric matrix plus $i$ times a real antisymmetric matrix, and every real antisymmetric $2\times2$ matrix is a multiple of $J$. We may therefore write
\begin{align}
Q_j=M_j-i\kappa_jJ,\qquad \mu_j=\sqrt{\det M_j},\label{eq:Mk}
\end{align}
with $M_j$ real symmetric and $\kappa_j$ real. We will see shortly that $\kappa_j$ is a gain and $\mu_j$ a noise. The share weighs the one against the other, and it is normalized so that a neighbor that holds a partner mode of the node without loss or noise has the value one, while a neighbor that is uncorrelated with the node has a value of at most zero.

\begin{definition}[Share]\label{def:share}
The share of neighbor $Y_j$ is
\begin{align}
\zeta_j=\frac{1+\kappa_j-\mu_j}{2},\label{eq:share}
\end{align}
and its positive part is $\zeta_j^+=\max(\zeta_j,0)$.
\end{definition}

The share is an accounting quantity for the strong correlations of a pair. It is not an entanglement or steering measure, and it should not be confused with the Gaussian steerability of Ref.~\cite{Kogias2015} or with the transmission of a channel. For a single pair it coincides with the fractional extendibility parameter of Ref.~\cite{Ahmed2026}, as we show in Sec.~\ref{sec:hier}. Since the definition needs only the covariance matrix, the share can be computed from homodyne data. Its meaning becomes clear in the purified picture of Sec.~\ref{sec:purif}, where the neighbor is the output of a channel fed by the partner modes and by vacuum. The next lemma shows that $Q_j$ depends only on this channel.

\begin{lemma}[Channel form]\label{lem:channel}
In the purified picture,
\begin{align}
Q_j=N_jN_j^T-iK_j\Omega_BK_j^T.\label{eq:Qchannel}
\end{align}
Hence $M_j=N_jN_j^T$ is the noise added by the channel from the partners to $Y_j$, and $\kappa_j$, defined by $K_j\Omega_BK_j^T=\kappa_jJ$, is the part of the neighbor's commutator supplied by the partners. The covariance matrix of the node does not enter.
\end{lemma}

\begin{proof}
Let $V_{AB}$ be the covariance matrix of the $k$ two-mode squeezed vacua. Since the state of $A$ and $B$ is pure, $V_{AB}+i\Omega_{AB}$ has rank $2k$, half its size~\cite{Serafini2017}. Its upper left block $V_A+i\Omega_A$ is positive definite and already has rank $2k$, so the Schur complement with respect to this block vanishes,
\begin{align}
V_B+i\Omega_B-V_{BA}(V_A+i\Omega_A)^{-1}V_{AB}=0.\label{eq:pureschur}
\end{align}
By Eq.~\eqref{eq:channel}, and since the vacuum inputs have covariance $\iden$ and are uncorrelated with everything else,
\begin{align}
V_{Y_j}&=K_jV_BK_j^T+N_jN_j^T,\nonumber\\
V_{Y_jA}&=K_jV_{BA}.\label{eq:VYchannel}
\end{align}
Inserting this into Eq.~\eqref{eq:Q} and using Eq.~\eqref{eq:pureschur},
\begin{align}
Q_j&=K_j\left[V_B-V_{BA}(V_A+i\Omega_A)^{-1}V_{AB}\right]K_j^T+N_jN_j^T\nonumber\\
&=-iK_j\Omega_BK_j^T+N_jN_j^T.\label{eq:Qchannelproof}
\end{align}
The node's covariance matrix has dropped out, and comparing with Eq.~\eqref{eq:Mk} identifies $M_j=N_jN_j^T$ and $K_j\Omega_BK_j^T=\kappa_jJ$. This completes the proof.
\end{proof}

Combined with Eq.~\eqref{eq:commsplit}, the lemma divides the neighbor's unit commutator into a part $\kappa_j$ that comes from the partners and a part $1-\kappa_j$ that comes from the environment. For a lossy channel $\kappa_j$ is the transmission, and for an amplifier it is the gain. The share rewards this gain and subtracts the noise $\mu_j=\sqrt{\det M_j}$ that comes with it. A channel can therefore not raise the share by amplifying the signal, because amplification brings noise in exact proportion, as the examples below will make explicit.

Before using the share we record two basic facts. It never exceeds one, so a single neighbor can take at most one full unit, and it does not depend on the local mode basis of either party, as any meaningful accounting of correlations requires.

\begin{lemma}[Basic properties]\label{lem:props}
The following holds for every neighbor $Y_j$.
\begin{itemize}
\item[(i)] $M_j\geq0$ and $\mu_j\geq|1-\kappa_j|$. Consequently $\zeta_j\leq1$.
\item[(ii)] $\zeta_j$ is invariant under Gaussian unitaries of the node and under Gaussian unitaries of the neighbor.
\end{itemize}
\end{lemma}

\begin{proof}
(i) The uncertainty relation~\eqref{eq:unc} for the node and $Y_j$ reads $V^{(AY_j)}+i(\Omega_A\oplus J)\geq0$. Its Schur complement with respect to $V_A+i\Omega_A$ is $Q_j+iJ=M_j+i(1-\kappa_j)J$, which is therefore positive semidefinite by Eq.~\eqref{eq:schur1}. Lemma~\ref{lem:2x2} gives $\tr M_j\geq0$ and $\det M_j\geq(1-\kappa_j)^2\geq0$, so both eigenvalues of $M_j$ are nonnegative and $\mu_j\geq|1-\kappa_j|$. Then
\begin{align}
\zeta_j\leq\frac{1+\kappa_j-|1-\kappa_j|}{2}\leq1.\label{eq:zetaleq1}
\end{align}
(ii) A symplectic matrix $S_A$ of the node maps $V_A\mapsto S_AV_AS_A^T$ and $V_{AY_j}\mapsto S_AV_{AY_j}$. Since $S_A\Omega_AS_A^T=\Omega_A$, we have $S_AV_AS_A^T+i\Omega_A=S_A(V_A+i\Omega_A)S_A^T$, and $S_A$ cancels in Eq.~\eqref{eq:Q}. A symplectic matrix $S_Y$ of the neighbor maps $Q_j\mapsto S_YQ_jS_Y^T$, hence $M_j\mapsto S_YM_jS_Y^T$ and $\kappa_jJ\mapsto\kappa_jS_YJS_Y^T=\kappa_jJ$. Since $\det S_Y=1$, $\mu_j$ is unchanged as well, and so is $\zeta_j$. This completes the proof.
\end{proof}

A quantity that accounts for a resource should not grow under the operations that are considered free. In Gaussian networks the natural free operations are local Gaussian channels, which a node or a neighbor can apply with squeezers, beam splitters, Gaussian ancilla modes and the discarding of modes. The following proposition, proven in Appendix~\ref{app:mono}, shows that the positive part of the share does not grow under these operations, and it quantifies how loss and amplification at a neighbor reduce it.

\begin{proposition}[Monotonicity]\label{prop:mono}
The positive part $\zeta_j^+$ is invariant under Gaussian unitaries of the node and of the neighbor. It does not increase under Gaussian channels of the node, or under Gaussian channels of the neighbor that map one mode to one mode. A pure-loss channel of transmission $\lambda$ at the neighbor maps $\zeta_j$ to at most $\lambda\zeta_j$, and a quantum-limited amplifier of gain $G\geq1$ at the neighbor maps it to at most $1-G(1-\zeta_j)$.
\end{proposition}

Here a Gaussian channel is a trace-preserving completely positive map that takes Gaussian states to Gaussian states. A channel of the neighbor with several output modes is covered as well, in the following sense. A Gaussian unitary on the outputs followed by discarding all outputs but one is again a Gaussian channel from one mode to one mode, so no output mode, in any local mode basis, has a positive share larger than that of the input. By Proposition~\ref{prop:multimode} in Sec.~\ref{sec:gsteer}, the outputs together can then steer the node by Gaussian measurements only if the input could. Conditional operations that keep only some outcomes of a measurement, such as homodyne conditioning, are not trace preserving and are not covered. The proposition therefore does not make the share a general resource monotone, and we make no claim for non-Gaussian operations. The share can also be negative. In the channel picture this happens when the noise exceeds the gain by more than one vacuum unit, $\mu_j>1+\kappa_j$, as for a thermal-loss channel with $\eta<N/(N+1)$, which is entanglement breaking, or for the phase-conjugate output of an amplifier. Such a channel delivers no quantum correlation with the node at all, since $\zeta_j\leq0$ means separability (Proposition~\ref{prop:hier}). Unlike the usual quantifier of Gaussian steerability~\cite{Kogias2015}, the share does not depend on the squeezing of the node. By Lemma~\ref{lem:channel} it is a property of the channel that links the partner modes to the neighbor.

\subsection{Two variational characterizations}\label{sec:var}

Definition~\ref{def:share} gives the share in closed form, but a closed form is hard to add over neighbors. A sum is much easier to control when every term is an optimum over vectors, because the vectors that belong to different neighbors can then be combined into a single object whose properties limit the sum. We therefore write the share as an optimum in two equivalent ways, both proven in Appendix~\ref{app:var}. The first is a minimization over complex vectors of the neighbor's phase space.

\begin{proposition}[Variational form]\label{prop:var}
For every neighbor,
\begin{align}
1-2\zeta_j=\inf\left\{z^\dagger Q_jz\;:\;z\in\mathbb C^2,\ z^\dagger iJz=1\right\}.\label{eq:var}
\end{align}
\end{proposition}

The first form is compact, but it hides the physics. The second form expresses the same optimum through the mode operator of the neighbor, and it shows directly why amplification cannot raise the share. For this form we use the purified picture. Choose a local Gaussian unitary of the neighbor, and let $\hat c_j$ be the annihilation operator of $Y_j$ after it. By Eq.~\eqref{eq:channel} it is a linear combination of annihilation and creation operators of the inputs of $S$, which we number so that $c=1,\dots,k$ are the partner modes and $c=k+1,\dots,n$ are the vacuum inputs,
\begin{align}
\hat c_j=\sum_{c=1}^n\left(\alpha_{c}\,\hat b_c+\beta_{c}\,\hat b_c^\dagger\right).\label{eq:modeop}
\end{align}
The commutator $[\hat c_j,\hat c_j^\dagger]=1$ gives $\sum_c(|\alpha_c|^2-|\beta_c|^2)=1$.

\begin{proposition}[Mode-operator form]\label{prop:modeop}
For every neighbor,
\begin{align}
\zeta_j=\sup\Big\{\sum_{c\leq k}|\alpha_c|^2-\sum_{c=1}^n|\beta_c|^2\Big\},\label{eq:modeopform}
\end{align}
where the supremum runs over all local Gaussian unitaries of the neighbor, that is, over all its mode operators $\hat c_j$.
\end{proposition}

The first sum is the weight of annihilation operators that the neighbor takes from the node's partners, the part of its mode that is a faithful copy of the signal. The second sum is the weight of creation operators, which is the price of amplification. In a passive linear network no creation operators appear, and the share equals the transmission $\sum_{c\leq k}|\alpha_c|^2$ from the partner modes. Amplification increases the annihilation weight, but by the commutator it must add creation weight as well, and the share discounts it.

\subsection{Examples}\label{sec:examples}

To build intuition, we now compute the share for channels that occur in every optical network. Consider a single two-mode squeezed vacuum of a node mode and a partner mode, followed by a single-mode Gaussian channel from the partner to the neighbor. Such a channel maps the neighbor's covariance as $V_Y=XV_BX^T+Y$, with real $2\times2$ matrices $X$ and $Y\geq0$, and the correlations as $V_{YA}=XV_{BA}$~\cite{Serafini2017,Holevo2001}. Repeating the computation of Lemma~\ref{lem:channel} with $K_j=X$ gives $\kappa_j=\det X$ and $M_j=Y$. Table~\ref{tab:examples} lists the resulting shares, all of which are independent of the squeezing.

\begin{table}[htbp]
\caption{\justifying\small Gain $\kappa_j$, noise $\mu_j$ and share $\zeta_j$ for one partner mode sent to the neighbor through common Gaussian channels, namely pure loss and thermal loss of transmission $\eta$ with mean thermal photon number $N$, an amplifier of gain $G\geq1$, the same amplifier followed by a balanced beam splitter (one output shown), classical additive noise of variance $n_{\rm add}$, and the conjugate (idler) output of the amplifier.}
\label{tab:examples}
\centering\small\setlength{\tabcolsep}{2.5pt}
\begin{tabular}{lccc}
\hline\hline
Channel & $\kappa_j$ & $\mu_j$ & $\zeta_j$\\
\hline
Pure loss & $\eta$ & $1-\eta$ & $\eta$\\
Thermal loss & $\eta$ & $(1-\eta)(2N{+}1)$ & $\eta-(1-\eta)N$\\
Amplifier & $G$ & $G-1$ & $1$\\
Amplifier, split & $G/2$ & $G/2$ & $1/2$\\
Additive noise & $1$ & $n_{\rm add}$ & $1-n_{\rm add}/2$\\
Conjugate output & $1-G$ & $G$ & $1-G$\\
\hline\hline
\end{tabular}
\end{table}

\begin{figure*}[t]
\centering
\includegraphics[width=\textwidth]{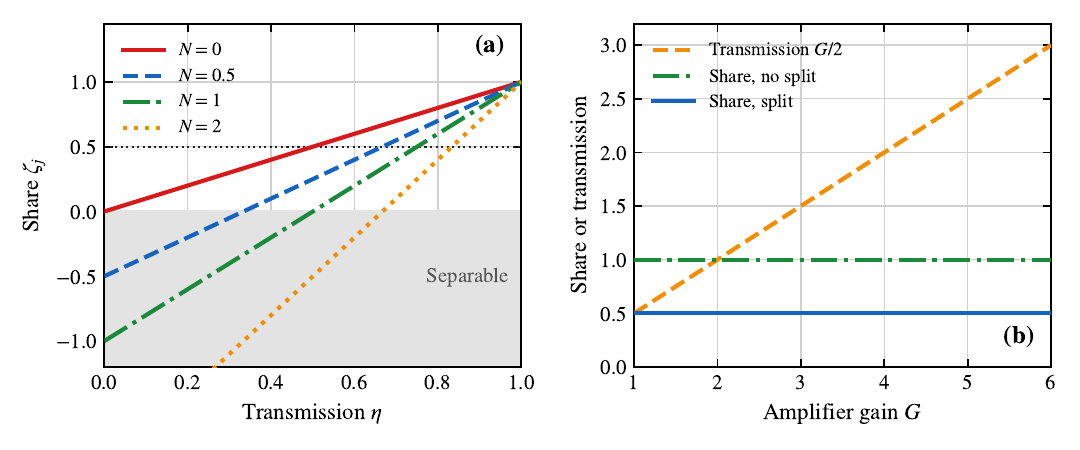}
\caption{\justifying\small Shares of simple channels. (a) Share of a neighbor that obtains a partner mode through a thermal-loss channel of transmission $\eta$ and thermal photon number $N$, for $N=0$ (red solid line), $N=0.5$ (blue dashed line), $N=1$ (green dash-dotted line) and $N=2$ (orange dotted line). The gray region $\zeta_j\leq0$ contains the separable states, and the black dotted line $\zeta_j=1/2$ is the threshold for Gaussian steering and, as a necessary condition, for two-setting steering and Bell nonlocality and for teleportation beyond the Gaussian no-cloning value (Proposition~\ref{prop:hier}, Theorems~\ref{thm:count} and \ref{thm:tel}). (b) A partner mode amplified with gain $G$ and split by a balanced beam splitter between two neighbors. Each neighbor's signal transmission $G/2$ (orange dashed line) exceeds one half for every $G>1$, while its share (blue solid line) equals one half for every gain. The green dash-dotted line is the share, equal to one, of a neighbor that keeps the whole amplified mode.}
\label{fig:examples}
\end{figure*}

For thermal loss, the share $\eta-(1-\eta)N$ decreases linearly with the thermal photon number [Fig.~\ref{fig:examples}(a)]. It is positive exactly when $\eta>N/(N+1)$, the known entanglement-breaking threshold of this channel~\cite{Holevo2001,Caruso2006}, in agreement with Proposition~\ref{prop:hier} below. The amplifier example resolves the second intuition of the introduction [Fig.~\ref{fig:examples}(b)]. An amplifier of gain $G$ followed by a balanced beam splitter gives each of two neighbors the signal transmission $G/2$, which exceeds one half as soon as $G>1$. If transmission were what counts, both neighbors would hold a strong correlation with a single mode, in conflict with the strict monogamy of single-mode steering~\cite{Reid2013,Ji2015}. The noise of the amplifier compensates its gain exactly. Each output has $\kappa_j=\mu_j=G/2$ and therefore $\zeta_j=1/2$, for every gain. Amplification can raise the transmission without limit, but not the share. The conjugate output of an amplifier receives a phase-conjugated copy of the partner, for which $\kappa_j<0$ and $\zeta_j<0$.

\subsection{One number for many correlations}\label{sec:hier}

We introduced the share through the steering condition, but it decides much more. Several known criteria for one pair of the node and a neighbor are members of the one-parameter family of matrix inequalities
\begin{align}
V^{(AY_j)}+i\left(\Omega_A\oplus tJ\right)\geq0,\qquad t\in\R.\label{eq:family}
\end{align}
The value $t=1$ is the uncertainty relation. The value $t=0$ is the condition that $Y_j$ cannot steer the node by Gaussian measurements~\cite{Wiseman2007}. For a single-mode neighbor, $t=2/n-1$ is the condition that the state is $n$-extendible on $Y_j$~\cite{Lami2019}, meaning that it is the common reduced state of a state of the node and $n$ copies of the neighbor. Finally, $t=-1$ is the condition that the partial transpose with respect to the neighbor is positive, which for Gaussian states with a single-mode party is equivalent to separability~\cite{Simon2000,Werner2001}. The share is the critical parameter of this whole family, as the following proposition shows.

\begin{proposition}[One number for many correlations]\label{prop:hier}
For every $t\leq1$, the inequality~\eqref{eq:family} holds if and only if
\begin{align}
\zeta_j\leq\frac{1+t}{2}.\label{eq:hier}
\end{align}
Consequently, by the criteria cited above,
\begin{itemize}
\item[(a)] $\rho_{AY_j}$ is entangled if and only if $\zeta_j>0$~\cite{Simon2000,Werner2001},
\item[(b)] $Y_j$ can steer the node by Gaussian measurements if and only if $\zeta_j>1/2$~\cite{Wiseman2007},
\item[(c)] $\rho_{AY_j}$ is $n$-extendible on $Y_j$ if and only if $\zeta_j\leq1/n$~\cite{Lami2019}.
\end{itemize}
\end{proposition}

\begin{proof}
The Schur complement of the left side of Eq.~\eqref{eq:family} with respect to $V_A+i\Omega_A$ is
\begin{align}
Q_j+itJ=M_j-i(\kappa_j-t)J.\label{eq:hierschur}
\end{align}
By Eq.~\eqref{eq:schur1}, Lemma~\ref{lem:2x2} and $M_j\geq0$, the inequality~\eqref{eq:family} holds if and only if $\mu_j\geq|\kappa_j-t|$, that is, if and only if
\begin{align}
\kappa_j-\mu_j\leq t\leq\kappa_j+\mu_j.\label{eq:interval}
\end{align}
By Lemma~\ref{lem:props}(i), $\kappa_j+\mu_j\geq\kappa_j+|1-\kappa_j|\geq1$, so the upper limit is automatically satisfied for $t\leq1$. The lower limit reads $t\geq\kappa_j-\mu_j=2\zeta_j-1$, which is Eq.~\eqref{eq:hier}. Statements (a), (b) and (c) follow by setting $t=-1$, $t=0$ and $t=2/n-1$ and invoking the cited criteria. This completes the proof.
\end{proof}

The proposition places every neighbor on a single scale. A share above zero means entanglement, a share above one half means steering by Gaussian measurements, and a share above $1/n$ means that the pair cannot be extended to $n$ copies of the neighbor. The smallest admissible value of $t$ is $t_j^*=2\zeta_j-1$, so the share is an affine reparametrization of the critical weight of the neighbor's symplectic form in the family~\eqref{eq:family}. Statements (a), (b) and (c) are the known criteria of Refs.~\cite{Simon2000,Werner2001,Wiseman2007,Lami2019}, rewritten in one variable. We state them in this form because the bound of the next section limits exactly this variable. Statement (c) refers to extensions by arbitrary states, not only by Gaussian states. This rests on Theorem~1 of Ref.~\cite{Lami2019}, by which a Gaussian state is $n$-extendible if and only if it has a Gaussian $n$-extension, so that the Gaussian criterion decides extendibility in general. This point matters in Sec.~\ref{sec:count}, where extensions are used to exclude nonlocality for arbitrary measurements. Two remarks extend the reach of the proposition beyond Gaussian states. Every quantum covariance matrix is the covariance matrix of some Gaussian state, and violating the partial-transpose condition on the covariance matrix implies entanglement for every state~\cite{Simon2000}. Hence $\zeta_j>0$, computed from the covariance matrix of an arbitrary state, certifies entanglement, although the converse requires Gaussianity.

The family~\eqref{eq:family} can also be read for real values of $n\geq1$ in the extendibility condition $t=2/n-1$. A related notion of fractional extendibility was introduced recently in a study of Gaussian quantum repeaters~\cite{Ahmed2026}, where a state of the node and a single-mode neighbor is called $1/\gamma$-extendible, for $\gamma\in[0,1]$, if $V^{(AY_j)}\geq i\Omega_A\oplus[(1-\gamma)\Delta+i\gamma J]$ for some single-mode covariance matrix $\Delta\geq iJ$. Using the Minkowski determinant inequality one finds that this holds if and only if $\zeta_j\leq\gamma$, so $\zeta_j^+$ is the smallest such $\gamma$ (Appendix~\ref{app:mono}). The share therefore coincides with the fractional extendibility parameter of a single pair. In Ref.~\cite{Ahmed2026} this parameter is used for the links of a repeater chain, one pair at a time. A bound on the sum of the parameters of many neighbors with different reduced states is, to our knowledge, not part of that framework or of any earlier work. It is the subject of the next section.

%=====================================================================
\section{\texorpdfstring{The shares add up to at most $k$}{The shares add up to at most k}}\label{sec:budget}
%=====================================================================

The share of one neighbor can be as large as one, by Lemma~\ref{lem:props}, and Proposition~\ref{prop:hier} tells which correlations each value allows. On its own this says nothing about how many neighbors can have large shares at the same time. In this section we show that the shares of all neighbors are bound together. We prove that they add up to at most the number $k$ of modes of the node, explain why no argument about single pairs could give this, compare it with the known monogamy inequality for a single mode, show that every set of shares that respects the bound occurs, and construct a network that reaches the bound. This network will be our main example in the rest of the paper.

\subsection{The share bound}\label{sec:budgetstate}

All neighbors draw their signal from the same $k$ partner modes, so their shares cannot be independent. The question is how the commutation relations of the neighbors limit them jointly. The following theorem, the main technical result of this paper, gives the answer.

\begin{theorem}[Share bound]\label{thm:budget}
For every Gaussian state of a node with $k$ modes and single-mode neighbors, and for every set $\mathcal S$ of neighbors,
\begin{align}
\sum_{j\in\mathcal S}\zeta_j\leq k.\label{eq:budget}
\end{align}
In particular $\sum_j\zeta_j^+\leq k$, where the sum runs over all neighbors.
\end{theorem}

The full proof is given in Appendix~\ref{app:budget}, and its logic is short. By Proposition~\ref{prop:modeop}, each share is the annihilation weight from the partners minus the total creation weight of a suitable mode operator of the neighbor. Collect the coefficients $\alpha_c$ and $\beta_c$ of the chosen mode operators of all neighbors in $\mathcal S$ as the columns of two matrices $\mathsf A$ and $\mathsf B$. The mode operators of different neighbors commute with each other and with each other's adjoints, and together with their normalization this gives the single matrix identity
\begin{align}
\mathsf A^\dagger\mathsf A-\mathsf B^\dagger\mathsf B=\iden.\label{eq:ABmain}
\end{align}
The sum of the shares is the weight of $\mathsf A$ in the $k$ partner rows minus the total weight of $\mathsf B$. The first term is the trace of $\mathsf A\mathsf A^\dagger$ compressed to a $k$-dimensional subspace, which by the maximum principle of Ky Fan~\cite{Fan1949,Bhatia1997} is at most the sum of the $k$ largest eigenvalues of $\mathsf A\mathsf A^\dagger$. By Eq.~\eqref{eq:ABmain}, these eigenvalues are one plus eigenvalues of $\mathsf B^\dagger\mathsf B$. Every extra unit of weight that amplification gives to $\mathsf A$ is thus paid for by weight in $\mathsf B$, which is subtracted, and what remains is at most $k$.

The bound cannot be obtained from statements about individual pairs. Each pair constrains only its own share, through Lemma~\ref{lem:props}, which allows every share up to one. It is the joint commutation relation~\eqref{eq:ABmain} of all neighbors that couples their different reduced states and limits the total. Extendibility, by contrast, constrains a single pair through the existence of identical copies. For $m$ neighbors with the same reduced state, extendibility gives $\zeta_j\leq1/m$ for every number of modes of the node, so identical neighbors never steer the node together. The same holds for the fractional extendibility of Ref.~\cite{Ahmed2026}, which equals the share of a single pair (Sec.~\ref{sec:hier}). Single-pair conditions of this kind allow every pair any value in $[0,1]$, whatever the reduced states of the other pairs are, so they cannot add up to a joint constraint. Composing extensions does not help either, because an extension of one pair produces copies of that neighbor and not the other neighbors, whose reduced states differ. The joint constraint comes from Eq.~\eqref{eq:ABmain}, which involves all neighbors at once. The share bound allows the average share $k/m$, and the network of Sec.~\ref{sec:network} reaches it with reduced states that differ in the collective mode of the node they address. As explained in Sec.~\ref{sec:setting}, the number $k$ in Eq.~\eqref{eq:budget} counts the modes of the node that are not in a pure state, which is also the number of partner modes in the purified picture.

For $k=1$ the bound reads $\zeta_1+\zeta_2\leq1$ for any two neighbors, so at most one of them steers a single mode, as also follows from the monogamy inequality $\det\Sigma_1\det\Sigma_2\geq1$~\cite{Reid2013,Ji2015}. Here $\Sigma_j=V_A-V_{AY_j}V_{Y_j}^{-1}V_{Y_jA}$ is the conditional covariance matrix of the single node mode given neighbor $Y_j$, which describes how well $Y_j$ predicts the quadratures of the node (Sec.~\ref{sec:herald}). The two inequalities constrain different quantities. The determinant inequality bounds how well two neighbors can jointly predict the quadratures of a single mode, while the share bound limits the gains and noises of the channels from the node's partner modes to the neighbors. For $k=1$ both give the same count, at most one steering neighbor, and the share bound is also related to the noise bounds of asymmetric Gaussian cloning~\cite{Cerf2000,Fiurasek2001}. For $k\geq2$ no inequality of the determinant type is known that counts steering neighbors, and this is where the share bound gives new information.

For a passive network the bound has an elementary reason. If the neighbors' modes are obtained from the partner modes and vacuum modes by an isometry $T$ of a beam-splitter network, neighbor $Y_j$ obtains the transmission $\eta_j=\sum_l|T_{jl}|^2$ from the partners, and since the columns of $T$ that belong to the partners are orthonormal, $\sum_j\eta_j\leq k$. Theorem~\ref{thm:budget} shows that no active Gaussian element can beat this accounting. Like Proposition~\ref{prop:hier}, the bound depends only on the covariance matrix, and since every quantum covariance matrix is that of a Gaussian state, Eq.~\eqref{eq:budget} holds for the covariance matrices of arbitrary states. The theorem also applies to any collection of single-mode outputs. For a neighbor that holds several modes it bounds the shares of its individual modes. For steering by Gaussian measurements these capture everything, as Proposition~\ref{prop:multimode} in Sec.~\ref{sec:gsteer} shows, while for other correlations a multimode neighbor may hold more than its modes reveal one at a time (Sec.~\ref{sec:open}).

\subsection{Every allowed set of shares occurs}\label{sec:realize}

An upper bound is most useful when it cannot be improved. The next proposition shows that the share bound is not only an upper bound but describes exactly the set of shares between zero and one that Gaussian states allow. Negative shares need no attention, since a neighbor with a negative share is not even entangled with the node and never contributes to a count.

\begin{proposition}[Every allowed set of shares occurs]\label{prop:realize}
For every collection of numbers $\zeta_1,\dots,\zeta_m\in[0,1]$ with $\sum_j\zeta_j\leq k$, there is a Gaussian state of a node with $k$ modes and $m$ single-mode neighbors, made of $k$ two-mode squeezed vacua and a passive linear-optical network, in which neighbor $Y_j$ has the share $\zeta_j$.
\end{proposition}

\begin{proof}
Add $k$ further output modes, which are discarded, and assign to each of them the number $(k-\sum_j\zeta_j)/k\in[0,1]$, so that all $m+k$ numbers lie in $[0,1]$ and add up to $k$. By the Schur-Horn theorem~\cite{Horn1954,Bhatia1997}, for every vector of numbers in $[0,1]$ that add up to $k$ there is a real orthogonal projector of rank $k$ with these numbers on its diagonal. Writing this projector as $TT^T$, the $(m+k)\times k$ matrix $T$ has orthonormal columns, and the squared norm of its $j$th row is the $j$th number. A passive network with isometry $T$, completed by vacuum inputs to an orthogonal transformation, takes the partner modes of $k$ two-mode squeezed vacua and delivers to neighbor $Y_j$ a collective partner mode through a pure-loss channel of transmission $\sum_l T_{jl}^2=\zeta_j$. By Table~\ref{tab:examples} and the argument of Lemma~\ref{lem:network} below, the share of $Y_j$ is then $\zeta_j$, for every $j$. This completes the proof.
\end{proof}

\subsection{A network that reaches the bound}\label{sec:network}

Proposition~\ref{prop:realize} guarantees that the bound can be reached, but its construction depends on the target shares and is not explicit. To show that the counts derived below are reached, we need one explicit family of states in which many neighbors have equal shares above a threshold. Harmonic tight frames provide such a family, and all our constructions belong to it.

Each node mode $\hat a_l$, $l=1,\dots,k$, forms a two-mode squeezed vacuum of squeezing $r$ with a partner mode $\hat b_l$. For $k\geq2$ let $m=2k-1$, $\vartheta_j=2\pi j/m$ for $j=1,\dots,m$, and define unit vectors $\bvec u_j\in\R^k$ as follows. For even $k$, with $p=k/2$,
\begin{align}
\bvec u_j=\sqrt{\tfrac2k}\left(\cos\vartheta_j,\sin\vartheta_j,\dots,\cos p\vartheta_j,\sin p\vartheta_j\right)^T,\label{eq:harmeven}
\end{align}
and for odd $k$, with $p=(k-1)/2$,
\begin{align}
\bvec u_j=\sqrt{\tfrac2k}\left(\tfrac{1}{\sqrt2},\cos\vartheta_j,\sin\vartheta_j,\dots,\cos p\vartheta_j,\sin p\vartheta_j\right)^T.\label{eq:harmodd}
\end{align}
Each vector has norm one, because it contains $p$ pairs $(\cos,\sin)$ of unit length, plus the constant entry of weight $1/2$ for odd $k$. These vectors form a harmonic tight frame~\cite{Waldron2018}, which means
\begin{align}
\sum_{j=1}^m\bvec u_j\bvec u_j^T=\frac mk\,\iden_k.\label{eq:frame}
\end{align}
To verify Eq.~\eqref{eq:frame}, note that the sums $\sum_j\cos q\vartheta_j$ and $\sum_j\sin q\vartheta_j$ vanish for every integer $q$ that is not a multiple of $m$. All products of two entries of $\bvec u_j$ reduce, by the product formulas for sines and cosines, to such sums with $q=r\pm s$, where $r,s\leq p$ are the frequencies of the two entries. Since $r+s\leq2p\leq k<m$, the only surviving sums are those with $q=0$, which give $m/2$ for $\cos^2$ and $\sin^2$ entries and $m/2$ for the squared constant entry $1/\sqrt2$. Multiplying by $2/k$ gives Eq.~\eqref{eq:frame}. For $k=1$ we take $m=1$ and $u_1=1$. For $k=2$ the three vectors are unit vectors in the plane at angles $120^\circ$, $240^\circ$ and $360^\circ$, which are equivalent, up to sign flips of the vectors, to $0^\circ$, $60^\circ$ and $120^\circ$. Such sign flips are local phase-space rotations of the neighbors and leave all shares unchanged.

Physically, a tight frame describes the most even way to spread $k$ modes over $m>k$ outputs. Every output receives the same fraction $k/m$ of the signal, and the frame identity~\eqref{eq:frame} says that no direction in the space of node modes is favored. Harmonic frames arise from the $m$-port discrete Fourier transform, the symmetric multiport splitter of linear optics, restricted to $k$ of its inputs and written in real quadratures, up to a passive transformation of the partner modes. For $k=2$ this is a balanced three-port splitter, a tritter. Any other equal-norm tight frame of $m$ vectors would serve the same purpose.

By Eq.~\eqref{eq:frame}, the $m\times k$ matrix with rows $\sqrt{k/m}\,\bvec u_j^T$ has orthonormal columns. Completing it to an $m\times m$ orthogonal matrix, we obtain a beam-splitter network that takes the $k$ partner modes and $m-k$ vacuum modes as inputs and delivers to neighbor $Y_j$ the mode
\begin{align}
\hat y_j=\sqrt\eta\,\hat\beta_j+\sqrt{1-\eta}\,\hat\epsilon_j,\qquad \hat\beta_j=\sum_{l=1}^ku_{j,l}\hat b_l,\label{eq:netmode}
\end{align}
where $\hat\epsilon_j$ is a normalized combination of the vacuum inputs and
\begin{align}
\eta=\frac km=\frac{k}{2k-1}.\label{eq:eta}
\end{align}
We call this the tight-frame network. The node mode that corresponds to neighbor $Y_j$ is the collective mode $\hat g_j=\sum_lu_{j,l}\hat a_l$. The network is useful because every pair of the node and one neighbor has a simple reduced state, as the next lemma shows.

\begin{lemma}[Tight-frame network]\label{lem:network}
In the tight-frame network, the reduced state of the node and neighbor $Y_j$ is a two-mode squeezed vacuum of $\hat g_j$ and $\hat\beta_j$ in which $\hat\beta_j$ has passed through a pure-loss channel of transmission $\eta$, times a thermal state of the node modes orthogonal to $\hat g_j$. In particular $\zeta_j=\eta$ for every $j$, and $\sum_j\zeta_j=k$, so the bound~\eqref{eq:budget} is reached.
\end{lemma}

\begin{proof}
The state of the node and the partners is proportional to $\exp(\lambda\sum_l\hat a_l^\dagger\hat b_l^\dagger)\ket0$. Let $R$ be a real orthogonal $k\times k$ matrix whose first row is $\bvec u_j^T$, and rotate both the $\hat a_l$ and the $\hat b_l$ with $R$. The sum $\sum_l\hat a_l^\dagger\hat b_l^\dagger$ is invariant under this simultaneous rotation, because $R^TR=\iden$. In the rotated basis the first node mode is $\hat g_j$, paired with $\hat\beta_j$, and every other node mode is paired with a partner mode orthogonal to $\hat\beta_j$. Neighbor $Y_j$ obtains only $\hat\beta_j$ and vacuum, so it is uncorrelated with these other pairs, and the reduced state of the corresponding node modes is thermal. The pair $(\hat g_j,Y_j)$ is a two-mode squeezed vacuum followed by loss, whose share is $\eta$ (Table~\ref{tab:examples}). Since all $m$ neighbors have the same share and $m\eta=k$, the shares add up to $k$, which completes the proof.
\end{proof}

Every neighbor of the tight-frame network thus holds a share $k/(2k-1)$, slightly more than one half. We now show what this means for steering.

%=====================================================================
\section{\texorpdfstring{Exactly $2k-1$ neighbors can steer the node}{Exactly 2k-1 neighbors can steer the node}}\label{sec:gsteer}\label{sec:gcount}
%=====================================================================

We can now answer the question of the introduction for steering by Gaussian measurements. Two facts are in place. Proposition~\ref{prop:hier}(b) says that a neighbor steers the node by Gaussian measurements exactly when its share exceeds one half, and Theorem~\ref{thm:budget} says that all shares together are at most $k$. Every steering neighbor therefore needs more than half of a unit, while only $k$ units exist, and fewer than $2k$ neighbors can steer the node. The tight-frame network shows that this simple accounting is exact.

\begin{theorem}[Gaussian steering count]\label{thm:gsteer}
In every Gaussian state of a node with $k$ modes and single-mode neighbors, at most $2k-1$ neighbors can steer the node by Gaussian measurements. In the tight-frame network of Sec.~\ref{sec:network}, all $2k-1$ neighbors steer the node by Gaussian measurements, for every squeezing $r>0$.
\end{theorem}

\begin{proof}
If $Y_j$ steers the node by Gaussian measurements, then $\zeta_j>1/2$ by Proposition~\ref{prop:hier}(b). Let $N$ neighbors have this property. Summing their shares and using Theorem~\ref{thm:budget},
\begin{align}
\frac{N}{2}<\sum_{j}\zeta_j\leq k,\label{eq:gcountproof}
\end{align}
so $N<2k$, which means $N\leq2k-1$. In the tight-frame network, Lemma~\ref{lem:network} gives $\zeta_j=k/(2k-1)>1/2$ for every neighbor and every $r>0$, so Proposition~\ref{prop:hier}(b) gives Gaussian steering of the node by all $2k-1$ neighbors. The bound is therefore attained, which completes the proof.
\end{proof}

Before discussing the theorem, we make precise what it counts, since the number $2k-1$ might suggest two measurement settings. Steering by Gaussian measurements allows the neighbor every Gaussian measurement, homodyne detection of any quadrature as well as heterodyne and general-dyne detection, and any number of them. The criterion of Proposition~\ref{prop:hier}(b) covers all of these at once~\cite{Wiseman2007}. A neighbor that measures many different quadratures therefore gains nothing, and $2k-1$ is the limit for every strategy built from Gaussian measurements. The same number reappears in Sec.~\ref{sec:count} for two settings of arbitrary, possibly non-Gaussian, measurements. This is not a coincidence. A pair with $\zeta_j\leq1/2$ is unsteerable by Gaussian measurements and, by Proposition~\ref{prop:hier}(c), also $2$-extendible, which excludes steering and Bell nonlocality with two settings of any kind~\cite{Lami2019,Terhal2003}. Gaussian measurements with any number of settings and arbitrary measurements with two settings thus face the same threshold, and a neighbor can exceed it only with non-Gaussian measurements and at least three settings.

The theorem settles both intuitions of the introduction. The first intuition held that a node with $k$ modes should serve about $k$ neighbors, one for each mode. This is wrong by almost a factor of two. A node with two modes serves three steering neighbors, a node with three modes serves five, and in general $2k-1$ neighbors steer a node that could serve only $k$ of them mode by mode. The second intuition held that amplification might serve arbitrarily many neighbors, since it makes every pair look strongly correlated. This is wrong as well. Amplification raises the share of a neighbor only together with the noise that it adds, as the examples of Sec.~\ref{sec:examples} show, and Theorem~\ref{thm:budget} holds for every Gaussian operation. The count $2k-1$ is a fixed number, independent of squeezing, gain and every other parameter of the state.

The count also has a direct physical meaning. Neighbor $Y_j$ steers the node by Gaussian measurements exactly when homodyne measurements on $Y_j$ predict the two quadratures of some collective mode of the node with errors whose product beats the uncertainty principle of that mode~\cite{Reid1989,Wiseman2007}. This form of the criterion is stated precisely in Eq.~\eqref{eq:steerSigma} below. Theorem~\ref{thm:gsteer} therefore says that at most $2k-1$ single-mode neighbors can each predict a collective mode of a $k$-mode node better than the uncertainty principle allows. In the tight-frame network the neighbors predict different, overlapping collective modes $\hat g_j$, and no two of them predict the same one. As explained in Sec.~\ref{sec:setting}, the count concerns pairwise properties of one global state, tested in separate runs.

The count also connects with known results in two ways. Steering from a neighbor to the node is the resource of one-sided device-independent protocols in which the devices of the node are trusted~\cite{Branciard2012}, so the theorem also limits how many neighbors with homodyne detectors can run such protocols with one node. For $k=1$ the theorem is the strict monogamy of Gaussian steering of a single mode~\cite{Reid2013,Ji2015}. For $k\geq2$ strict monogamy is known to fail~\cite{Lami2016,Xiang2017,Deng2017}, and the theorem tells exactly how far it fails.

The maximal count is not a fragile feature of an idealized state. If every neighbor of the tight-frame network suffers an additional loss of transmission $\lambda_{\rm ex}$, each pair remains a two-mode squeezed vacuum seen through loss, now of transmission $\lambda_{\rm ex}k/(2k-1)$, and all $2k-1$ neighbors keep steering the node if and only if $\lambda_{\rm ex}>1-1/(2k)$. The network thus tolerates an additional loss of $25\%$, $17\%$ and $10\%$ for $k=2$, $3$ and $5$. The margin shrinks with $k$, because the shares of the tight-frame network approach one half from above, but it is substantial for the small nodes that are relevant in practice.

So far every neighbor has held a single mode. Real neighbors in a network often hold several modes, and one may ask whether a multimode neighbor could steer the node in ways that none of its modes can on its own, which would put it outside the share bound. For steering by Gaussian measurements this cannot happen, as the next proposition shows. A multimode neighbor steers the node exactly when one of its modes, in a suitable local mode basis, does so on its own, and the count of Theorem~\ref{thm:gsteer} therefore holds for neighbors of any size.

\begin{proposition}[Neighbors with several modes]\label{prop:multimode}
A neighbor $Y$ with any number $\ell$ of modes steers the node by Gaussian measurements if and only if, after some Gaussian unitary of $Y$, one of its modes has a share above one half. Consequently, in every Gaussian state of a node with $k$ modes and neighbors with any numbers of modes, at most $2k-1$ neighbors can steer the node by Gaussian measurements.
\end{proposition}

\begin{proof}
Define the $2\ell\times2\ell$ matrix $Q_Y=V_Y-V_{YA}(V_A+i\Omega_A)^{-1}V_{AY}$ as in Eq.~\eqref{eq:Q}. By the criterion of Ref.~\cite{Wiseman2007} and Eq.~\eqref{eq:schur1}, $Y$ cannot steer the node by Gaussian measurements if and only if $Q_Y\geq0$. The uncertainty relation $V^{(AY)}+i(\Omega_A\oplus\Omega_Y)\geq0$ has the Schur complement $Q_Y+i\Omega_Y$ with respect to $V_A+i\Omega_A$, so $Q_Y+i\Omega_Y\geq0$ and
\begin{align}
z^\dagger Q_Yz\geq-z^\dagger i\Omega_Yz\label{eq:multibound}
\end{align}
for every $z\in\mathbb C^{2\ell}$.

Suppose that $Y$ steers the node. Then $z^\dagger Q_Yz<0$ for some $z$, and Eq.~\eqref{eq:multibound} forces $z^\dagger i\Omega_Yz>0$. We rescale $z$ so that $z^\dagger i\Omega_Yz=1$ and write $z=(\bvec f+i\bvec e)/\sqrt2$ with real vectors $\bvec e$ and $\bvec f$. A short calculation gives $z^\dagger i\Omega_Yz=\bvec e^T\Omega_Y\bvec f$, so $\bvec e^T\Omega_Y\bvec f=1$. The quadratures $\bvec e\cdot\hat{\bvec r}_Y$ and $\bvec f\cdot\hat{\bvec r}_Y$ therefore have the commutator $2i$ of a canonical pair, and since every canonical pair can be completed to a symplectic basis, there is a Gaussian unitary of $Y$ after which they are the quadratures $\hat x_1$ and $\hat p_1$ of its first mode $Y_1$. Under a local symplectic matrix $S$ of $Y$, the matrix $Q_Y$ becomes $SQ_YS^T$ and $z$ becomes $S^{-T}z$, so $z^\dagger Q_Yz$ and $z^\dagger i\Omega_Yz$ do not change. In the new basis $z$ has nonzero entries only for $Y_1$. Let $w\in\mathbb C^2$ collect these entries. Then $w^\dagger iJw=1$, and since the block of $Q_Y$ that belongs to $Y_1$ is the matrix $Q_1$ of Eq.~\eqref{eq:Q} for the pair of the node and $Y_1$, we have $z^\dagger Q_Yz=w^\dagger Q_1w$. Proposition~\ref{prop:var} now gives
\begin{align}
1-2\zeta_{Y_1}\leq w^\dagger Q_1w=z^\dagger Q_Yz<0,
\end{align}
so $\zeta_{Y_1}>1/2$. Conversely, if a mode of $Y$ has a share above one half after some Gaussian unitary of $Y$, it steers the node on its own by Proposition~\ref{prop:hier}(b), and $Y$ can apply the same unitary and the same measurements.

For the count, choose for every steering neighbor one such mode, after that neighbor's own Gaussian unitary, and discard all other modes of the neighbors. The node and the chosen modes are in a Gaussian state with single-mode neighbors, each with a share above one half, and Theorem~\ref{thm:budget} gives $N/2<k$ for their number $N$, as in Eq.~\eqref{eq:gcountproof}. This completes the proof.
\end{proof}

Steering by Gaussian measurements is not only a correlation to be counted. It is also exactly the resource that allows a heralded non-Gaussian event at the node to make a distant neighbor Wigner negative~\cite{Walschaers2020,WPT2020}. The next section shows that the count $2k-1$ carries over to this task, and that it can be reached in a single run.

%=====================================================================
\section{\texorpdfstring{One heralded event makes $2k-1$ neighbors Wigner negative}{One heralded event makes 2k-1 neighbors Wigner negative}}\label{sec:wigner}
%=====================================================================

Processes in which all states, operations and measurements have nonnegative Wigner functions can be simulated efficiently on a classical computer~\cite{Mari2012}, so a quantum advantage with light needs Wigner negativity somewhere. Wigner negative states are hard to make, and they are usually produced by a heralded non-Gaussian event, such as the detection of a photon, at one location. Steering by Gaussian measurements is the correlation that lets such an event act at a distance. When a neighbor steers the node, a heralded event at the node can make the neighbor Wigner negative, and without steering it cannot~\cite{Walschaers2020,WPT2020}. Theorem~\ref{thm:gsteer} therefore bounds how many neighbors can be made negative by operations of the node chosen separately for each of them. In this section we show that the bound $2k-1$ holds for remote Wigner negativity as well, and that a single heralded event reaches it.

\subsection{The question}\label{sec:wignerq}

Suppose now that the node applies one heralded non-Gaussian operation, for instance photon subtraction or photon counting on some of its modes, while the neighbors only hold their modes or apply Gaussian operations. We ask how many neighbors obtain a Wigner negative state in this way. The question differs from Theorem~\ref{thm:gsteer} in one important respect. The count of steering neighbors concerns pairwise properties tested in separate runs. Here a single operation acts once, and we ask how many neighbors become Wigner negative in the same run. We show that Gaussian steering decides whether one neighbor can be made negative, that the share bound limits the number of neighbors that one operation can make negative, and that this limit is reached.

\subsection{Heralded operations and quasi-states}\label{sec:herald}

To answer this question we need a description of what a heralded operation at the node does to a distant neighbor. We first recall Wigner functions and heralded operations, then describe the node conditioned on a phase-space point of the neighbor, and finally lift this conditional description to an operator, the quasi-state. The quasi-state reduces every question about remote Wigner negativity to the sign of one expectation value.

\emph{Wigner functions.} For a state $\rho$ of $N$ modes, the Wigner function $W_\rho(\bvec r)$ is a real function of the phase-space point $\bvec r\in\R^{2N}$ that integrates to one and reproduces expectation values through the trace rule $\tr(\rho\hat A)=\int W_\rho(\bvec r)A_W(\bvec r)\,d\bvec r$, where $A_W$ is the Weyl symbol of the operator $\hat A$~\cite{Serafini2017,Cahill1969,Royer1977}. We extend $W$ linearly to all trace-class operators. A Gaussian state with first moments $\bvec\mu$ and covariance matrix $V$ has the Wigner function $\mathcal N(\bvec r;\bvec\mu,V)$, where
\begin{align}
\mathcal N(\bvec r;\bvec\mu,\Sigma)=\frac{\exp\left[-\tfrac12(\bvec r-\bvec\mu)^T\Sigma^{-1}(\bvec r-\bvec\mu)\right]}{(2\pi)^N\sqrt{\det\Sigma}}\label{eq:gaussfun}
\end{align}
denotes the normalized Gaussian function with mean $\bvec\mu$ and covariance $\Sigma$. A state is Wigner negative if its Wigner function takes a negative value somewhere.

\emph{Heralded operations.} A heralded operation of the node is described by the Kraus operators $\hat K_\ell$ that are kept when the heralding outcome occurs, possibly after ancilla modes have been measured. The neighbors are affected only through the positive operator
\begin{align}
\hat E=\sum_\ell\hat K_\ell^\dagger\hat K_\ell\geq0,\label{eq:effect}
\end{align}
which we call the effect of the operation. For a physical measurement $0\leq\hat E\leq\iden$. Ideal subtraction of one photon from a mode with annihilation operator $\hat a$ has the single Kraus operator $\hat a$ and the effect $\hat a^\dagger\hat a$, which is the limit of a weakly reflecting tap followed by a click. After a successful herald, the unnormalized state of neighbor $Y_j$ is
\begin{align}
\tilde\rho_{Y_j}=\tr_A\left[\left(\hat E\otimes\iden\right)\rho_{AY_j}\right],\label{eq:heralded}
\end{align}
and dividing by the success probability $\tr[(\hat E\otimes\iden)\rho_{AY_j}]$ does not change the sign of its Wigner function. If the node applies a Gaussian unitary $\hat U$ before the operation, the effect becomes $\hat U^\dagger\hat E\hat U$, which is again an effect.

\emph{Conditional quantities.} A heralded operation acts on the node, but its effect is seen at the neighbor. The link between the two is the description of the node conditioned on the neighbor. For the pair of the node and neighbor $Y_j$ define the matrices
\begin{align}
\Gamma_j&=V_{AY_j}V_{Y_j}^{-1},\nonumber\\
\Sigma_j&=V_A-V_{AY_j}V_{Y_j}^{-1}V_{Y_jA}.\label{eq:conditional}
\end{align}
The Wigner function of the Gaussian pair factorizes as~\cite{Serafini2017,WPT2020}
\begin{align}
W_{\rho_{AY_j}}(\bvec\alpha,\bvec y)=\mathcal N(\bvec y;0,V_{Y_j})\,\mathcal N(\bvec\alpha;\Gamma_j\bvec y,\Sigma_j),\label{eq:factor}
\end{align}
where $\bvec\alpha$ and $\bvec y$ are phase-space points of the node and of the neighbor. Conditioned on the point $\bvec y$ of the neighbor, the node is described by a Gaussian function with mean $\Gamma_j\bvec y$ and covariance matrix $\Sigma_j$. Neighbor $Y_j$ steers the node by Gaussian measurements if and only if its predictions beat the uncertainty principle of the node~\cite{Wiseman2007,Reid1989},
\begin{align}
\Sigma_j+i\Omega_A\not\geq0.\label{eq:steerSigma}
\end{align}
The Schur complement of $V^{(AY_j)}+i(\Omega_A\oplus0)$ with respect to the positive definite block $V_{Y_j}$ is exactly $\Sigma_j+i\Omega_A$, so by Eq.~\eqref{eq:schur1} and Proposition~\ref{prop:hier}(b) the condition~\eqref{eq:steerSigma} is equivalent to $\zeta_j>1/2$.

\emph{Quasi-states.} When the neighbor steers the node, the Gaussian function $\mathcal N(\bvec\alpha;\Gamma_j\bvec y,\Sigma_j)$ is not the Wigner function of a state, but it is still the Wigner function of a Hermitian operator, which we now construct. For $q\in(-1,1)$ define the single-mode operator
\begin{align}
\rho_q=(1-q)\sum_{n=0}^\infty q^n\ketbra{n}{n}.\label{eq:rhoq}
\end{align}
For $q\geq0$ this is a thermal state. For $q<0$ its weights alternate in sign, so it has unit trace but is not positive. Its Wigner function is $\mathcal N(\bvec r;0,d\,\iden_2)$ with $d=(1+q)/(1-q)$, and its mean photon number is $q/(1-q)=(d-1)/2$ (Lemma~\ref{lem:rhoq} in Appendix~\ref{app:quasi}). The map $q\mapsto d$ is a bijection of $(-1,1)$ onto $(0,\infty)$, so a Gaussian function with variance below the vacuum level, $d<1$, corresponds to negative $q$. The operators $\rho_q$ thus give every single-mode Gaussian function in normal form an operator. A general Gaussian function of $k$ modes is brought to such a normal form by a symplectic transformation and a shift, so applying the corresponding Gaussian unitary and displacement to a product of operators $\rho_q$ lifts every Gaussian function to an operator. This leads to the following definition.

\begin{definition}[Quasi-state]\label{def:quasi}
Let $\Sigma$ be a real symmetric positive definite $2k\times2k$ matrix with Williamson form $\Sigma=S\,\mathrm{diag}(d_1,d_1,\dots,d_k,d_k)\,S^T$, where $S$ is symplectic and $d_l>0$ are the symplectic eigenvalues~\cite{Williamson1936,Simon1994}, and let $\bvec\mu\in\R^{2k}$. The quasi-state with first moments $\bvec\mu$ and covariance matrix $\Sigma$ is
\begin{align}
\tau(\bvec\mu,\Sigma)&=\hat D(\bvec\mu)\,\hat U_S\left(\bigotimes_{l=1}^k\rho_{q_l}\right)\hat U_S^\dagger\,\hat D(\bvec\mu)^\dagger,\nonumber\\
q_l&=\frac{d_l-1}{d_l+1},\label{eq:quasi}
\end{align}
where $\hat U_S$ is the Gaussian unitary of $S$ and $\hat D(\bvec\mu)$ is the displacement by $\bvec\mu$.
\end{definition}

The operator $\tau(\bvec\mu,\Sigma)$ is Hermitian and trace class, has unit trace and the Wigner function $\mathcal N(\cdot\,;\bvec\mu,\Sigma)$, and it is positive, and then a Gaussian state, if and only if $\Sigma+i\Omega\geq0$ (Lemma~\ref{lem:quasi}). With quasi-states at hand, the effect of a heralded operation on a neighbor takes a simple form. Combining the factorization~\eqref{eq:factor} with the trace rule, the heralded Wigner function of the neighbor at a point $\bvec y$ becomes a positive Gaussian weight times the expectation value of the effect in the quasi-state of the node conditioned on $\bvec y$.

\begin{theorem}[Quasi-state representation]\label{thm:quasirep}
For a Gaussian state of the node and neighbor $Y_j$ and every bounded effect $\hat E$ of the node, and every effect that is a polynomial in the mode operators of the node, the Wigner function of the unnormalized heralded state~\eqref{eq:heralded} is
\begin{align}
W_{\tilde\rho_{Y_j}}(\bvec y)=\mathcal N(\bvec y;0,V_{Y_j})\,\tr\left[\hat E\,\tau(\Gamma_j\bvec y,\Sigma_j)\right].\label{eq:quasirep}
\end{align}
\end{theorem}

The proof is given in Appendix~\ref{app:quasi}. Physical heralding measurements have bounded effects, $0\leq\hat E\leq\iden$. Polynomial effects such as $\hat a^\dagger\hat a$ for ideal photon subtraction are included because they are the weak-tap limit of physical operations, up to a factor that tends to zero and does not affect signs, and because the expectation values of polynomials in quasi-states converge, the Fock weights of $\rho_q$ decaying geometrically. The operator $\tau(\Gamma_j\bvec y,\Sigma_j)$ is the state of the node conditioned on the phase-space point $\bvec y$ of the neighbor, and it is the operator form of the conditional Wigner functions of Ref.~\cite{WPT2020}. The Gaussian prefactor in Eq.~\eqref{eq:quasirep} is positive, so the neighbor is Wigner negative at $\bvec y$ exactly when the expectation value of the effect in the conditional quasi-state is negative.

\subsection{Steering decides whether one neighbor can be made negative}\label{sec:onereceiver}

The representation~\eqref{eq:quasirep} turns the question of remote Wigner negativity for one neighbor into a question about the positivity of the conditional quasi-state. The following proposition collects the two known directions~\cite{Walschaers2020,WPT2020} in this language.

\begin{proposition}[One neighbor]\label{prop:onereceiver}
\begin{itemize}
\item[(i)] If $\zeta_j\leq1/2$, the heralded state of $Y_j$ has a nonnegative Wigner function for every effect of the node, with or without Gaussian preprocessing, and it remains nonnegative under Gaussian operations of the neighbor.
\item[(ii)] If $\zeta_j>1/2$, there is a Gaussian unitary of the node followed by the subtraction of one photon from one mode that makes $Y_j$ Wigner negative at the origin of its phase space.
\end{itemize}
\end{proposition}

\begin{proof}
(i) If $\zeta_j\leq1/2$, then $\Sigma_j+i\Omega_A\geq0$, so every quasi-state $\tau(\Gamma_j\bvec y,\Sigma_j)$ is a Gaussian state and $\tr[\hat E\tau]\geq0$ for every positive $\hat E$. By Eq.~\eqref{eq:quasirep} the Wigner function is nonnegative everywhere. A Gaussian preprocessing with symplectic matrix $S$ replaces $\Sigma_j$ by $S\Sigma_jS^T$, and $S\Sigma_jS^T+i\Omega_A=S(\Sigma_j+i\Omega_A)S^T\geq0$, so the argument applies again. Gaussian operations of the neighbor act on Wigner functions by linear symplectic maps, convolutions with Gaussian functions and Gaussian conditioning, all of which preserve nonnegativity~\cite{Serafini2017,Mari2012}.

(ii) If $\zeta_j>1/2$, some symplectic eigenvalue $d_1$ of $\Sigma_j$ is smaller than one~\cite{Simon1994}. Let $S$ be the symplectic matrix of its Williamson form, and let the node apply $\hat U_S^\dagger$ and subtract one photon from its first mode, which gives the effect $\hat E=\hat U_S\hat a_1^\dagger\hat a_1\hat U_S^\dagger$. At $\bvec y=0$ the conditional mean vanishes, and by Definition~\ref{def:quasi}
\begin{align}
\tr\left[\hat E\,\tau(0,\Sigma_j)\right]=\tr\left(\rho_{q_1}\hat a_1^\dagger\hat a_1\right)=\frac{d_1-1}{2}<0.\label{eq:onerec}
\end{align}
By Eq.~\eqref{eq:quasirep} the Wigner function of $Y_j$ is negative at the origin, so the neighbor is Wigner negative. This completes the proof.
\end{proof}

Part (i) is the necessity statement of Ref.~\cite{WPT2020}, and part (ii) reproduces the sufficiency result of Ref.~\cite{Walschaers2020}. Steering by Gaussian measurements is thus necessary and sufficient for a neighbor to be made Wigner negative by some heralded operation of the node, which is why the corresponding bar in Fig.~\ref{fig:scheme}(b) is solid. The condition cannot be bypassed with exact catalysis either. A neighbor that cannot steer the node by Gaussian measurements cannot be made Wigner negative even when the node may use an ancilla that is returned exactly in its initial state~\cite{Chakrabarty2026Catalysis}. A count of steering neighbors therefore bounds the number of neighbors that can be made negative one at a time. For one operation acting in one run the argument needs one more step, because an operation may act on only some of the modes of the node.

\subsection{Counting negative neighbors}\label{sec:negcount}

An operation often acts on only a few modes of the node, possibly on collective modes. We say that an effect acts on $k'$ modes if $k'$ is the smallest number for which there is a Gaussian unitary $\hat U$ of the node such that, after $\hat U$, the effect acts nontrivially only on a set $A_1$ of $k'\leq k$ modes, that is, $\hat E=\hat U^\dagger(\hat E_1\otimes\iden)\hat U$ with $\hat E_1$ acting on $A_1$. This number depends only on the effect, since it is defined by a minimum over all Gaussian unitaries. By cyclicity of the partial trace, the heralded state of $Y_j$ is then obtained from the reduced Gaussian state of $A_1$ and $Y_j$ after the unitary, with the effect $\hat E_1$. Proposition~\ref{prop:onereceiver}(i) applied to this reduced state shows that a negative neighbor must steer $A_1$, and the share bound for a node with $k'$ modes then gives the following count.

\begin{corollary}[Negative neighbors]\label{cor:negativity}
Let the node and the single-mode neighbors be in a joint Gaussian state, and let only the node apply a non-Gaussian operation. If the effect of a heralded operation acts nontrivially on $k'$ modes of the node, in some mode basis reached by a Gaussian unitary, the operation makes at most $2k'-1\leq2k-1$ neighbors Wigner negative. The same holds when the neighbors apply Gaussian operations, adaptive or not, conditioned on classical messages from the node.
\end{corollary}

\begin{proof}
Each negative neighbor steers $A_1$ by Gaussian measurements, so its share with respect to the node $A_1$ exceeds one half by Proposition~\ref{prop:hier}(b). Theorem~\ref{thm:gsteer} for the reduced state of $A_1$ and the neighbors gives at most $2k'-1$ such neighbors. Gaussian operations of the neighbors, also when conditioned on earlier outcomes, preserve the nonnegativity of Wigner functions and of their mixtures, so they cannot make a nonsteering neighbor negative. Hence at most $2k'-1\leq2k-1$ neighbors become negative, which completes the proof.
\end{proof}

The corollary says that the number of negative neighbors is limited by the number of modes on which the operation acts, not by the size of the node. The case $k'=1$ is worth stating on its own, because it covers the most common non-Gaussian operations, such as photon subtraction from one mode.

\begin{proposition}[Single-mode operations]\label{prop:singlemode}
An operation that acts on a single mode of the node, in some mode basis reached by a Gaussian unitary, creates Wigner negativity at no more than one neighbor. This includes the subtraction of any number of photons from one collective mode $\sum_lc_l\hat a_l$ of the node.
\end{proposition}

\begin{proof}
This is Corollary~\ref{cor:negativity} with $k'=1$. For the collective mode, a passive Gaussian unitary maps $\sum_lc_l\hat a_l$ to $\hat a_1$, and the subtraction of $p$ photons has the effect $(\hat a_1^\dagger)^p\hat a_1^p$, which acts on one mode, so the corollary applies with $k'=1$ and gives at most one negative neighbor. This completes the proof.
\end{proof}

For photon subtraction this agrees with the monogamy of remotely generated negativity found in Ref.~\cite{Xiang2022}, and it also follows from the strict monogamy of single-mode steering~\cite{Reid2013,Ji2015} together with Ref.~\cite{WPT2020}. To make two neighbors negative with one Gaussian resource, the effect must therefore act nontrivially on at least two modes in every symplectic basis. Several remarks fix the scope of the corollary. First, the classical messages may come from the node only. If a neighbor conditions its state on the outcomes of measurements made by other neighbors, the node and those neighbors act together as a larger steered party, and the count must be applied to this larger party instead. For example, let neighbor $Y_2$ measure its mode by heterodyne detection and announce its outcome, and let $Y_1$ keep only the runs with a chosen outcome. The state of $Y_1$ is then heralded by an effect that acts on the $k'$ modes of $A_1$ and on the mode of $Y_2$, so Corollary~\ref{cor:negativity} applied to the party $A_1Y_2$ bounds the number of negative neighbors among the others by $2k'+1$ instead of $2k'-1$. Second, the corollary counts neighbors whose own Wigner function becomes negative. It is silent about joint Wigner negativity of groups of neighbors, which may appear even when every single neighbor remains nonnegative. Corollary~\ref{cor:negativity} also does not apply when the neighbors themselves perform non-Gaussian operations, since then the neighbors are no longer limited by the steering they receive. The corollary also holds for neighbors with several modes. Its proof uses only that a negative neighbor steers $A_1$ by Gaussian measurements, which holds for neighbors with any number of modes~\cite{WPT2020}, and Proposition~\ref{prop:multimode} then bounds their number by $2k'-1$. The corollary does cover every strategy of the node within one run. A sequence of heralded operations, possibly chosen adaptively and possibly with Gaussian operations in between, keeps only those runs in which all heralds occur, and its overall action on the neighbors is again described by a single effect $\hat E$.

\subsection{One operation for all neighbors}\label{sec:oneforall}

That $2k-1$ neighbors can steer the node means that each of them could be made negative by an operation chosen for it. Whether one operation can make all of them negative at once is a different question, because all neighbors depend on the same effect $\hat E$, and an effect that suits one neighbor may spoil another. By Eq.~\eqref{eq:quasirep}, neighbor $Y_j$ is negative at a chosen point $\bvec y_j$ exactly when $\tr(\hat E\tau_j)<0$ for the quasi-state $\tau_j=\tau(\Gamma_j\bvec y_j,\Sigma_j)$. What we need is therefore a criterion for the existence of one operator that has a negative expectation value in several quasi-states at once. The following theorem gives such a criterion. Its proof, based on the minimax theorem of von Neumann~\cite{vonNeumann1928}, is given in Appendix~\ref{app:crit}.

\begin{theorem}[Simultaneous negativity]\label{thm:criterion}
Let $\bvec y_1,\dots,\bvec y_m$ be phase-space points of the neighbors and $\tau_j=\tau(\Gamma_j\bvec y_j,\Sigma_j)$. There is an effect $0\leq\hat E\leq\iden$ with $W_{Y_j}(\bvec y_j)<0$ for all $j$ if and only if no convex combination $\sum_jp_j\tau_j$ is positive semidefinite.
\end{theorem}

Since the theorem quantifies over all effects, it includes any Gaussian unitary or displacement that the node applies before a non-Gaussian measurement. A displacement of the node by $\bvec\delta$, for instance, replaces $\hat E$ by $\hat D(\bvec\delta)^\dagger\hat E\hat D(\bvec\delta)$, which is equivalent to shifting the means of all quasi-states $\tau_j$ by $\bvec\delta$. In impossibility proofs the theorem is used in one direction, since exhibiting one positive mixture shows that no operation activates the chosen neighbors at the chosen points. In constructions it is used in the other direction, and in a truncated Fock space the search for an effect is a semidefinite program. Steering of every neighbor is necessary for the condition, because a nonsteering neighbor has a positive $\tau_j$, but it is not obviously sufficient. We now show that in the tight-frame network, which reaches the share bound, an explicit effect works.

\subsection{One projection reaches the count}\label{sec:saturation}

Consider the tight-frame network of Sec.~\ref{sec:network} with $m=2k-1$ neighbors, transmission $\eta=k/(2k-1)$ and squeezing $r>0$. By Lemma~\ref{lem:network}, neighbor $Y_j$ is correlated only with the collective mode $\hat g_j=\sum_lu_{j,l}\hat a_l$ of the node, and Lemma~\ref{lem:framesteer} in Appendix~\ref{app:sat} shows that its conditional quasi-state at the origin is
\begin{align}
\tau(0,\Sigma_j)&=\hat U_j\left(\rho_{q_d}\otimes\rho_{q_v}^{\otimes(k-1)}\right)\hat U_j^\dagger,\nonumber\\
q_v&=\tanh^2r,\qquad q_d=-\frac{q_v}{2k-1},\label{eq:taumain}
\end{align}
where $\hat U_j$ is a passive unitary that maps the first mode to $\hat g_j$. The steered mode carries a negative parameter $q_d$, the other modes are thermal, and the ratio $q_d/q_v=-1/(2k-1)$ does not depend on the squeezing. The node projects onto the state with one photon in each steered mode,
\begin{align}
\ket{\Phi_k}=\frac{1}{\sqrt{\mathcal N_k}}\prod_{j=1}^{2k-1}\hat g_j^\dagger\ket0,\label{eq:Phi}
\end{align}
where $\mathcal N_k>0$ normalizes the state, which contains exactly $2k-1$ photons. This state is the natural candidate. Proposition~\ref{prop:onereceiver}(ii) shows that one photon in the steered mode makes one neighbor negative, and $\ket{\Phi_k}$ places one photon in the steered mode of every neighbor at once, while the symmetry of the frame treats all neighbors alike. The following theorem shows that this projection indeed makes all neighbors negative in the same run.

\begin{theorem}[Saturation]\label{thm:saturation}
For every $k\geq1$ and every squeezing $r>0$, projecting the node of the tight-frame network onto $\ket{\Phi_k}$ makes all $2k-1$ neighbors Wigner negative in the same run, with
\begin{align}
W_{Y_j}(0)\leq-\frac{1}{2\pi(2k-1)^2}.\label{eq:Wbound}
\end{align}
The heralded state of every neighbor does not depend on the squeezing.
\end{theorem}

The proof is given in Appendix~\ref{app:sat}, and the mechanism is transparent. Projecting the node's halves of the two-mode squeezed vacua onto $\ket{\Phi_k}$ prepares the partner modes in the same state with $\hat a_l$ replaced by $\hat b_l$, whatever the squeezing is, because the two-mode squeezed vacua are perfectly correlated in photon number. The network is therefore a heralded source of this state followed by a passive splitter. Neighbor $Y_j$ obtains the collective partner mode $\hat\beta_j$, whose photon statistics are those of $\hat g_j$ in $\ket{\Phi_k}$, through a loss channel of transmission $\eta$. The reduced state of one mode of a state with a fixed total photon number is diagonal in the Fock basis. At the origin the Wigner function of a single-mode state equals $1/(2\pi)$ times its photon-number parity, and a loss channel of transmission $\eta$ maps the parity $(-1)^n$ of an $n$-photon state to
\begin{align}
\sum_{l=0}^n\binom nl\eta^l(1-\eta)^{n-l}(-1)^l&=(1-2\eta)^n=(-\varrho)^n,\nonumber\\
\varrho&=\frac{1}{2k-1}.\label{eq:parity}
\end{align}
If $p_n$ denotes the probability of $n$ photons in the mode $\hat g_j$ of $\ket{\Phi_k}$, then
\begin{align}
W_{Y_j}(0)=\frac{X_k}{2\pi},\qquad X_k=\sum_np_n(-\varrho)^n.\label{eq:Xmain}
\end{align}
The factor $\hat g_j^\dagger$ in Eq.~\eqref{eq:Phi} guarantees $p_0=0$. The tight-frame identity~\eqref{eq:frame} together with the cyclic symmetry of the harmonic frame gives every steered mode the mean photon number $2-1/k$, and together with $p_0=0$ this forces $p_1\geq1/k$ (Lemma~\ref{lem:stats}). Since $|(-\varrho)^n|\leq\varrho^2$ for $n\geq2$, the one-photon term dominates,
\begin{align}
X_k&\leq-\varrho\,p_1+\varrho^2(1-p_1)\nonumber\\
&\leq-\frac{\varrho}{k}+\varrho^2\left(1-\frac1k\right)=-\varrho^2,\label{eq:Xbound}
\end{align}
where the last step uses $1-(k-1)\varrho=k\varrho$.

\begin{figure*}[t]
\centering
\includegraphics[width=\textwidth]{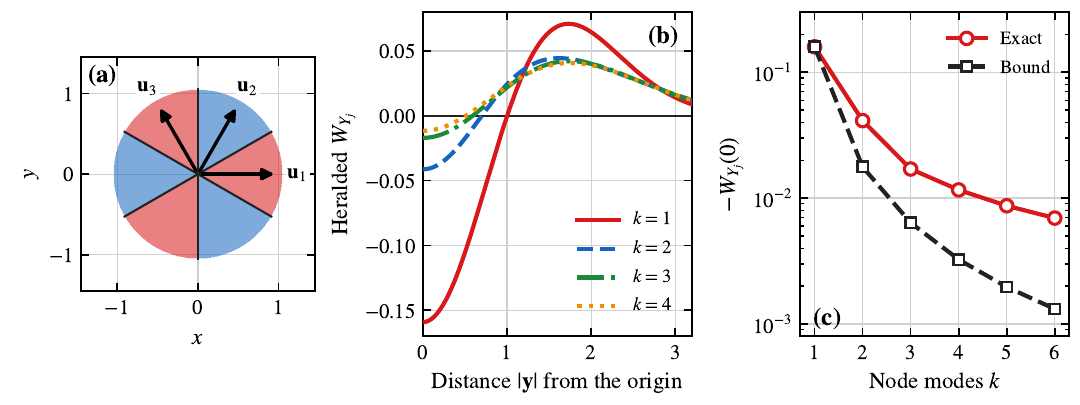}
\caption{\justifying\small Remote Wigner negativity in the tight-frame network. (a) Sign pattern of the cubic polynomial $\prod_j(\bvec u_j\cdot\bvec x)=-(x^3-3xy^2)/4$ in the plane $\bvec x=(x,y)$, with negative values in red and positive values in blue. Replacing $x$ and $y$ by $\hat a_1^\dagger$ and $\hat a_2^\dagger$ gives the state $\ket{\Phi_2}$. The black arrows are the frame vectors $\bvec u_j$ of the steered modes $\hat g_j$, and the polynomial vanishes on the black lines perpendicular to them. A rotation by $60^\circ$ maps one arrow to the next and only flips the sign, so the state looks the same from every neighbor. (b) Heralded Wigner function of every neighbor as a function of the distance $|\bvec y|$ from the origin, after the projection onto $\ket{\Phi_k}$, for $k=1$ (red solid line), $k=2$ (blue dashed line), $k=3$ (green dash-dotted line) and $k=4$ (orange dotted line), with $2k-1$ neighbors and any squeezing. The black horizontal line marks zero. (c) Value $-W_{Y_j}(0)$ at the origin on a logarithmic scale, exact (red open circles) and the bound~\eqref{eq:Wbound} (black open squares), for $k=1,\dots,6$.}
\label{fig:wigner}
\end{figure*}

For $k=2$ the picture is fully explicit. The steered modes are $\hat g_j=\cos\theta_j\hat a_1+\sin\theta_j\hat a_2$ with $\theta_j=0,\pi/3,2\pi/3$, up to irrelevant signs, and in terms of the circular modes $\hat a_\pm=(\hat a_1\pm i\hat a_2)/\sqrt2$ the projected state is the three-photon NOON state
\begin{align}
\ket{\Phi_2}&=\frac{1}{\sqrt{12}}\left(\hat a_+^{\dagger3}+\hat a_-^{\dagger3}\right)\ket0\nonumber\\
&=\frac12\ket{3,0}-\frac{\sqrt3}{2}\ket{1,2},\label{eq:noon}
\end{align}
up to a global sign. The modes $\hat a_\pm$ acquire the phases $e^{\pm i\theta}$ under a rotation of the node modes by the angle $\theta$, so the rotation by $60^\circ$ that maps one steered mode to the next only changes the sign of $\ket{\Phi_2}$, and the state looks the same from every neighbor. Figure~\ref{fig:wigner}(a) shows this symmetry through the cubic polynomial $\prod_j(\bvec u_j\cdot\bvec x)=-(x^3-3xy^2)/4$, which gives $\ket{\Phi_2}$ when $x$ and $y$ are replaced by $\hat a_1^\dagger$ and $\hat a_2^\dagger$. In the frame of every neighbor, $\ket{\Phi_2}$ has one photon in the steered mode with probability $3/4$ and three photons with probability $1/4$, never an even number, and every neighbor obtains
\begin{align}
W_{Y_j}(0)&=\frac{1}{2\pi}\left[\frac34\left(-\frac13\right)+\frac14\left(-\frac13\right)^3\right]\nonumber\\
&=-\frac{7}{54\pi}\approx-0.041,\label{eq:three}
\end{align}
about a quarter of the value $-1/(2\pi)\approx-0.159$ of a single photon. For $k=3,\dots,6$ the exact values are $W_{Y_j}(0)=-0.0170$, $-0.0116$, $-0.0087$ and $-0.0069$. The bound~\eqref{eq:Wbound} is not tight, since $p_1$ equals $0.583$, $0.543$, $0.518$ and $0.504$ for $k=3,\dots,6$, well above $1/k$. Figure~\ref{fig:wigner}(b) shows the heralded Wigner functions as functions of the distance from the origin, and Fig.~\ref{fig:wigner}(c) compares the exact values at the origin with the bound.

It is worth stating what is and is not distributed here. No non-Gaussian state is copied or sent to the neighbors. The heralded event at the node creates negativity at every neighbor anew, through the Gaussian correlations that each neighbor already shares with the node, and what the node distributes is the information that the event occurred. The network also shows that the negativity cannot be produced mode by mode. For $k=2$ the neighbors at $60^\circ$ and $120^\circ$ steer neither physical mode $\hat a_1$ nor $\hat a_2$ on its own, and for $k\geq3$ no neighbor steers any single mode $\hat a_l$, for any squeezing (Appendix~\ref{app:sat}). By Proposition~\ref{prop:onereceiver}(i), no operation on one physical mode of the node can make these neighbors negative. The optimal neighbors draw on overlapping collective modes, an instance of collective steering~\cite{He2013,Armstrong2015,Sun2026}, and only an operation on the whole node, not photon subtraction on a single mode, can use all of them. Collective non-Gaussian operations have been studied before, for instance photon subtraction from collective modes of a fully connected three-mode squeezed vacuum~\cite{Sun2026}. The projection onto $\ket{\Phi_k}$ differs in that it places one photon in the steered mode of every neighbor at once, which is what allows one heralded event to activate the maximal number $2k-1$ of neighbors.

\subsection{Linear optics, rates and imperfections}\label{sec:rates}

Theorem~\ref{thm:saturation} is a statement about an ideal projection. We now show how the projection can be implemented with linear optics, how often it succeeds, and how robust the negativity is against detector inefficiency and loss. The projection onto $\ket{\Phi_k}$ requires no new resource. The node sends its $k$ modes, together with $k-1$ vacuum modes, through the same passive network that distributes the partner modes, and keeps the runs in which photon-number-resolving detectors find exactly one photon in each of the $2k-1$ outputs. Restricted to vacuum ancillas, the creation operator of output $j$ becomes $\sqrt\eta\,\hat g_j^\dagger$, so the effect of this event is proportional to $\ketbra{\Phi_k}{\Phi_k}$, with the prefactor $\eta^{2k-1}\mathcal N_k$, which equals $4/9$ for $k=2$ and $0.195$ for $k=3$. The normalization constants themselves are $\mathcal N_k=3/2$, $2.510$, $4.245$, $7.243$ and $11.78$ for $k=2,\dots,6$, computed from the Fock expansion of Eq.~\eqref{eq:Phi}~\cite{Github_code}. The heralded states of the neighbors are those of Theorem~\ref{thm:saturation}.

The squeezing sets only the rate. The ideal projection succeeds with probability $(1-q_v)^kq_v^{2k-1}$, the weight of $\ket{\Phi_k}$ in the thermal reduced state of the node. It is largest at $q_v=(2k-1)/(3k-1)$, where it equals $3.5\%$ for $k=2$, at about $9\,$dB of squeezing, and $0.5\%$ for $k=3$. With the prefactor of the linear-optical realization these become $1.5\%$ and $0.1\%$. The rate decreases exponentially with $k$, and the projection involves $2k-1$ photons, so the network proves tightness rather than offering a fast source.

Photon-number-resolving detectors of efficiency $\eta_s$ act as a loss channel of transmission $\eta_s$ on the node's modes before an ideal projection, because uniform loss commutes with the passive network. This maps the symplectic eigenvalues of the conditional covariance matrix to $\eta_sd+1-\eta_s<1$ and $\eta_sv+1-\eta_s$, the ratio $q_d/q_v$ stays negative and becomes smaller in modulus, and all neighbors remain Wigner negative for every $\eta_s>0$. For $k=2$ at the optimal squeezing, $W_{Y_j}(0)=-0.035$ for $\eta_s=0.9$ and $-0.024$ for $\eta_s=0.7$. Dark counts can produce false heralds. A false herald requires a dark count in at least one of the $2k-1$ detectors, so for dark-count probabilities $p_{\rm dc}$ per detection window it occurs at most at a rate of order $(2k-1)p_{\rm dc}$. Superconducting nanowire detectors have dark-count rates of order $1$ to $100\,$Hz~\cite{EsmaeilZadeh2021}, which for detection windows of a few nanoseconds gives $p_{\rm dc}\sim10^{-7}$. Even for the conservative value $p_{\rm dc}=10^{-5}$ and $k=2$, the false-herald rate is below $10^{-4}$, compared with a success rate of $1.5\%$, so false heralds make up less than one percent of the heralded runs and change $W_{Y_j}(0)$ by less than $1/\pi$ times this fraction, a correction below $3\times10^{-3}$, small compared with $|W_{Y_j}(0)|\approx0.04$. An additional loss of transmission $\lambda$ at a neighbor changes its share to $\lambda k/(2k-1)$, in line with Proposition~\ref{prop:mono}, and the parity factor of Eq.~\eqref{eq:parity} to $(1-2\lambda\eta)^n$. For $\lambda\leq1-1/(2k)$ the neighbor no longer steers the node and cannot be negative at all. For larger $\lambda$ it remains negative at the origin. Then $0<2\lambda\eta-1\leq\varrho$, and the first line of Eq.~\eqref{eq:Xbound} with $\varrho$ replaced by $\varrho'=2\lambda\eta-1$ reads $-\varrho'[p_1-\varrho'(1-p_1)]$, which is negative because, for $k\geq2$, $p_1\geq1/k$ gives $p_1/(1-p_1)\geq1/(k-1)>\varrho\geq\varrho'$. For $k=1$ we have $p_1=1$, and the expression is negative at once. The tolerable additional loss is $25\%$, $17\%$ and $10\%$ for $k=2$, $3$ and $5$, the same margin as for steering and CHSH violation. Three different quantities thus characterize the construction. The number of negative neighbors, $2k-1$, is the maximum allowed by Corollary~\ref{cor:negativity}. The negativity of each neighbor at the origin is at least $1/(2\pi(2k-1)^2)$ and decreases with $k$. The success probability decreases exponentially with $k$. The theorem reaches the maximum of the first quantity, and the second and third show that it is a statement of principle for large $k$ and a practical target only for small $k$. The heralded states can be certified by homodyne tomography~\cite{Lvovsky2001,Lvovsky2009} or by moment-based witnesses of Wigner negativity~\cite{Mallick2025,Chakrabarty2026Operational}.

\section{Arbitrary measurements and Bell nonlocality}\label{sec:count}
%=====================================================================

The counts so far assume that the neighbors measure quadratures. This assumption is natural for steering, but it hides two possibilities. Gaussian measurements can never reveal Bell nonlocality of a Gaussian state, and non-Gaussian measurements can reveal steering that Gaussian measurements miss~\cite{Ji2016}. A neighbor with a photon counter might therefore be served where a neighbor with a homodyne detector is not, and device-independent tasks need Bell nonlocality in the first place. In this section we allow arbitrary measurements of both parties and fix only the number of settings of the neighbors. The share still decides the count, because its third threshold, extendibility, is exactly what forbids nonlocality with a given number of settings. We show that at most $nk-1$ neighbors can be nonlocal with the node with $n$ settings, that the count $2k-1$ is reached for two settings by a CHSH violation of every neighbor of the tight-frame network, and that the count $3k-1$ is reached for three-setting steering in every case that we have certified.

\subsection{Steering and Bell tests with any measurement}\label{sec:defs}

We first recall what it means for a neighbor to steer the node, or to be Bell nonlocal with it, when the measurements are arbitrary. In a Bell test between the node and neighbor $Y_j$, the node chooses a setting $x$ and obtains an outcome $a$, and the neighbor chooses a setting $y$ and obtains an outcome $b$. Quantum theory predicts
\begin{align}
p(a,b|x,y)=\tr\left[\rho_{AY_j}\left(A_{a|x}\otimes B_{b|y}\right)\right],\label{eq:pab}
\end{align}
where $\{A_{a|x}\}_a$ and $\{B_{b|y}\}_b$ are positive operator-valued measures for each setting. The probabilities admit a local hidden-variable model if there are a distribution $q(\lambda)$ and response functions $q(a|x,\lambda)$ and $q(b|y,\lambda)$ such that
\begin{align}
p(a,b|x,y)=\sum_\lambda q(\lambda)\,q(a|x,\lambda)\,q(b|y,\lambda).\label{eq:lhvdef}
\end{align}
In a steering test from $Y_j$ to the node, only the neighbor measures, and the node's conditional states form the assemblage
\begin{align}
\sigma_{b|y}=\tr_{Y_j}\left[\rho_{AY_j}\left(\iden\otimes B_{b|y}\right)\right].\label{eq:assemblage}
\end{align}
It admits a local hidden-state model if there are positive operators $\sigma_\lambda$ on the node and response functions $q(b|y,\lambda)$ with $\sigma_{b|y}=\sum_\lambda q(b|y,\lambda)\sigma_\lambda$~\cite{Wiseman2007,Uola2020}.

Whether a model of either kind exists depends on the number of settings of the neighbor, since more settings give the neighbor more ways to reveal nonclassical correlations. Our counts depend on this number, and the following definition makes it explicit.

\begin{definition}\label{def:nl}
Neighbor $Y_j$ is \emph{$n$-setting Bell nonlocal} with the node if there are measurements of the node, with any number of settings and outcomes, and measurements of $Y_j$ with $n$ settings and any number of outcomes, such that the probabilities~\eqref{eq:pab} admit no local hidden-variable model. It \emph{$n$-setting steers} the node if there are measurements of $Y_j$ with $n$ settings such that the assemblage~\eqref{eq:assemblage} admits no local hidden-state model.
\end{definition}

A Gaussian state measured with Gaussian measurements admits a local hidden-variable model, since its Wigner function is a nonnegative distribution that reproduces all outcomes~\cite{Banaszek1998}. Bell nonlocality of Gaussian states therefore requires non-Gaussian measurements, such as displaced parity or photon counting~\cite{Banaszek1998,Banaszek1999,Jeong2003}. Definition~\ref{def:nl} allows arbitrary measurements and fixes only the number of settings of the neighbor.

The link between these operational notions and the share is extendibility. If the state of the node and a neighbor is the reduced state of a larger state with $n$ copies of the neighbor, each of the $n$ settings can be measured on its own copy. The outcomes of all settings then exist at once, which is exactly a classical model. The following lemma makes this precise. It is the extendibility argument of Ref.~\cite{Terhal2003}, extended in the obvious way to steering, and we include the short proof for completeness.

\begin{lemma}[Extendibility excludes nonlocality]\label{lem:tds}
If $\rho_{AY_j}$ is $n$-extendible on $Y_j$, then every Bell test in which $Y_j$ uses $n$ settings admits a local hidden-variable model, and every assemblage produced by $n$ settings of $Y_j$ admits a local hidden-state model, whatever the measurements are.
\end{lemma}

\begin{proof}
Let $\sigma$ be a state of the node and copies $Y_j^{(1)},\dots,Y_j^{(n)}$ whose reduction to the node and any single copy equals $\rho_{AY_j}$. Imagine that setting $y$ is measured on copy $Y_j^{(y)}$, for all $y=1,\dots,n$ at once, and let $\bvec b=(b_1,\dots,b_n)$ collect the outcomes. Define the positive operators on the node
\begin{align}
\sigma_{\bvec b}=\tr_{Y^{(1)}\cdots Y^{(n)}}\left[\sigma\left(\iden\otimes B_{b_1|1}\otimes\cdots\otimes B_{b_n|n}\right)\right].\label{eq:sigmab}
\end{align}
Summing over all $b_{y'}$ with $y'\neq y$ replaces the corresponding measurement operators by the identity, so the marginal property of $\sigma$ gives
\begin{align}
\sigma_{b|y}=\sum_{\bvec b}\delta_{b,b_y}\,\sigma_{\bvec b}.\label{eq:lhs}
\end{align}
This is a local hidden-state model with hidden variable $\bvec b$, hidden states $\sigma_{\bvec b}$ and deterministic responses $q(b|y,\bvec b)=\delta_{b,b_y}$. For the Bell test, measuring setting $x$ on the node gives
\begin{align}
p(a,b|x,y)=\sum_{\bvec b}\tr\left(A_{a|x}\sigma_{\bvec b}\right)\delta_{b,b_y},\label{eq:lhv}
\end{align}
which has the form~\eqref{eq:lhvdef} with $q(\bvec b)=\tr\sigma_{\bvec b}$ and $q(a|x,\bvec b)=\tr(A_{a|x}\sigma_{\bvec b})/q(\bvec b)$, a valid response function because $\sum_aA_{a|x}=\iden$. Both models exist for all measurements, which completes the proof.
\end{proof}

\subsection{The counts}\label{sec:counts}

Lemma~\ref{lem:tds} says that a neighbor that is $n$-setting nonlocal with the node, or $n$-setting steers it, cannot hold an $n$-extendible state, and Proposition~\ref{prop:hier}(c) translates this into $\zeta_j>1/n$. Every such neighbor therefore needs more than $1/n$ of a unit, and the share bound provides only $k$ units. This gives the following counts.

\begin{theorem}[Counts for $n$ settings]\label{thm:count}
In every Gaussian state of a node with $k$ modes and single-mode neighbors, at most $nk-1$ neighbors can $n$-setting steer the node, and at most $nk-1$ neighbors can be $n$-setting Bell nonlocal with it, for arbitrary measurements of both parties.
\end{theorem}

\begin{proof}
If $Y_j$ is $n$-setting Bell nonlocal with the node or $n$-setting steers it, then $\rho_{AY_j}$ is not $n$-extendible on $Y_j$ by Lemma~\ref{lem:tds}, so $\zeta_j>1/n$ by Proposition~\ref{prop:hier}(c). Let $N$ neighbors have $\zeta_j>1/n$. Summing over them and using Theorem~\ref{thm:budget},
\begin{align}
\frac{N}{n}<\sum_{j}\zeta_j\leq k,\label{eq:countproof}
\end{align}
so $N<nk$, which means $N\leq nk-1$. This completes the proof.
\end{proof}

For $n=2$ the count $2k-1$ coincides with that of Theorem~\ref{thm:gsteer}, now for arbitrary measurements. Two settings of any measurement therefore serve no more neighbors than Gaussian measurements with any number of settings, as anticipated in Sec.~\ref{sec:gcount}, and the count grows beyond $2k-1$ only for $n\geq3$. For $k=1$ and $n=2$ it also follows from strict single-mode monogamy together with the equivalence of Gaussian unsteerability and $2$-extendibility~\cite{Lami2019}. Three remarks clarify what the theorem does and does not say. First, the number of settings is fixed only at the neighbors, while the node may use any number of settings and outcomes, which makes the count stronger. It is an upper bound for every $n$, but we do not know in general whether it is attained, and the tightness results below depend on the number of settings and on the construction. Second, the counts hold for every measurement, but they depend on fixing the number of settings of the neighbors. Non-Gaussian measurements are necessary for any Bell violation of a Gaussian state, and they cannot raise the number of two-setting nonlocal neighbors above $2k-1$, but with an unbounded number of settings the theorem gives no finite count. Third, any steering of the node that non-Gaussian measurements reveal and Gaussian measurements miss~\cite{Ji2016} must use at least three settings at the neighbor. A neighbor that cannot steer the node by Gaussian measurements has $\zeta_j\leq1/2$, so by Proposition~\ref{prop:hier}(c) and Lemma~\ref{lem:tds} it cannot steer it with two settings of any kind.

\subsection{Two settings, a CHSH violation for every neighbor}\label{sec:tight2}

An upper bound that holds for arbitrary measurements is only as strong as the best measurements that reach it. For two settings the tight-frame network provides the answer. Its neighbors already steer the node by Gaussian measurements (Theorem~\ref{thm:gsteer}), but Gaussian measurements can never violate a Bell inequality with a Gaussian state, so the Bell tests need non-Gaussian observables. We use dichotomic observables that act as a qubit observable on the vacuum and one-photon states of a mode and as the identity on all higher photon numbers,
\begin{align}
O_\theta=\left(\cos\theta\,\sigma_z+\sin\theta\,\sigma_x\right)\oplus\iden_{n\geq2},\label{eq:obs}
\end{align}
where $\sigma_z\ket0=\ket0$, $\sigma_z\ket1=-\ket1$, and $\sigma_x$ exchanges $\ket0$ and $\ket1$. These are legitimate observables with outcomes $\pm1$. They serve to prove that the bound is mathematically attainable, and we discuss experimentally more accessible measurements afterwards. With these observables every neighbor of the network violates the CHSH inequality, as the following theorem states.

\begin{theorem}[CHSH violation by every neighbor]\label{thm:tight}
In the tight-frame network with $\tanh^2r=1/[8(2k-1)]$, every neighbor $Y_j$ violates the CHSH inequality with the node. The node measures $O_\theta$ on the collective mode $\hat g_j$ with $\theta\in\{0,\pi/2\}$, the neighbor measures $O_\phi$ with $\phi\in\{\phi_0,-\phi_0\}$, where $\phi_0$ is given in Eq.~\eqref{eq:phi0}, and the CHSH value satisfies
\begin{align}
\mathcal B_j\geq2+\frac{1}{8(2k-1)^2}.\label{eq:chshk}
\end{align}
Consequently, all $2k-1$ neighbors are two-setting Bell nonlocal with the node and two-setting steer it, and the count of Theorem~\ref{thm:count} for $n=2$ is attained.
\end{theorem}

\begin{proof}
The CHSH bound~\eqref{eq:chshk} is proven in Appendix~\ref{app:chsh}. A CHSH violation with two settings of the neighbor implies that the neighbor two-setting steers the node, because a local hidden-state model for the neighbor's two settings would, combined with any measurements of the node on the hidden states, give a local hidden-variable model for the CHSH test~\cite{Wiseman2007}. Hence all $2k-1$ neighbors are two-setting nonlocal and two-setting steer the node, which completes the proof.
\end{proof}

The CHSH part rests on a statement about a single two-mode squeezed vacuum whose second mode passes through a loss channel of transmission $\eta$, which by Lemma~\ref{lem:network} is the reduced state of every pair. For weak squeezing $\lambda=\tanh r$, this state is close to the two-qubit state $\ket{0,0}+\lambda\sqrt\eta\ket{1,1}$, mixed with a small admixture $\lambda^2(1-\eta)$ of $\ket{1,0}$ in which the neighbor's photon was lost. With the observables~\eqref{eq:obs}, the correlations are those of this effective two-qubit state up to corrections of order $\lambda^4$. The correlation along $z$ is $1+\lambda^2(2\eta-1)$ and the correlation along $x$ is $2\lambda\sqrt\eta$, up to normalization, and the CHSH value exceeds $2$ at order $\lambda^2$ by an amount proportional to $2\eta-1$. Appendix~\ref{app:chsh} turns this into the bound
\begin{align}
\mathcal B\geq2+\frac{(2\eta-1)^2}{8}\qquad\text{at}\quad\lambda^2=\frac{2\eta-1}{8},\label{eq:chshbound}
\end{align}
valid for every $\eta\in(1/2,1]$. Equation~\eqref{eq:chshk} is the case $\eta=k/(2k-1)$, for which $2\eta-1=1/(2k-1)$. For this family of states, Proposition~\ref{prop:hier}(c) and Lemma~\ref{lem:tds} forbid any two-setting violation for $\eta\leq1/2$, so for the family of two-mode squeezed vacua seen through loss the condition $\zeta_j>1/2$ is necessary and sufficient for two-setting Bell nonlocality. Whether this holds for general Gaussian states is open (Sec.~\ref{sec:open}).

Since the bound~\eqref{eq:chshbound} holds for every $\eta>1/2$, the CHSH violation of all $2k-1$ neighbors survives an additional loss of transmission $\lambda_{\rm ex}>1-1/(2k)$ at every neighbor, the same margin as for Gaussian steering (Sec.~\ref{sec:gcount}), provided the squeezing is adjusted to the new transmission.

\begin{figure*}[t]
\centering
\includegraphics[width=\textwidth]{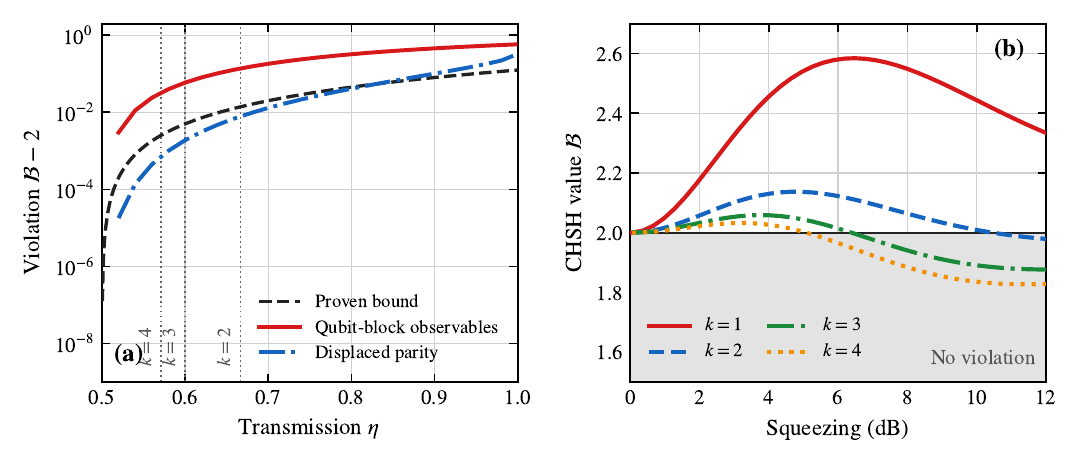}
\caption{\justifying\small CHSH violation. (a) Violation $\mathcal B-2$ for a two-mode squeezed vacuum whose second mode passes through a loss channel of transmission $\eta$, which is the reduced state of every pair in the tight-frame network with $\eta=k/(2k-1)$. The red solid line uses the qubit-block observables of Eq.~\eqref{eq:obs} and the blue dash-dotted line uses displaced parity, both numerically optimized over squeezing and settings on a grid of transmissions. The black dashed curve is the proven lower bound~\eqref{eq:chshbound} for the qubit-block observables. Gray dotted vertical lines mark $\eta=k/(2k-1)$ for $k=2,3,4$. No violation of any two-setting Bell inequality is possible for $\eta\leq1/2$. (b) CHSH value of every neighbor of the tight-frame network as a function of the squeezing, with the observables of Eq.~\eqref{eq:obs} and numerically optimized settings, for $k=1$ (red solid line), $k=2$ (blue dashed line), $k=3$ (green dash-dotted line) and $k=4$ (orange dotted line). In the gray region $\mathcal B\leq2$ there is no violation.}
\label{fig:bell}
\end{figure*}

Figure~\ref{fig:bell}(a) compares the proven bound~\eqref{eq:chshbound} with numerically optimized CHSH values~\cite{Github_code}. With the observables~\eqref{eq:obs} and optimized squeezing, the violation is several times larger than the bound. For $k=2$ the bound gives $\mathcal B_j\geq2.014$, the settings of Theorem~\ref{thm:tight} give $\mathcal B_j=2.046$, and optimizing the squeezing gives $\mathcal B_j=2.137$ at about $4.6\,$dB. Figure~\ref{fig:bell}(b) shows the CHSH value of the network neighbors as a function of the squeezing. The violation is largest at moderate squeezing, about $6.5$, $4.6$, $3.7$ and $3.1\,$dB for $k=1,2,3,4$, with maximal values $2.584$, $2.137$, $2.059$ and $2.032$. For these observables strong squeezing is detrimental, because it populates photon numbers on which they act trivially.

We stress the distinction between three statements. The upper bounds of Theorem~\ref{thm:count} hold for all measurements. Their attainability for two settings, Theorem~\ref{thm:tight}, is a mathematical statement that uses the idealized observables~\eqref{eq:obs}, which are not directly available in optical experiments. An experimentally practical scheme that reaches the same count with a comfortable margin is not established in this work. As a step in this direction, Fig.~\ref{fig:bell}(a) shows displaced-parity measurements~\cite{Banaszek1998,Banaszek1999}, whose expectation value in a Gaussian state is proportional to the Wigner function at the displaced point. Optimizing the squeezing and the four displacements numerically, we find a violation for every transmission above one half that we tested, down to $\eta=0.52$, but it is much smaller than for the observables~\eqref{eq:obs}, about $0.008$ at $\eta=2/3$ ($k=2$). These numerical results are illustrations and do not carry the analytic bound~\eqref{eq:chshbound}.

\subsection{Three and four settings}\label{sec:tight3}

With more settings the price of a neighbor drops from one half to $1/n$, and the count rises from $2k-1$ to $nk-1$. The network of Theorem~\ref{thm:tight} has only $2k-1$ neighbors and cannot reach this larger count. Whether other states attain it depends on specific states and measurements, and we study it numerically in this subsection~\cite{Github_code}. A natural candidate is again a frame network. For $m=nk-1$ neighbors, a harmonic tight frame of $m$ vectors in $\R^k$, constructed as in Eqs.~\eqref{eq:harmeven} and \eqref{eq:harmodd} with $\vartheta_j=2\pi j/m$, splits the $k$ partner modes equally, so that every pair is again a two-mode squeezed vacuum seen through loss, now with transmission
\begin{align}
\eta_n=\frac{k}{nk-1}>\frac1n.\label{eq:etan}
\end{align}
The count $nk-1$ is therefore attained if every two-mode squeezed vacuum seen through loss with $\eta>1/n$ can be $n$-setting nonlocal for a suitable squeezing. By Proposition~\ref{prop:hier}(c) and Lemma~\ref{lem:tds}, $\eta>1/n$ is necessary. We have tested numerically whether it is sufficient, for steering with $n=3$ and $n=4$ and for Bell nonlocality with $n=3$.

\emph{Method for steering.} The neighbor measures qubit-block observables of the form~\eqref{eq:obs}, extended to all directions on the Bloch sphere, along the three Pauli directions for $n=3$ and along the four directions of a regular tetrahedron for $n=4$. From the resulting assemblage of the node mode we compute the steering robustness~\cite{Piani2015,Cavalcanti2017}, defined as $\min\tr\sum_\lambda\sigma_\lambda-1$ over positive operators $\sigma_\lambda$, one for each deterministic assignment $\lambda$ of outcomes to the $n$ settings, subject to $\sum_\lambda\delta_{b,\lambda_y}\sigma_\lambda\geq\sigma_{b|y}$ for all $b$ and $y$. It is a semidefinite program, and it is positive exactly when the assemblage admits no local hidden-state model. We truncate the two-mode squeezed vacuum at three photons before the loss. This truncation is a projection of the node's mode onto at most three photons, and since it acts on the node only, it commutes with the loss on the neighbor's side. The truncated state is therefore obtained from the original state by a local filtering of the steered node alone, after which the neighbor's mode also contains at most three photons and no further truncation is needed. A filtering of the steered party maps local hidden-state models to local hidden-state models, so a positive robustness of the truncated assemblage certifies $n$-setting steering of the original state, with the same measurements of the neighbor. Raising the cutoff from three to five photons changes the robustness only in the fourth significant digit. We maximize the robustness over the squeezing.

\begin{figure*}[t]
\centering
\includegraphics[width=\textwidth]{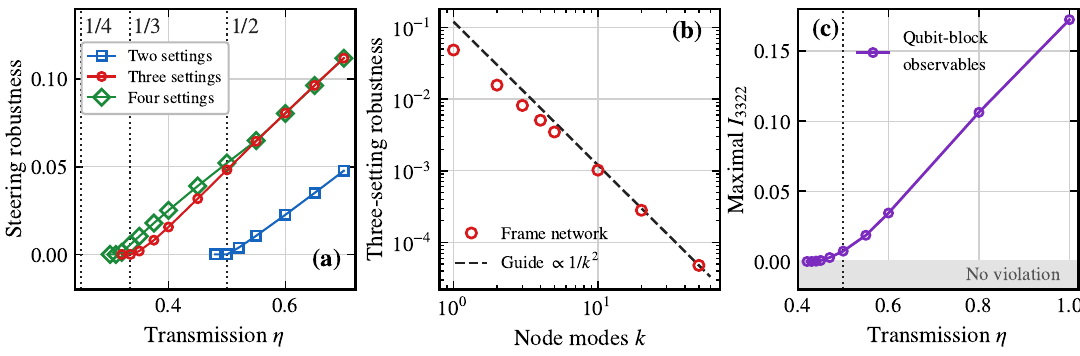}
\caption{\justifying\small Beyond two settings, for a two-mode squeezed vacuum whose second mode passes through a loss channel of transmission $\eta$, optimized over the squeezing. (a) Steering robustness of the node's assemblage when the neighbor measures qubit-block observables along two directions (blue open squares), the three Pauli directions (red open circles) or the four directions of a regular tetrahedron (green open diamonds). Black dotted vertical lines mark $\eta=1/4$, $1/3$ and $1/2$. Steering with $n$ settings is impossible for $\eta\leq1/n$ (Theorem~\ref{thm:count}). (b) Three-setting steering robustness at the transmission $\eta=k/(3k-1)$ of the frame network with $3k-1$ neighbors, for $k=1,\dots,5$, $10$, $20$ and $50$ (red open circles). The black dashed line is a guide to the eye proportional to $1/k^2$. The robustness is positive for every $k$ tested, and certified by a dual steering inequality, so all $3k-1$ neighbors steer the node. (c) Maximal value of the Collins-Gisin expression $I_{3322}$, whose local bound is $0$, with qubit-block observables for both parties (purple open circles). The gray region contains no violation, and the black dotted line marks $\eta=1/2$, below which no two-setting Bell inequality can be violated. The values near the threshold, which are too small to be seen on this scale, are given in Sec.~\ref{sec:tight3}.}
\label{fig:many}
\end{figure*}

\emph{Results for steering.} The results are shown in Fig.~\ref{fig:many}(a). For two settings the robustness vanishes, to solver precision, for $\eta\leq1/2$ and is positive above, as required by Theorem~\ref{thm:count}. For three settings it is positive for every transmission tested above $1/3$, down to $\eta=0.3335$, and it vanishes at $\eta=0.332$ and below. The threshold therefore lies between $0.332$ and $0.3335$, close to the bound $1/3$. The value at $\eta=0.3335$ is too small to be certified, so we do not claim that the threshold equals $1/3$. Figure~\ref{fig:many}(b) shows the robustness at the network transmissions $\eta_3=k/(3k-1)$ of Eq.~\eqref{eq:etan}. It is positive for all values of $k$ tested, $k=1,\dots,5$, $10$, $20$ and $50$, and it decreases roughly as $1/k^2$, from $0.048$ for $k=1$ and $0.016$ for $k=2$ to $5\times10^{-5}$ for $k=50$, with an optimal squeezing that decreases with $k$. For these values of $k$, all $3k-1$ neighbors of the frame network three-setting steer the node, so the count $3k-1$ of Theorem~\ref{thm:count} is attained for three-setting steering. The robustness is small and shrinks with $k$, so the effect is fragile. To make these values independent of the accuracy of the solver, we certified each of them. The dual variables of the semidefinite program define a steering inequality $\sum_{b,y}\tr(F_{b|y}\sigma_{b|y})\leq1$, which holds for every assemblage with a local hidden-state model whenever $F_{b|y}\geq0$ and $\sum_yF_{\lambda_y|y}\leq\iden$ for every assignment $\lambda$. We made the extracted $F_{b|y}$ exactly positive definite, rescaled them so that the second condition holds, and evaluated both conditions and the left side of the inequality in $50$-digit arithmetic. The programs, the certification procedure and the values for every $k$, together with the squeezing used, are listed in Appendix~\ref{app:sdp}. The certified violations agree with the robustness values to the digits quoted for $k\leq20$, and for $k=50$ the certified violation is $4.7\times10^{-5}$, compared with a robustness of $4.8\times10^{-5}$. For every tested $k$ the attainment of the count $3k-1$ is therefore a computer-assisted proof, although we have no proof for all $k$. For $k=1$ this means that two neighbors can each three-setting steer a single node mode, which is impossible with two settings or with Gaussian measurements. For four settings along a tetrahedron, the robustness vanishes below a transmission between $0.31$ and $0.32$, so these measurements do not reach the bound $1/4$. Qubit-block observables effectively act on a qubit, and measurements that use higher photon numbers may be needed for $n\geq4$.

\emph{Bell nonlocality with three settings.} For Bell tests a truncation of the Fock space would amount to postselection, so we use a cutoff of seven photons per mode, at which the truncation error is negligible for the squeezings considered. Both parties measure qubit-block observables along arbitrary directions, and we maximize the Collins-Gisin inequality $I_{3322}$~\cite{Collins2004}, whose local bound is $0$, over the six directions, over the assignment of the two roles in the inequality to the node and the neighbor, and over the squeezing. The results are shown in Fig.~\ref{fig:many}(c). The inequality is violated for every transmission down to $\eta=0.44$, where the maximal value is $2.3\times10^{-4}$, it is marginally violated at $\eta=0.43$, by $4\times10^{-6}$, and we find no violation at $\eta=0.42$. With three settings, Bell nonlocality therefore appears for transmissions between about $0.43$ and $1/2$, where two-setting Bell nonlocality is impossible, but with these observables and this inequality it does not reach the bound $1/3$. This already changes the counts. For $k=1$, a partner mode split equally between two neighbors gives each of them $\eta=1/2$, and both violate $I_{3322}$ with the single node mode, so the count $3k-1=2$ for three-setting Bell nonlocality is attained for $k=1$. For $k\geq2$, equal splitting among $m$ neighbors with $k/m\geq0.44$, for instance $m=4$ for $k=2$, gives more three-setting nonlocal neighbors than the two-setting count $2k-1$, while the count $3k-1$ remains out of reach of this construction.

%=====================================================================
\section{Teleportation and entanglement}\label{sec:tel}
%=====================================================================

Steering and Bell nonlocality are certified by tests, and remote Wigner negativity prepares states at the neighbors. A node in a network also has to receive quantum states from its neighbors, and teleportation is the standard way to send them. It gives a third task in which steering appears as a threshold. A neighbor that teleports an unknown coherent state into the node with a fidelity above the Gaussian no-cloning value shows that its link to the node outperforms any strategy in which a Gaussian cloner distributes copies, and such a fidelity requires steering~\cite{He2015}. In this section we bound the fidelity of every neighbor by its share and turn the bound into the same count $2k-1$. We then turn to entanglement, the weakest correlation of the hierarchy, and show that it is the one correlation that the share bound does not count.

\subsection{Protocol class and fidelity}\label{sec:telprot}

We count the neighbors that can use their link to teleport a coherent state into the node. Throughout this section we restrict to the following class of protocols, which contains the standard continuous-variable protocol~\cite{Braunstein1998,Furusawa1998}. All statements of this section refer to this class, and we do not claim bounds for more general strategies.

\emph{Step 1, local preparation.} The neighbor holds an unknown coherent state in an input mode with quadratures $(\hat x_{\rm in},\hat p_{\rm in})$. It may first apply a Gaussian unitary to its own mode, which turns its quadratures into $(\hat x_{Y}',\hat p_{Y}')=(\bvec s_1\cdot\hat{\bvec r}_{Y_j},\bvec s_2\cdot\hat{\bvec r}_{Y_j})$, where $\bvec s_1$ and $\bvec s_2$ are the rows of a symplectic matrix, so $\bvec s_1^TJ\bvec s_2=1$.

\emph{Step 2, measurement.} The neighbor mixes the input and its mode on a balanced beam splitter and measures the commuting combinations $\hat x_{\rm in}-\hat x_Y'$ and $\hat p_{\rm in}+\hat p_Y'$, up to a common factor $1/\sqrt2$ that we absorb. It sends the outcomes to the node.

\emph{Step 3, correction.} The node applies a Gaussian unitary to its modes and selects one mode, with quadratures $(\bvec u\cdot\hat{\bvec r}_A,\bvec u'\cdot\hat{\bvec r}_A)$ and $\bvec u^T\Omega_A\bvec u'=1$. It displaces this mode by the received outcomes with unit gain.

The output quadratures are then
\begin{align}
\hat x_{\rm out}&=\hat x_{\rm in}+\hat X,&\hat X&=\bvec u\cdot\hat{\bvec r}_A+\bvec t\cdot\hat{\bvec r}_{Y_j},\nonumber\\
\hat p_{\rm out}&=\hat p_{\rm in}+\hat P,&\hat P&=\bvec u'\cdot\hat{\bvec r}_A+\bvec t'\cdot\hat{\bvec r}_{Y_j},\label{eq:XP}
\end{align}
with $\bvec t=-\bvec s_1$ and $\bvec t'=\bvec s_2$, so that $\bvec t^TJ\bvec t'=-1$. The noise operators $\hat X$ and $\hat P$ commute, since $[\hat X,\hat P]=2i(\bvec u^T\Omega_A\bvec u'+\bvec t^TJ\bvec t')=0$, and they are independent of the input. Conversely, every pair of vectors with $\bvec u^T\Omega_A\bvec u'=1$ and $\bvec t^TJ\bvec t'=-1$ arises from some choice of the two local unitaries, because every canonical pair of quadratures can be completed to a symplectic basis. We call such pairs admissible. Let $\Xi_j$ be the $2\times2$ covariance matrix of $(\hat X,\hat P)$ in the joint state. Averaged over the outcomes, the protocol adds Gaussian classical noise with covariance $\Xi_j$ to the input~\cite{Braunstein1998,Serafini2017}. A coherent state has covariance $\iden$, so the output has covariance $\iden+\Xi_j$ and the same mean as the input. The fidelity of two single-mode Gaussian states with equal means, one of them pure, is $2/\sqrt{\det(V_1+V_2)}$~\cite{Serafini2017}, which gives
\begin{align}
F=\frac{2}{\sqrt{\det(2\iden+\Xi_j)}},\label{eq:F}
\end{align}
the same for every coherent input. We write $F_j$ for the supremum of $F$ over all protocols of this class.

The restriction to unit gain is not a loss of generality for the task we consider. The benchmarks below refer to the teleportation of all coherent states, without prior knowledge of the amplitude~\cite{Braunstein2000}. If the node instead displaces its mode by a linear function $G$ of the outcomes, the output has the mean $G\bvec\alpha$ for an input of mean $\bvec\alpha$, and its fidelity with the input is
\begin{align}
F=\frac{2\exp\left[-\tfrac12\bvec\delta^T(\iden+V_{\rm out})^{-1}\bvec\delta\right]}{\sqrt{\det(\iden+V_{\rm out})}},\quad\bvec\delta=(G-\iden)\bvec\alpha,\label{eq:Fgain}
\end{align}
where $V_{\rm out}$ is the output covariance matrix, which does not depend on $\bvec\alpha$~\cite{Serafini2017}. Unless $G=\iden$, this fidelity tends to zero for large amplitudes. Unit gain is therefore necessary for a protocol whose fidelity does not vanish on the set of all coherent states. What remains specific to our class is that the neighbor measures a pair of homodyne quadratures after a Gaussian unitary and that the node corrects by displacing one of its modes. Without entanglement the fidelity cannot exceed $1/2$~\cite{Braunstein2000}. The value $2/3$ is the fidelity of the optimal Gaussian $1\to2$ cloner of coherent states~\cite{Cerf2000,Grosshans2001}, and we call it the Gaussian no-cloning value. A fidelity above it guarantees that no other party holds a better copy produced by a Gaussian cloner. Non-Gaussian cloners reach about $0.6826$~\cite{Cerf2005}. All counts below concern the threshold $2/3$ and therefore also hold for the higher threshold $0.6826$.

\subsection{Bound and count}\label{sec:telbound}

By Eq.~\eqref{eq:F}, the fidelity depends on the joint state only through the noise matrix $\Xi_j$, and the noise operators $\hat X$ and $\hat P$ combine a quadrature pair of the node with a quadrature pair of the neighbor. This is the same kind of combination that appears in the variational form of Proposition~\ref{prop:var}, which therefore bounds the noise from below by the share. The result is the following bound.

\begin{theorem}[Teleportation]\label{thm:tel}
Every protocol of the class of Sec.~\ref{sec:telprot} from neighbor $Y_j$ into the node has fidelity
\begin{align}
F\leq\frac{1}{2-\zeta_j}.\label{eq:telbound}
\end{align}
\end{theorem}

The proof, given in Appendix~\ref{app:tel}, has three steps. The arithmetic-geometric mean inequality first reduces the fidelity~\eqref{eq:F} to the determinant of the noise matrix $\Xi_j$. A local symplectic transformation of the pair $(\hat X,\hat P)$ then reduces the determinant to a trace, and the trace is finally bounded from below by the variational form of Proposition~\ref{prop:var}, which brings in the share.

The bound of Theorem~\ref{thm:tel} concerns a single neighbor. Combined with the share bound it limits how many neighbors can teleport into the node with high fidelity, in the same way as for steering.

\begin{corollary}\label{cor:tel}
Within the protocol class of Sec.~\ref{sec:telprot}, a neighbor can teleport coherent states into the node with fidelity above the Gaussian no-cloning value $2/3$ only if $\zeta_j>1/2$. In every Gaussian state at most $2k-1$ neighbors can do so, and the optimal fidelities satisfy
\begin{align}
\sum_{j:\,F_j>1/2}\left(2-\frac{1}{F_j}\right)\leq k.\label{eq:telbudget}
\end{align}
\end{corollary}

\begin{proof}
Theorem~\ref{thm:tel} gives $2-1/F_j\leq\zeta_j$ for every neighbor. A fidelity above $2/3$ thus requires $\zeta_j>1/2$, and Theorem~\ref{thm:budget} gives the count as in Eq.~\eqref{eq:gcountproof}. For the neighbors with $F_j>1/2$ the left side of Eq.~\eqref{eq:telbudget} is a sum of positive numbers, each at most $\zeta_j^+$, so it is at most $k$ by Theorem~\ref{thm:budget}. This completes the proof.
\end{proof}

By Proposition~\ref{prop:hier}(b), a fidelity above $2/3$ in this class requires Gaussian steering of the node by the neighbor, in agreement with the known observation that teleportation beyond the no-cloning value requires steering~\cite{He2015}. Theorem~\ref{thm:tel} adds a quantitative bound in a quantity independent of the node's squeezing, and the share bound turns it into a count. The bound is also consistent with cloning. For $k=1$ and $m$ neighbors with equal shares $1/m$, Eq.~\eqref{eq:telbound} gives $F\leq m/(2m-1)$, which is the fidelity of the optimal Gaussian $1\to m$ cloner of coherent states~\cite{Cerf2000,Braunstein2001}. Since non-Gaussian cloners can do better~\cite{Cerf2005}, more general teleportation strategies, with non-unit gain, general Gaussian operations or non-Gaussian operations, may change counts of this type, and Theorem~\ref{thm:tel} should not be read as a statement about them. The direction also matters. We teleport from many neighbors into one multimode node, whereas teleportation from a single-mode sender to many neighbors is strictly monogamous~\cite{Lee2016}.

\subsection{Tightness}\label{sec:teltight}

It remains to show that the count of Corollary~\ref{cor:tel} is attained. The bound of Theorem~\ref{thm:tel} is reached by the standard protocol when the neighbor sees a partner mode through loss and the squeezing is chosen well, and this suffices for the tight-frame network.

\begin{proposition}\label{prop:teltight}
For a two-mode squeezed vacuum whose second mode passes through a loss channel of transmission $\eta<1$, the protocol with $\hat X=\hat x_A-\hat x_Y$ and $\hat P=\hat p_A+\hat p_Y$ reaches $F=1/(2-\eta)=1/(2-\zeta_j)$ at $\cosh2r=(1+\eta)/(1-\eta)$. In the tight-frame network with $k\geq2$ and $\cosh2r=(3k-1)/(k-1)$, every neighbor therefore teleports into the node with
\begin{align}
F=\frac{2k-1}{3k-2}>\frac23,\label{eq:Fk}
\end{align}
so all $2k-1$ neighbors beat the Gaussian no-cloning value, and the count of Corollary~\ref{cor:tel} is attained.
\end{proposition}

\begin{proof}
With $v=\cosh2r$ and $s=\sinh2r$, the covariance matrix of the pair has the blocks $V_A=v\iden$, $V_Y=w\iden$ with $w=\eta v+1-\eta$, and $V_{AY}=cZ$ with $c=\sqrt\eta\,s$. Hence $\langle\hat X^2\rangle=\langle\hat P^2\rangle=v+w-2c$, while $\langle\hat X\hat P+\hat P\hat X\rangle=0$. The noise matrix is $\Xi=n\iden$ with
\begin{align}
n=v(1+\eta)-2\sqrt{\eta(v^2-1)}+1-\eta.\label{eq:nlos}
\end{align}
Setting $dn/dv=0$ gives $v/\sqrt{v^2-1}=(1+\eta)/(2\sqrt\eta)$, whose solution is $v=(1+\eta)/(1-\eta)$, with $\sqrt{v^2-1}=2\sqrt\eta/(1-\eta)$. There
\begin{align}
n=\frac{(1+\eta)^2-4\eta}{1-\eta}+1-\eta=2(1-\eta),\label{eq:nmin}
\end{align}
and Eq.~\eqref{eq:F} gives $F=2/(4-2\eta)$. In the tight-frame network the node uses the mode $\hat g_j$, and by Lemma~\ref{lem:network} the pair is of this form with $\eta=k/(2k-1)$. Then $(1+\eta)/(1-\eta)=(3k-1)/(k-1)$ and $1/(2-\eta)=(2k-1)/(3k-2)$. Finally, $3(2k-1)=6k-3>6k-4=2(3k-2)$, so $F>2/3$ for every neighbor, which completes the proof.
\end{proof}

\begin{figure*}[t]
\centering
\includegraphics[width=\textwidth]{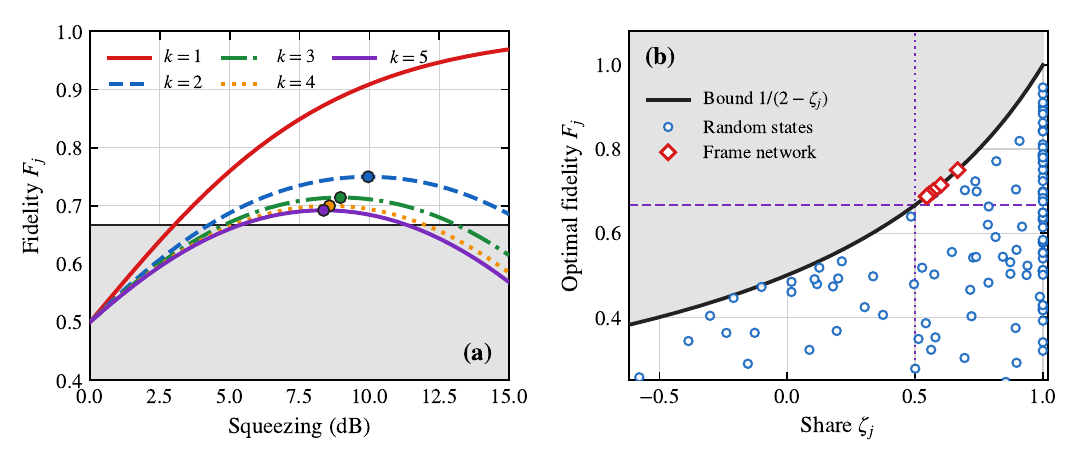}
\caption{\justifying\small Teleportation of coherent states into the node with the unit-gain protocols of Sec.~\ref{sec:telprot}. (a) Fidelity of every neighbor of the tight-frame network as a function of the squeezing, Eq.~\eqref{eq:nlos}, for $k=1$ (red solid line), $k=2$ (blue dashed line), $k=3$ (green dash-dotted line), $k=4$ (orange dotted line) and $k=5$ (purple solid line). The black horizontal line is the Gaussian no-cloning value $2/3$, and fidelities in the gray region below it do not beat it. Filled circles mark the optimal squeezing of Proposition~\ref{prop:teltight}, where the fidelity equals the bound $1/(2-\zeta_j)$. (b) Optimal fidelity within the protocol class versus share, for random Gaussian states (blue open circles) and for the neighbors of the tight-frame network at optimal squeezing for $k=2,\dots,6$ (red open diamonds). The black curve is the bound of Theorem~\ref{thm:tel}, and the gray region above it is excluded. The purple dashed line is the Gaussian no-cloning value $2/3$ and the purple dotted line marks $\zeta_j=1/2$.}
\label{fig:tel}
\end{figure*}

Figure~\ref{fig:tel}(a) shows the fidelity of the network neighbors as a function of the squeezing. For $k=2$ the three neighbors reach $F=3/4$ at about $10\,$dB, for $k=3$ the five neighbors reach $F=5/7$ at about $9\,$dB, and for $k=4$ and $5$ the optimal squeezings are about $8.6$ and $8.4\,$dB. For $k=1$ the single neighbor approaches $F=1$ as the squeezing grows. Unlike the Bell tests, which favor weak squeezing, teleportation needs strong squeezing, but not arbitrarily strong, because too much squeezing makes the node's mode much noisier than the attenuated mode of the neighbor. With an additional loss $\lambda_{\rm ex}$ at every neighbor, all $2k-1$ neighbors can still beat $2/3$ if and only if $\lambda_{\rm ex}>1-1/(2k)$, since Proposition~\ref{prop:teltight} applies with $\eta$ replaced by $\lambda_{\rm ex}k/(2k-1)$. Figure~\ref{fig:tel}(b) compares the bound with the optimal fidelity of random Gaussian states~\cite{Github_code}. For general channels the bound need not be reached within this protocol class, but the count $2k-1$ is.

\subsection{Entanglement reaches every neighbor}\label{sec:ent}

Every count so far came from a threshold on the share, one half for steering, Wigner negativity and teleportation and $1/n$ for $n$ settings. Entanglement has the threshold zero, and this changes the picture completely. The share bound does not limit the number of entangled neighbors, but it does limit how strongly entangled, in the sense of the share, many neighbors can be.

\begin{proposition}[Entanglement]\label{prop:ent}
For every $m$ there are Gaussian states of a node with $k$ modes and $m$ single-mode neighbors in which all neighbors are entangled with the node. For every $\epsilon>0$, at most $k/\epsilon$ neighbors have $\zeta_j\geq\epsilon$.
\end{proposition}

\begin{proof}
For $m\leq k$ give each neighbor one partner mode. For $m>k$ split the $k$ partner modes equally among the $m$ neighbors with a passive network, for instance with an equal-norm tight frame of $m$ vectors as in Sec.~\ref{sec:network}. By the argument of Lemma~\ref{lem:network}, each neighbor then sees a pure-loss channel of transmission $k/m$ from a collective partner mode, so $\zeta_j=k/m>0$, and all neighbors are entangled with the node by Proposition~\ref{prop:hier}(a). The second statement follows from Theorem~\ref{thm:budget}, because the shares of the neighbors in question add up to at least their number times $\epsilon$, and at most $k$. This completes the proof.
\end{proof}

The share bound therefore does not make entanglement monogamous in the counting sense. Neighbors with $0<\zeta_j\leq1/2$ are entangled with the node, but they can neither steer it by Gaussian measurements, nor steer it or violate a Bell inequality with it using two settings, nor be made Wigner negative by the node, and there can be arbitrarily many of them. What the share bound limits is the number of neighbors with a share above any fixed value, and with it every task whose threshold is a fixed share.

\begin{figure*}[t]
\centering
\includegraphics[width=\textwidth]{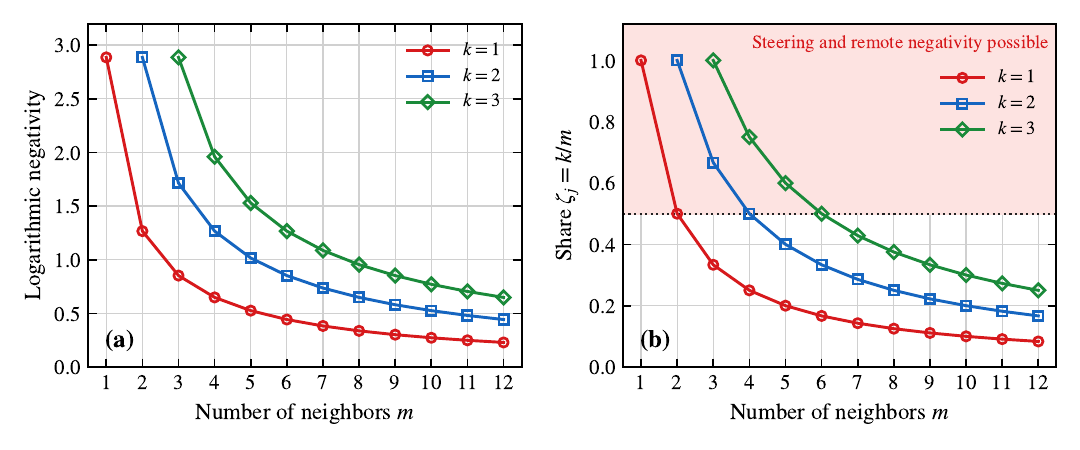}
\caption{\justifying\small A node with $k$ modes, each in a two-mode squeezed vacuum with squeezing $r=1$ ($8.7\,$dB), whose partner modes are split equally among $m$ neighbors, for $k=1$ (red open circles), $k=2$ (blue open squares) and $k=3$ (green open diamonds), connected by lines as a guide to the eye. (a) Logarithmic negativity between the node and each neighbor, Eq.~\eqref{eq:LN}. It stays positive for every $m$, as required by Proposition~\ref{prop:hier}(a) since $\zeta_j=k/m>0$. (b) Share $\zeta_j=k/m$ of each neighbor. Markers above the black dotted line $\zeta_j=1/2$ lie in the pink region, where Gaussian steering and remote Wigner negativity are possible and where two-setting steering, Bell nonlocality and teleportation beyond the Gaussian no-cloning value are allowed. For each $k$ the markers in this region end at $m=2k-1$.}
\label{fig:ent}
\end{figure*}

Figure~\ref{fig:ent} illustrates Proposition~\ref{prop:ent} and the counting for equal splitting. Every pair is then a two-mode squeezed vacuum followed by loss $\eta=k/m$, whose logarithmic negativity is~\cite{Adesso2007,Serafini2017}
\begin{align}
E_N=\max\left(0,-\log_2\tilde\nu_-\right),\label{eq:LN}
\end{align}
where $\tilde\nu_-=\big[v+w-\sqrt{(v-w)^2+4c^2}\big]/2$ is the smallest symplectic eigenvalue of the partially transposed covariance matrix, with $w=\eta v+1-\eta$ and $c=\sqrt\eta\,s$. The negativity decreases with $m$ but never vanishes [Fig.~\ref{fig:ent}(a)], while only $2k-1$ neighbors have shares above one half [Fig.~\ref{fig:ent}(b)].

Entanglement can thus reach every neighbor, while every stronger correlation of the hierarchy is limited to a number fixed by $k$. The next section places these counts among earlier results and collects them in one table.
%=====================================================================
\section{Discussion}\label{sec:disc}
%=====================================================================

All counts are now derived and shown to be reached. In this section we place them in context. We separate the known ingredients of this work from its new results, collect all counts in one table, discuss how a first experiment could test them, and close with open questions.

\subsection{What is known and what is new}\label{sec:related}

The results of this paper fall into three levels. The first is the share bound of Theorem~\ref{thm:budget}, together with its realizability and the monotonicity of the share. The second consists of its operational consequences, the exact counts, the single heralded projection that reaches the negativity count in one run, the criterion for simultaneous negativity and the teleportation bound. The third consists of known single-pair results, which we only rewrite in the variable $\zeta_j$. Table~\ref{tab:known} lists the known ingredients and the new results side by side. The single-pair criteria for separability~\cite{Simon2000,Werner2001}, Gaussian steering~\cite{Wiseman2007} and Gaussian extendibility~\cite{Lami2019}, the exclusion of Bell nonlocality by extendibility~\cite{Terhal2003}, the role of steering for remote Wigner negativity~\cite{Walschaers2020,WPT2020} and for teleportation beyond the no-cloning value~\cite{He2015}, and the fractional extendibility of a single pair~\cite{Ahmed2026} are known. In Propositions~\ref{prop:hier} and \ref{prop:onereceiver} we only rewrite them in the variable $\zeta_j$, and Theorem~\ref{thm:quasirep} expresses the conditional Wigner functions of Ref.~\cite{WPT2020} as expectation values of Hilbert-space operators. Strict monogamy of steering of a single mode~\cite{Reid2013,Ji2015} gives the case $k=1$ of our steering and Bell counts, and the monogamy of remotely generated negativity under photon subtraction~\cite{Xiang2022} gives a special case of Proposition~\ref{prop:singlemode}. That multimode steered parties escape strict monogamy is known theoretically~\cite{Ji2015,Lami2016,Xiang2017} and experimentally~\cite{Deng2017}. These works prove strict monogamy for a steered party with one mode, give counterexamples for steered parties with two or more modes, and bound sums of steering quantifiers, but none of them bounds the number of parties that can steer a party with $k\geq2$ modes.

What is new is the share bound of Theorem~\ref{thm:budget}, which constrains neighbors with different reduced states at once, together with the facts that every set of shares allowed by the bound occurs (Proposition~\ref{prop:realize}) and that the share does not increase under local Gaussian channels (Proposition~\ref{prop:mono}). From the share bound follow the counts for $k\geq2$ in Theorems~\ref{thm:gsteer} and \ref{thm:count}, Corollary~\ref{cor:negativity} and Corollary~\ref{cor:tel}. Their tightness for two settings is shown in Theorems~\ref{thm:gsteer} and \ref{thm:tight} and Proposition~\ref{prop:teltight}, numerically, with certified steering inequalities, for three-setting steering in Sec.~\ref{sec:tight3}, and in the strongest operational form by the single heralded projection of Theorem~\ref{thm:saturation}, which makes all $2k-1$ neighbors Wigner negative in the same run. The criterion of Theorem~\ref{thm:criterion} for activating several neighbors with one operation and the quantitative teleportation bound of Theorem~\ref{thm:tel} are new as well. The question of how far multimode nodes escape strict steering monogamy therefore has an exact answer. A node with $k$ modes can be steered by Gaussian measurements by up to $2k-1$ single-mode parties, and not more.

\begin{table*}[t]
\caption{\justifying\small Known ingredients and new results. The left part lists known results used in this work, the right part the new results.}
\label{tab:known}
\centering\small\setlength{\tabcolsep}{4pt}
\begin{tabular}{llll}
\hline\hline
Known result & Source & New result & Where\\
\hline
Separability of $1\times N$ Gaussian states & \cite{Simon2000,Werner2001} & Share bound $\sum_j\zeta_j\leq k$ & Thm.~\ref{thm:budget}\\
Gaussian steering criterion & \cite{Wiseman2007} & Monotonicity and realizability & Props.~\ref{prop:mono}, \ref{prop:realize}\\
Gaussian extendibility criterion & \cite{Lami2019} & Counts $2k-1$ and $nk-1$ for $k\geq2$ & Thms.~\ref{thm:gsteer}, \ref{thm:count}\\
Extendibility excludes Bell nonlocality & \cite{Terhal2003} & Tightness for two settings & Thms.~\ref{thm:gsteer}, \ref{thm:tight}\\
One mode is steered by at most one party & \cite{Reid2013,Ji2015} & Three-setting steering count, numerically & Sec.~\ref{sec:tight3}\\
Multimode parties escape strict monogamy & \cite{Lami2016,Xiang2017,Deng2017} & At most $2k-1$ negative neighbors & Cor.~\ref{cor:negativity}\\
Steering is needed for remote negativity & \cite{WPT2020} & One projection negates all $2k-1$ & Thm.~\ref{thm:saturation}\\
Steering suffices for one neighbor & \cite{Walschaers2020} & Criterion for simultaneous negativity & Thm.~\ref{thm:criterion}\\
One subtraction negates one remote mode & \cite{Xiang2022} & Teleportation bound $1/(2-\zeta_j)$ and count & Thm.~\ref{thm:tel}, Cor.~\ref{cor:tel}\\
Teleportation above $2/3$ requires steering & \cite{He2015} &  & \\
\hline\hline
\end{tabular}
\end{table*}

Monogamy relations for Gaussian steering and entanglement quantifiers~\cite{Kogias2015,Xiang2017,Lami2016,Hiroshima2007,Adesso2007} bound sums of nonlinear quantities and, apart from the single-mode case, do not count parties. The difference is visible in a simple example. For a neighbor that sees one partner mode through a loss channel of transmission $\eta>1/2$, the Gaussian steerability of Ref.~\cite{Kogias2015} never exceeds $\ln[\eta/(1-\eta)]$, which vanishes at the threshold. Neighbors just above one half add almost nothing to any sum of such measures, yet each of them uses more than half a unit of the total $k$. Extendibility constrains one pair at a time, and since every share in $[0,1]$ occurs for a single pair, no constraint on single pairs can give the share bound. Remote generation of Wigner negativity through steering has been studied for one neighbor~\cite{Walschaers2020,Liu2022}, in a general framework of conditional operations~\cite{WPT2020,Walschaers2023}, and for multipartite photon subtraction~\cite{Xiang2022}. Non-Gaussian resources cannot be broadcast with Gaussian operations~\cite{Chatterjee2026}, but our task is of a different kind, since no non-Gaussian state is copied and negativity is created anew at every neighbor. To our knowledge, the simultaneous activation of several neighbors by one operation, its limit $2k-1$ and a construction that reaches it have not been considered before.

\subsection{A hierarchy of counts}\label{sec:hiertab}

Table~\ref{tab:hier} and Fig.~\ref{fig:scheme}(b) collect the counts. The type of statement differs from row to row. For entanglement, Gaussian steering and the existence of an operation that makes one neighbor Wigner negative, the share gives necessary and sufficient conditions. For steering and Bell nonlocality with $n$ settings and arbitrary measurements, and for teleportation in the stated protocol class, it gives necessary conditions only. For a Gaussian state, a measured covariance matrix therefore decides entanglement and Gaussian steering of each neighbor exactly, it tells exactly which neighbors can be made Wigner negative by some heralded operation of the node, and for the other correlations it can only exclude them. The exclusions assume that the state is Gaussian. A positive share certifies entanglement for every state with the measured covariance matrix~\cite{Simon2000}, but a non-Gaussian state with the same covariance matrix may hold correlations that the share of a Gaussian state excludes.

\begin{table*}[t]
\caption{\justifying\small Correlations between a node of $k$ modes and single-mode neighbors in one Gaussian state. The counts for Gaussian steering and remote negativity hold for neighbors with any number of modes (Proposition~\ref{prop:multimode}). The condition on the share is necessary and sufficient (N\&S) for entanglement, Gaussian steering and the existence of a heralded operation that makes one neighbor Wigner negative, and necessary (N) otherwise. The fourth column gives the largest number of neighbors that can hold the correlation, and the last column states whether this number is attained, where $3^*$ means numerically, with certified steering inequalities, for all $k\leq50$ tested $3^\dagger$ means numerically for $k=1$ only, and $2^\ddagger$ means with the idealized observables of Eq.~\eqref{eq:obs}. Gaussian steering allows all Gaussian measurements, with any number of settings, and for $n=2$ the counts of the $n$-setting rows equal $2k-1$. Remote negativity refers to one heralded operation in one run, and teleportation to the unit-gain protocol class of Sec.~\ref{sec:telprot}.}
\label{tab:hier}
\centering\small\setlength{\tabcolsep}{8pt}
\begin{tabular}{lcccc}
\hline\hline
Correlation & Condition & Type & Count & Attained\\
\hline
Entanglement & $\zeta_j>0$ & N\&S & unbounded & yes\\
Gaussian steering & $\zeta_j>1/2$ & N\&S & $\leq2k-1$ & yes\\
Remote negativity & $\zeta_j>1/2$ & N\&S & $\leq2k-1$ & yes\\
$n$-setting steering & $\zeta_j>1/n$ & N & $\leq nk-1$ & $n=2,3^*$\\
$n$-setting Bell & $\zeta_j>1/n$ & N & $\leq nk-1$ & $n=2^\ddagger,3^\dagger$\\
Teleportation, $F>2/3$ & $\zeta_j>1/2$ & N & $\leq2k-1$ & yes\\
\hline\hline
\end{tabular}
\end{table*}

The number $2k-1$ appears so often because five different tasks have the same threshold, a share above one half. Gaussian steering with any number of settings, remote Wigner negativity, two-setting steering, two-setting Bell nonlocality and teleportation beyond the Gaussian no-cloning value all need it. Only non-Gaussian measurements with $n\geq3$ settings lower the threshold to $1/n$ and allow more neighbors. Read from top to bottom, the table shows one pattern. Monogamy in a multimode network is neither strict nor absent. It is an accounting rule, with a total set by the number of modes of the node and a price set by the threshold of each correlation. Entanglement costs nothing and reaches every neighbor. Steering by Gaussian measurements, remote Wigner negativity, two-setting nonlocality and teleportation beyond the no-cloning value each cost more than half a unit, and they are limited to $2k-1$ neighbors. Three settings lower the price to more than one third of a unit, and for steering the numerics of Sec.~\ref{sec:tight3} show that this lower price is attained for every $k$ we tested.

\subsection{Experimental perspective}\label{sec:exp}

The upper bounds and the conditions of Proposition~\ref{prop:hier} depend only on the covariance matrix, which can be measured by homodyne detection in the platforms mentioned in the introduction, so for a Gaussian state the exclusions of Table~\ref{tab:hier} can be checked without performing the non-Gaussian experiments. In practice this requires some confidence that the prepared state is Gaussian, which can be tested independently, for instance through higher moments of the homodyne data. The tightness constructions use two-mode squeezed vacua, beam splitters, homodyne detection with feedforward for teleportation, and photon-number-resolving detection for the heralded projection, all of which are standard~\cite{Furusawa1998,Yonezawa2004,Ourjoumtsev2006,NeergaardNielsen2006}. The heralded projection onto $\ket{\Phi_k}$ is the most demanding element, since its rate falls exponentially with $k$, but for $k=2$ it succeeds in about $1.5\%$ of the runs with linear optics, and the resulting negativity survives realistic detector efficiencies (Sec.~\ref{sec:rates}). The CHSH tests are demanding as well. The observables~\eqref{eq:obs} are not directly available in optics, and displaced parity gives a violation of only about $0.008$ for $k=2$, which falls to zero as the transmission approaches one half (Fig.~\ref{fig:bell}). An optical Bell test of the full count is therefore out of reach at present, and steering and remote Wigner negativity are the realistic first targets. Optimizing measurements based on photon counting with displacements~\cite{Banaszek1999} for the tight-frame network is a natural experimental question.

Since the shares are estimated from measured covariance matrices, their statistical uncertainty matters. The share is a smooth function of the covariance matrix wherever $V_A+i\Omega_A$ is invertible, so the uncertainty of an estimated share follows from the covariance of the estimated matrix elements by standard error propagation or by bootstrapping the homodyne data. The inverse becomes ill conditioned only when a node mode is nearly pure, and such a mode is nearly uncorrelated with every neighbor and can be dropped (Sec.~\ref{sec:setting}). A count is then certified when each share lies above its threshold by several standard deviations. In the $k=2$ example below, the shares $2/3$ exceed the steering threshold $1/2$ by $1/6$, a margin that standard homodyne statistics resolve easily.

A first experiment, a proof of principle for the central steering and negativity claims rather than a test of all results of this paper, could address the smallest nontrivial case, $k=2$, in which three neighbors are linked to a two-mode node. The two-mode squeezed vacua and the three-output splitter are standard, and homodyne measurement of the covariance matrix certifies the three shares $\zeta_j=2/3$ and hence Gaussian steering of the node by all three neighbors, which no single-mode node allows. The heralded projection onto the NOON state~\eqref{eq:noon} needs three photon-number-resolving detectors and about $9\,$dB of squeezing, and it succeeds in about $1.5\%$ of the runs (Sec.~\ref{sec:rates}). Each neighbor then holds a state with photon-number parity $2\pi W_{Y_j}(0)=-7/27\approx-0.26$, or about $-0.22$ at a detector efficiency of $90\%$. A parity measurement with a photon-number-resolving detector yields $\pm1$ in every shot, so about $2\times10^2$ heralded events per neighbor resolve this value at three standard deviations from zero, which requires about $10^4$ runs. Loss before the neighbors' detectors reduces the negativity as described in Sec.~\ref{sec:rates}. Displaced-parity or photon-counting tests of the CHSH inequality~\cite{Banaszek1999} are more demanding, since the violations near the threshold are small.

\subsection{Open questions}\label{sec:open}

Several questions remain open. First, for three settings the attainment of the steering count $3k-1$ is certified numerically for each tested $k$ up to $k=50$, but not for all $k$, and an analytic proof, perhaps through an explicit steering inequality for the effective two-qubit state of Sec.~\ref{sec:tight2}, would be desirable. For four or more settings we do not know whether the steering counts $nk-1$ are attained, and for three-setting Bell nonlocality our construction reaches the count $3k-1$ only for $k=1$, because the $I_{3322}$ violation stops near $\eta\approx0.43$. Measurements beyond qubit-block observables, or other three-setting inequalities, may close these gaps. Second, beyond two-mode squeezed vacua seen through loss we do not know whether $\zeta_j>1/2$ suffices for two-setting Bell nonlocality of general Gaussian states. Third, all counts assume that the global state is Gaussian. The share bound depends only on the covariance matrix and therefore holds for every state, but the criteria that turn shares into counts are specific to Gaussian states, and we do not know whether the counts survive for non-Gaussian states with the same covariance matrix. Fourth, Proposition~\ref{prop:multimode} extends the counts for Gaussian steering and remote Wigner negativity to neighbors with several modes. For $n$-setting steering, Bell nonlocality and teleportation, the correlations of a multimode neighbor with the node are described by a matrix rather than a number, and the corresponding counts are not known. Fifth, the heralded projection of Theorem~\ref{thm:saturation} attains the count but with a small negativity and a rate that decreases exponentially with $k$, and the trade-off between the number of negative neighbors, their negativity and the success rate remains to be explored. Finally, the teleportation count may change for protocols outside the class of Sec.~\ref{sec:telprot}, in view of the advantage of non-Gaussian cloning~\cite{Cerf2005}.

%=====================================================================
\section{Conclusions}\label{sec:conc}
%=====================================================================

We asked how many neighbors a multimode Gaussian node can serve with steering, and the answer is exact. A node with $k$ modes can be steered by Gaussian measurements, with any number of settings, by at most $2k-1$ neighbors, whatever the number of modes each neighbor holds, and a linear-optical network built from a harmonic tight frame reaches this number for every squeezing. Neither the number of modes nor amplification sets the limit. Neighbors addressing different, overlapping collective modes of the node exceed the mode count by almost a factor of two, and amplification pays for every gain in correlation with noise.

The count follows from one bound. Every single-mode neighbor holds a share, computed from the covariance matrix by a Schur complement, and the shares of all neighbors add up to at most $k$, whatever squeezing, amplification or noise the state contains. Every set of shares allowed by this bound occurs with passive optics, and the share does not grow under local Gaussian channels. Because the share is the critical parameter of the known Gaussian criteria for separability, steering and extendibility, the bound turns each criterion into a count. Entanglement can reach arbitrarily many neighbors, at most $2k-1$ neighbors can steer the node by Gaussian measurements or teleport into it beyond the Gaussian no-cloning value with unit-gain protocols, and at most $nk-1$ can steer it or violate a Bell inequality with it using $n$ settings, whatever the measurements. For two settings this is again $2k-1$, so serving more neighbors requires non-Gaussian measurements with at least three settings.

The strongest consequence concerns Wigner negativity. One heralded operation at the node can make at most $2k-1$ neighbors Wigner negative, and a single projection of the node of the tight-frame network makes all $2k-1$ of them negative in the same run. Reaching this number requires an operation that acts on all $k$ modes of the node, and for $k\geq3$ no operation on a single physical mode can make even one of these neighbors negative, so collective non-Gaussian operations are a necessity rather than a refinement. The same network reaches the counts for two-setting steering and CHSH violation, and a frame network with $3k-1$ neighbors reaches the count $3k-1$ for three-setting steering for every $k$ we have certified. Its Bell violations are small and use idealized observables, and the projection becomes rare as $k$ grows, so these constructions establish that the counts are exact rather than offering practical sources. For a node with two modes, the tests of steering and remote Wigner negativity are within reach of current experiments, and the share bound gives a simple rule for allocating strong correlations in optical networks.

\paragraph*{Data and code availability.} All numerical data and figures of this paper are generated by a self-contained Jupyter notebook, which is openly available on GitHub~\cite{Github_code}. It reproduces every numerical value quoted in the text, the checks of Appendix~\ref{app:num}, the semidefinite programs and certificates of Sec.~\ref{sec:tight3} and Appendix~\ref{app:sdp}, and all seven figures, with all outputs stored in the notebook.

\appendix

%=====================================================================
\section{Proofs of Propositions~\ref{prop:var} and \ref{prop:modeop}}\label{app:var}
%=====================================================================

This appendix proves the two variational forms of the share stated in Sec.~\ref{sec:var}. The first reduces to a property of $2\times2$ matrices, and the second follows from the first by expanding the neighbor's quadratures in the input modes of the purified picture.

\subsection{Proposition~\ref{prop:var}}\label{app:var1}

Let $z\in\mathbb C^2$ with $z^\dagger iJz=1$. Then
\begin{align}
z^\dagger Q_jz=z^\dagger M_jz-\kappa_j\,z^\dagger iJz=z^\dagger M_jz-\kappa_j.\label{eq:A1}
\end{align}
It therefore suffices to show that the infimum of $z^\dagger M_jz$ over the constraint set equals $\mu_j$, because then the infimum in Eq.~\eqref{eq:var} is $\mu_j-\kappa_j=1-2\zeta_j$.

\emph{Case $\mu_j>0$.} Since $M_j\geq0$ by Lemma~\ref{lem:props}(i) and $\det M_j=\mu_j^2$, Lemma~\ref{lem:2x2} gives $M_j-i\mu_jJ\geq0$. Hence
\begin{align}
z^\dagger M_jz\geq\mu_j\,z^\dagger iJz=\mu_j.\label{eq:A2}
\end{align}
The matrix $M_j-i\mu_jJ$ has determinant $\det M_j-\mu_j^2=0$, so it has a nonzero null vector $z_0$, for which $z_0^\dagger M_jz_0=\mu_jz_0^\dagger iJz_0$. Since $M_j$ is positive definite, the left side is positive and so is $z_0^\dagger iJz_0$. Rescaling $z_0$ so that $z_0^\dagger iJz_0=1$, it satisfies the constraint and attains the bound~\eqref{eq:A2}.

\emph{Case $\mu_j=0$.} Then $M_j\geq0$ has a real unit null vector $\bvec u_0$. For $s>0$ put $z=s\bvec u_0+iJ\bvec u_0/(2s)$. Writing $z=\bvec x+i\bvec y$ with $\bvec x=s\bvec u_0$ and $\bvec y=J\bvec u_0/(2s)$, one finds
\begin{align}
z^\dagger iJz&=-2\,\bvec x^TJ\bvec y=-\bvec u_0^TJ^2\bvec u_0=1,\nonumber\\
z^\dagger M_jz&=\bvec x^TM_j\bvec x+\bvec y^TM_j\bvec y=\frac{\bvec u_0^TJ^TM_jJ\bvec u_0}{4s^2},\label{eq:A3}
\end{align}
which tends to zero as $s\to\infty$. Since $z^\dagger M_jz\geq0$ always, the infimum is $0=\mu_j$. In both cases the infimum of $z^\dagger M_jz$ equals $\mu_j$, which by Eq.~\eqref{eq:A1} proves Proposition~\ref{prop:var}.

\subsection{Proposition~\ref{prop:modeop}}\label{app:var2}

We use the purified picture. By Lemma~\ref{lem:channel}, $Q_j=R_jGR_j^T$, where $R_j=(K_j,N_j)$ collects the two rows of $S$ for $Y_j$ and
\begin{align}
G=\left(-i\Omega_B\right)\oplus\iden_{2(n-k)}\label{eq:Gin}
\end{align}
acts on the inputs, the $k$ partner modes first. Let $e=(1,-i)^T/\sqrt2$. For every mode,
\begin{align}
\iden_2=ee^\dagger+\bar e\bar e^\dagger,\qquad iJ=ee^\dagger-\bar e\bar e^\dagger,\label{eq:edec}
\end{align}
so each $2\times2$ block of $G$ equals $ee^\dagger+\bar e\bar e^\dagger$ for a vacuum input and $\bar e\bar e^\dagger-ee^\dagger$ for a partner input. Writing $R_{j,c}$ for the $2\times2$ block of $R_j$ that acts on input $c$, define for $z\in\mathbb C^2$
\begin{align}
\alpha_c(z)=e^\dagger R_{j,c}^Tz,\qquad\beta_c(z)=\bar e^\dagger R_{j,c}^Tz.\label{eq:abz}
\end{align}
Since $z^\dagger R_{j,c}e=\overline{e^\dagger R_{j,c}^Tz}$, and similarly for $\bar e$, Eq.~\eqref{eq:edec} gives
\begin{align}
z^\dagger Q_jz=\sum_{c>k}\left(|\alpha_c|^2+|\beta_c|^2\right)+\sum_{c\leq k}\left(|\beta_c|^2-|\alpha_c|^2\right),\label{eq:zQz}
\end{align}
and, since $R_j\Omega R_j^T=J$ by Eq.~\eqref{eq:commsplit},
\begin{align}
z^\dagger iJz=\sum_{c=1}^n\left(|\alpha_c|^2-|\beta_c|^2\right).\label{eq:ziJz}
\end{align}
On the constraint set $z^\dagger iJz=1$, eliminating $\sum_{c>k}|\alpha_c|^2$ from Eq.~\eqref{eq:zQz} with Eq.~\eqref{eq:ziJz} gives
\begin{align}
z^\dagger Q_jz=1+2\sum_{c=1}^n|\beta_c|^2-2\sum_{c\leq k}|\alpha_c|^2.\label{eq:zQz2}
\end{align}
Proposition~\ref{prop:var} therefore gives $\zeta_j=\sup_z\{\sum_{c\leq k}|\alpha_c(z)|^2-\sum_c|\beta_c(z)|^2\}$ over $z^\dagger iJz=1$.

It remains to identify these numbers with the coefficients of a mode operator. Let $S_Y$ be a local symplectic matrix of the neighbor with rows $\bvec s_1^T$ and $\bvec s_2^T$, and let $\hat c_j=(\hat x'+i\hat p')/2$ be the annihilation operator of its mode after $S_Y$, where $(\hat x',\hat p')^T=S_Y\hat{\bvec r}_{Y_j}$. Each input mode satisfies $(\hat x_c,\hat p_c)^T=\sqrt2(e\,\hat b_c+\bar e\,\hat b_c^\dagger)$, so Eq.~\eqref{eq:channel} gives
\begin{align}
\hat c_j=\sum_c\frac{(1,i)}{\sqrt2}S_YR_{j,c}\left(e\,\hat b_c+\bar e\,\hat b_c^\dagger\right).\label{eq:cexp}
\end{align}
The coefficients of $\hat b_c$ and $\hat b_c^\dagger$ are the complex conjugates of $\alpha_c(z)$ and $\beta_c(z)$ in Eq.~\eqref{eq:abz} for the vector
\begin{align}
z=\frac{S_Y^T(1,-i)^T}{\sqrt2}=\frac{\bvec s_1-i\bvec s_2}{\sqrt2}.\label{eq:zfromS}
\end{align}
Conversely, every $z$ with $z^\dagger iJz=1$ is of this form. Indeed, write $z=(\bvec s_1-i\bvec s_2)/\sqrt2$ with real vectors $\bvec s_1$ and $\bvec s_2$, which is always possible. A direct computation gives $z^\dagger iJz=\bvec s_1^TJ\bvec s_2$, so the constraint says that the real matrix with rows $\bvec s_1^T$ and $\bvec s_2^T$ has determinant $\bvec s_1^TJ\bvec s_2=1$, that is, it is symplectic. The two suprema therefore coincide, which proves Eq.~\eqref{eq:modeopform} and completes the proof of Proposition~\ref{prop:modeop}.

%=====================================================================
\section{Proof of Theorem~\ref{thm:budget}}\label{app:budget}
%=====================================================================

This appendix proves the share bound of Theorem~\ref{thm:budget}. The idea is to choose for every neighbor the optimal vector of Proposition~\ref{prop:var}, transfer all vectors to the inputs of the purified picture, where they become the Bogoliubov coefficients of commuting mode operators, and bound the resulting sum with the maximum principle of Ky Fan. We continue with the notation of Appendix~\ref{app:var2}. Let $S$ be the $2n\times2n$ symplectic matrix that produces the neighbors and the environment from the inputs, so that the rows $R_j$ of neighbor $Y_j$ are two rows of $S$.

\emph{Step 1, choice of vectors.} Fix a set $\mathcal S$ of $s$ neighbors and $\epsilon>0$. By Proposition~\ref{prop:var}, for each $j\in\mathcal S$ there is $z_j\in\mathbb C^2$ with $z_j^\dagger iJz_j=1$ and
\begin{align}
z_j^\dagger Q_jz_j<1-2\zeta_j+\epsilon.\label{eq:B1}
\end{align}
The $\epsilon$ is needed because the infimum in Proposition~\ref{prop:var} need not be attained. Let $\hat z_j\in\mathbb C^{2n}$ be the vector equal to $z_j$ in the two quadratures of output $Y_j$ and zero elsewhere, and let $\hat Z$ be the $2n\times s$ matrix with columns $\hat z_j$. Vectors supported on different outputs are orthogonal with respect to $i\Omega$, which is block diagonal, so
\begin{align}
\hat Z^\dagger i\Omega\hat Z=\iden_s.\label{eq:B2}
\end{align}

\emph{Step 2, transfer to the inputs.} Let $W=S^T\hat Z$. Since $S\Omega S^T=\Omega$, Eq.~\eqref{eq:B2} becomes $W^\dagger i\Omega W=\iden_s$. Since $Q_j=R_jGR_j^T$ and the $R_j$ are rows of $S$,
\begin{align}
\sum_{j\in\mathcal S}z_j^\dagger Q_jz_j=\tr\left(W^\dagger GW\right).\label{eq:B3}
\end{align}

\emph{Step 3, Bogoliubov coefficients.} Write $W_c$ for the two rows of $W$ that belong to input mode $c$, and define the row vectors $\bvec\alpha_c=e^\dagger W_c$ and $\bvec\beta_c=\bar e^\dagger W_c$ of length $s$. They are the rows of two $n\times s$ matrices $\mathsf A$ and $\mathsf B$, whose columns are the coefficients of Eq.~\eqref{eq:abz} for the neighbors in $\mathcal S$. Using Eq.~\eqref{eq:edec} block by block, $W^\dagger i\Omega W=\iden_s$ becomes
\begin{align}
\mathsf A^\dagger\mathsf A-\mathsf B^\dagger\mathsf B=\iden_s,\label{eq:B4}
\end{align}
which is the matrix form of the commutation relations of the neighbors' mode operators. With $P$ the projector onto the $k$ partner rows, Eq.~\eqref{eq:B3} becomes
\begin{align}
\tr\left(W^\dagger GW\right)&=\tr\mathsf A^\dagger\mathsf A+\tr\mathsf B^\dagger\mathsf B-2\tr\mathsf A^\dagger P\mathsf A\nonumber\\
&=s+2\tr\mathsf B^\dagger\mathsf B-2\tr\mathsf A^\dagger P\mathsf A,\label{eq:B5}
\end{align}
where the second line uses the trace of Eq.~\eqref{eq:B4}. Summing Eq.~\eqref{eq:B1} over $j\in\mathcal S$ and using Eqs.~\eqref{eq:B3} and \eqref{eq:B5},
\begin{align}
\sum_{j\in\mathcal S}\zeta_j<\tr\mathsf A^\dagger P\mathsf A-\tr\mathsf B^\dagger\mathsf B+\frac{s\epsilon}{2}.\label{eq:B6}
\end{align}

\emph{Step 4, the decisive matrix inequality.} The quantity $\tr\mathsf A^\dagger P\mathsf A=\tr(P\mathsf A\mathsf A^\dagger P)$ is the trace of the positive semidefinite $n\times n$ matrix $H=\mathsf A\mathsf A^\dagger$ compressed to the $k$-dimensional range of $P$. The maximum principle of Ky Fan~\cite{Fan1949,Bhatia1997} states that for a Hermitian matrix $H$ with eigenvalues $h_1\geq h_2\geq\cdots$ and every orthogonal projector $P$ of rank $k$,
\begin{align}
\tr(PHP)\leq\sum_{i=1}^kh_i.\label{eq:kyfanprinciple}
\end{align}
It follows from the variational characterization of eigenvalues, since $\tr(PHP)=\sum_{l=1}^k\bvec f_l^\dagger H\bvec f_l$ for any orthonormal basis $\bvec f_1,\dots,\bvec f_k$ of the range of $P$. Next, $\mathsf A\mathsf A^\dagger$ and $\mathsf A^\dagger\mathsf A$ have the same nonzero eigenvalues, because both are formed from the singular values of $\mathsf A$. By Eq.~\eqref{eq:B4}, $\mathsf A^\dagger\mathsf A=\iden_s+\mathsf B^\dagger\mathsf B$ has the eigenvalues $1+b_i$, $i=1,\dots,s$, where $b_i\geq0$ are the eigenvalues of $\mathsf B^\dagger\mathsf B$. Hence $\mathsf A\mathsf A^\dagger$ has at most $s$ nonzero eigenvalues, all of the form $1+b_i$, and its $k$ largest eigenvalues add up to at most $\sum_{i=1}^{\min(k,s)}(1+b_i)$. Therefore
\begin{align}
\tr\mathsf A^\dagger P\mathsf A\leq\sum_{i=1}^{\min(k,s)}(1+b_i)\leq k+\sum_{i=1}^sb_i=k+\tr\mathsf B^\dagger\mathsf B.\label{eq:B7}
\end{align}
Inserting Eq.~\eqref{eq:B7} into Eq.~\eqref{eq:B6} and letting $\epsilon\to0$ gives $\sum_{j\in\mathcal S}\zeta_j\leq k$. Choosing $\mathcal S$ as the set of neighbors with positive shares gives $\sum_j\zeta_j^+\leq k$. This completes the proof of Theorem~\ref{thm:budget}.

%=====================================================================
\section{Proof of the CHSH bound}\label{app:chsh}
%=====================================================================

This appendix proves the CHSH bound~\eqref{eq:chshbound} and hence Eq.~\eqref{eq:chshk}. The idea is to split the state into a part with at most one photon in the node mode, on which the observables act as qubit observables, and a remainder of weight $\lambda^4$, whose contribution is bounded crudely. By Lemma~\ref{lem:network} it suffices to consider the state $\ket\psi=\sqrt{1-\lambda^2}\sum_n\lambda^n\ket{n,n}$ of the node mode $\hat g_j$ and the partner mode, with the partner sent through a pure-loss channel of transmission $\eta\in(1/2,1]$. The node modes orthogonal to $\hat g_j$ are uncorrelated with the neighbor and are not measured. A loss channel maps $\ket n$ to a mixture of $\ket{n-l}$ with binomial weights $\binom nl\eta^{n-l}(1-\eta)^l$. It leaves the photon number of the node mode unchanged and does not increase that of the neighbor. Call the resulting state $\rho$.

\emph{Step 1, splitting the state.} The node observable $O_\theta$ is block diagonal with respect to the subspace with at most one photon in the node mode and the subspace with at least two photons, and it equals the identity on the latter. Hence the coherences of $\rho$ between these subspaces do not contribute to any correlation $E(\theta,\phi)=\tr[\rho(O_\theta\otimes O_\phi)]$. This uses only the block structure of the node observable, so no assumption on the neighbor observable is needed. The part of $\rho$ with at least two photons in the node mode has weight $(1-\lambda^2)\sum_{n\geq2}\lambda^{2n}=\lambda^4$, and on it the expectation value of $O_\theta\otimes O_\phi$ lies between $-1$ and $1$. Its contribution to every correlation therefore lies in $[-\lambda^4,\lambda^4]$.

\emph{Step 2, the qubit part.} In the remaining part the node mode has at most one photon, and so does the neighbor. Both observables act there as the qubit observables $\cos\theta\,\sigma_z+\sin\theta\,\sigma_x$. This part of $\rho$ equals $(1-\lambda^2)X$ with
\begin{align}
X&=\ketbra{\psi'}{\psi'}+\lambda^2(1-\eta)\ketbra{1,0}{1,0},\nonumber\\
\ket{\psi'}&=\ket{0,0}+\lambda\sqrt\eta\ket{1,1},\label{eq:C1}
\end{align}
where the second term describes the loss of the neighbor's photon. The correlations of $X$ in the $x$-$z$ plane follow from $\braket{\psi'|\sigma_z\otimes\sigma_z|\psi'}=1+\lambda^2\eta$, $\braket{1,0|\sigma_z\otimes\sigma_z|1,0}=-1$, $\braket{\psi'|\sigma_x\otimes\sigma_x|\psi'}=2\lambda\sqrt\eta$, $\braket{1,0|\sigma_x\otimes\sigma_x|1,0}=0$, and the vanishing of all mixed terms $\sigma_x\otimes\sigma_z$ and $\sigma_z\otimes\sigma_x$. This gives
\begin{align}
E_q(\theta,\phi)=t_z\cos\theta\cos\phi+t_x\sin\theta\sin\phi\label{eq:C2}
\end{align}
for the contribution of the qubit part, with
\begin{align}
t_z&=(1-\lambda^2)\left[1+\lambda^2(2\eta-1)\right],\nonumber\\
t_x&=2(1-\lambda^2)\lambda\sqrt\eta.\label{eq:C3}
\end{align}

\emph{Step 3, the CHSH combination.} Take $\theta\in\{0,\pi/2\}$ for the node and $\phi\in\{\phi_0,-\phi_0\}$ for the neighbor, with
\begin{align}
\tan\phi_0=\frac{t_x}{t_z}.\label{eq:phi0}
\end{align}
Then
\begin{align}
&E_q(0,\phi_0)+E_q(\tfrac\pi2,\phi_0)+E_q(0,-\phi_0)-E_q(\tfrac\pi2,-\phi_0)\nonumber\\
&\quad=2t_z\cos\phi_0+2t_x\sin\phi_0=2\sqrt{t_z^2+t_x^2}.\label{eq:C4}
\end{align}
Each of the four full correlations differs from $E_q$ by at most $\lambda^4$, by Step 1, so the CHSH value satisfies $\mathcal B\geq2\sqrt{t_z^2+t_x^2}-4\lambda^4$. We now write this in the variables $x=\lambda^2$ and $a=2\eta-1\in(0,1]$, so that $4\eta=2(1+a)$. Inserting Eq.~\eqref{eq:C3} gives $t_z^2+t_x^2=(1-x)^2\,Y$, where
\begin{align}
Y=(1+ax)^2+2(1+a)x.\label{eq:C5}
\end{align}
Hence
\begin{align}
\mathcal B\geq L(x),\qquad L(x)=2(1-x)\sqrt{Y}\,-\,4x^2,\label{eq:C6}
\end{align}
where the square root acts on $Y$ only and the term $4x^2$ is subtracted after multiplication.

\emph{Step 4, explicit bound.} Choose $x=a/8$ and write $Y=1+y$ with
\begin{align}
y=(2+4a)x+a^2x^2\geq0.\label{eq:C7}
\end{align}
For $y\geq0$ one has $\sqrt{1+y}\geq1+y/2-y^2/8$, which follows from the Taylor expansion of $\sqrt{1+y}$ with remainder, whose third derivative is positive. Since $1-x>0$, this gives
\begin{align}
L(a/8)\geq2(1-x)\left(1+\frac y2-\frac{y^2}{8}\right)-4x^2\Big|_{x=a/8}.\label{eq:C8}
\end{align}
The right side is a polynomial in $a$. A direct expansion shows that it equals $2+a^2/8+a^2q(a)$ with
\begin{align}
q(a)&=\frac{17}{64}-\frac{63a}{512}-\frac{5a^2}{128}+\frac{a^3}{256}-\frac{15a^4}{4096}\nonumber\\
&\quad+\frac{a^5}{2048}-\frac{a^6}{16384}+\frac{a^7}{131072}.\label{eq:C9}
\end{align}
For $a\in(0,1]$, dropping the positive terms and bounding each negative term by its value at $a=1$ gives
\begin{align}
q(a)\geq\frac{17}{64}-\frac{63}{512}-\frac{5}{128}-\frac{15}{4096}-\frac{1}{16384}>0.09.\label{eq:C10}
\end{align}
Hence $\mathcal B\geq2+a^2/8=2+(2\eta-1)^2/8$ at $\lambda^2=(2\eta-1)/8$, which is Eq.~\eqref{eq:chshbound}. As a check, at $a=1/7$, the value relevant for $k=4$, the exact $L(a/8)$ exceeds $2+a^2/8$ by $5.1\times10^{-3}$. For $\eta=k/(2k-1)$ we have $a=1/(2k-1)$ and $\tanh^2r=1/[8(2k-1)]$, which gives Eq.~\eqref{eq:chshk}. This completes the proof of the CHSH part of Theorem~\ref{thm:tight}.

%=====================================================================
\section{Proof of Proposition~\ref{prop:mono} and fractional extendibility}\label{app:mono}
%=====================================================================

This appendix proves Proposition~\ref{prop:mono} and the equivalence between the share and fractional extendibility stated in Sec.~\ref{sec:hier}. Both proofs use a form of the variational characterization that contains no inverse.

\emph{A form without inverses.} Combining Proposition~\ref{prop:var} with Eq.~\eqref{eq:schur2} for $G=V^{(AY_j)}+i(\Omega_A\oplus0)$ gives
\begin{align}
1-2\zeta_j=\inf\left\{\upsilon^\dagger\left[V^{(AY_j)}+i(\Omega_A\oplus0)\right]\upsilon\right\},\label{eq:infw}
\end{align}
where the infimum runs over all $\upsilon=a\oplus z$ with $a\in\mathbb C^{2k}$ and $z\in\mathbb C^2$ such that $z^\dagger iJz=1$. This expression contains no inverse. It agrees with Definition~\ref{def:share} whenever $V_A+i\Omega_A$ is invertible, and we use it to define the share also when the node has modes in a pure state.

\emph{Gaussian unitaries.} Lemma~\ref{lem:props}(ii) shows that $\zeta_j$ is invariant under Gaussian unitaries of the node and of the neighbor, and hence so is $\zeta_j^+$.

\emph{Gaussian channels of the node.} Every Gaussian channel of the node can be realized by adding ancilla modes $E$ in a Gaussian state with covariance matrix $V_E$, uncorrelated with everything else, applying a Gaussian unitary with symplectic matrix $S$ to the node and the ancillas, and discarding some modes~\cite{Serafini2017}. Let the remaining modes of the node be $A'$. Let $V^{(AEY_j)}$ be the covariance matrix of the node, the ancillas and $Y_j$ before the unitary. It contains $V^{(AY_j)}$ and $V_E$ as blocks and has no correlations between the ancillas and the other modes. The covariance matrix of $A'$ and $Y_j$ is a principal submatrix of
\begin{align}
\tilde V=(S\oplus\iden_2)\,V^{(AEY_j)}\,(S\oplus\iden_2)^T.\label{eq:Vtilde}
\end{align} In Eq.~\eqref{eq:infw} for the pair $A'Y_j$, the vector $a'$ ranges over the coordinates of $A'$ only. Embedding it into the coordinates of all modes of the node and the ancillas, with zeros on the discarded modes, shows that the infimum for $A'Y_j$ is at least the infimum of $\tilde w^\dagger[\tilde V+i(\Omega_{AE}\oplus0)]\tilde w$ over all vectors $\tilde w=\tilde a\oplus z$, with $\tilde a$ unrestricted. Since $S\Omega_{AE}S^T=\Omega_{AE}$,
\begin{align}
\tilde V+i(\Omega_{AE}\oplus0)&=(S\oplus\iden_2)\left[V^{(AEY_j)}+i(\Omega_{AE}\oplus0)\right]\nonumber\\
&\quad\times(S\oplus\iden_2)^T.\label{eq:Vtilde2}
\end{align} The substitution $b=(S\oplus\iden_2)^T\tilde w$ leaves $z$ unchanged and maps the unrestricted vectors $\tilde a$ onto all vectors of the node and the ancillas. Writing $b=a\oplus e\oplus z$, the quadratic form becomes
\begin{align}
&(a\oplus z)^\dagger\left[V^{(AY_j)}+i(\Omega_A\oplus0)\right](a\oplus z)\nonumber\\
&\quad+e^\dagger(V_E+i\Omega_E)e.\label{eq:splitform}
\end{align}
The second term is nonnegative by the uncertainty relation of the ancillas and vanishes for $e=0$, so the infimum over all $b$ equals $1-2\zeta_j$. Hence $1-2\zeta_j'\geq1-2\zeta_j$, that is, $\zeta_j'\leq\zeta_j$. If all modes of the node are discarded, the neighbor is uncorrelated with the node, $Q_j=V_{Y_j}$, $\kappa_j=0$ and $\zeta_j'=(1-\mu_j)/2\leq0$ by Lemma~\ref{lem:props}(i).

\emph{Gaussian channels of the neighbor.} A Gaussian channel from one mode to one mode acts as $V_{Y_j}\mapsto XV_{Y_j}X^T+Y$ and $V_{Y_jA}\mapsto XV_{Y_jA}$, with a real $2\times2$ matrix $X$ and a real symmetric $Y$ that satisfy $Y+i(J-XJX^T)\geq0$~\cite{Holevo2001,Serafini2017}. Since $XJX^T=tJ$ with $t=\det X$, this condition reads $Y+i(1-t)J\geq0$, which by Lemma~\ref{lem:2x2} means $Y\geq0$ and $\sqrt{\det Y}\geq|1-t|$. Equation~\eqref{eq:Q} gives $Q_j\mapsto XQ_jX^T+Y$, hence
\begin{align}
\kappa_j\mapsto t\kappa_j,\qquad M_j\mapsto XM_jX^T+Y.\label{eq:recchannel}
\end{align}
For positive semidefinite $2\times2$ matrices the Minkowski determinant inequality~\cite{Bhatia1997} states $\sqrt{\det(P+R)}\geq\sqrt{\det P}+\sqrt{\det R}$, so
\begin{align}
\mu_j'\geq|t|\mu_j+|1-t|.\label{eq:mink}
\end{align}
Inserting this into Eq.~\eqref{eq:share} gives
\begin{align}
\zeta_j'\leq\frac{1+t\kappa_j-|t|\mu_j-|1-t|}{2}.\label{eq:zetaprime}
\end{align}
For $0\leq t\leq1$ the right side equals $t(1+\kappa_j-\mu_j)/2=t\zeta_j$. For $t>1$ it equals $1-t/2+t(\kappa_j-\mu_j)/2=1-t(1-\zeta_j)$, using $\kappa_j-\mu_j=2\zeta_j-1$. For $t<0$ it equals $t(1+\kappa_j+\mu_j)/2$, and since $1+\kappa_j+\mu_j\geq1+\kappa_j+|1-\kappa_j|\geq2$, it is at most $t<0$. In all three cases $\max(\zeta_j',0)\leq\max(\zeta_j,0)$, because $t\zeta_j\leq\zeta_j^+$ for $0\leq t\leq1$, and $1-t(1-\zeta_j)\leq\zeta_j$ for $t>1$ since $(1-\zeta_j)(1-t)\leq0$. A pure-loss channel of transmission $\lambda$ has $t=\lambda$, and a quantum-limited amplifier of gain $G$ has $t=G$, which gives the last statement of Proposition~\ref{prop:mono}. Together with the invariance under Gaussian unitaries and the monotonicity under Gaussian channels of the node, this completes the proof of Proposition~\ref{prop:mono}.

\emph{Fractional extendibility.} Following Ref.~\cite{Ahmed2026}, the pair is $1/\gamma$-extendible on $Y_j$, for $\gamma\in[0,1]$, if there is a single-mode covariance matrix $\Delta\geq iJ$ with
\begin{align}
V^{(AY_j)}\geq i\Omega_A\oplus\left[(1-\gamma)\Delta+i\gamma J\right].\label{eq:fracdef}
\end{align}
We show that this holds if and only if $\zeta_j\leq\gamma$. Taking the complex conjugate and the Schur complement with respect to the block $V_A+i\Omega_A$, as in Proposition~\ref{prop:hier}, Eq.~\eqref{eq:fracdef} is equivalent to
\begin{align}
N_\gamma-i(\kappa_j-\gamma)J\geq0,\qquad N_\gamma=M_j-(1-\gamma)\Delta,\label{eq:fracN}
\end{align}
which by Lemma~\ref{lem:2x2} means $N_\gamma\geq0$ and $\sqrt{\det N_\gamma}\geq|\kappa_j-\gamma|$. If these hold, the Minkowski inequality and $\det\Delta\geq1$ give $\mu_j\geq\sqrt{\det N_\gamma}+1-\gamma\geq1-\gamma+|\kappa_j-\gamma|$. Conversely, if $\mu_j\geq1-\gamma+|\kappa_j-\gamma|$ and $\mu_j>0$, the choice $\Delta=M_j/\mu_j$ has $\det\Delta=1$, satisfies $\Delta\geq iJ$, and gives $N_\gamma=[1-(1-\gamma)/\mu_j]M_j\geq0$ with $\sqrt{\det N_\gamma}=\mu_j-(1-\gamma)\geq|\kappa_j-\gamma|$. If $\mu_j=0$, the condition forces $\gamma=1$, and then Eq.~\eqref{eq:fracdef} is the uncertainty relation. It remains to show that $\mu_j\geq1-\gamma+|\kappa_j-\gamma|$ is equivalent to $\zeta_j\leq\gamma$. For $\kappa_j\geq\gamma$ both read $\mu_j\geq1+\kappa_j-2\gamma$. For $\kappa_j<\gamma$ the condition reads $\mu_j\geq1-\kappa_j$, which always holds, and $\zeta_j\leq\kappa_j<\gamma$ holds as well, because $\mu_j\geq1-\kappa_j$. Hence the pair is $1/\gamma$-extendible exactly when $\zeta_j\leq\gamma$, and $\zeta_j^+$ is the smallest $\gamma$ for which this holds, as claimed in Sec.~\ref{sec:hier}.

%=====================================================================
\section{Quasi-states and the proof of Theorem~\ref{thm:quasirep}}\label{app:quasi}
%=====================================================================

This appendix establishes the properties of quasi-states used in Sec.~\ref{sec:herald} and proves the representation of Theorem~\ref{thm:quasirep}. We use the following standard facts about Wigner functions~\cite{Serafini2017,Cahill1969,Royer1977}. The trace rule $\tr(\hat X\hat A)=\int W_{\hat X}A_W\,d\bvec r$ holds for trace-class $\hat X$ and operators $\hat A$ for which the integral converges, and the identity has Weyl symbol one. The Weyl symbol of a tensor product is the product of the symbols. A Gaussian unitary with symplectic matrix $S$ and a displacement by $\bvec\delta$ map Wigner functions as $W(\bvec r)\mapsto W(S^{-1}(\bvec r-\bvec\delta))$. For a state $\rho$ of two groups of modes $A$ and $Y$ and an operator $\hat E$ on $A$, these facts give the partial trace rule
\begin{align}
W_{\tr_A[(\hat E\otimes\iden)\rho]}(\bvec y)=\int E_W(\bvec\alpha)\,W_\rho(\bvec\alpha,\bvec y)\,d\bvec\alpha.\label{eq:partialrule}
\end{align}
Finally, the characteristic function $\chi_{\hat X}(\bvec\xi)=\tr(\hat Xe^{i\bvec\xi^T\hat{\bvec r}})$ determines the Wigner function by a Fourier transform, and for one mode the diagonal Fock matrix elements of the Weyl operator are $e^{-|\bvec\xi|^2/2}L_n(|\bvec\xi|^2)$, where $L_n$ is the Laguerre polynomial and $|\bvec\xi|^2=\bvec\xi^T\bvec\xi$.

\begin{lemma}\label{lem:rhoq}
For $q\in(-1,1)$ the operator $\rho_q$ of Eq.~\eqref{eq:rhoq} is trace class, its Wigner function is $\mathcal N(\bvec r;0,d\iden_2)$ with $d=(1+q)/(1-q)$, and $\tr(\rho_q\hat n)=q/(1-q)=(d-1)/2$.
\end{lemma}

\begin{proof}
The trace norm is $\sum_n(1-q)|q|^n=(1-q)/(1-|q|)<\infty$. The characteristic function is
\begin{align}
\chi_{\rho_q}(\bvec\xi)=(1-q)\sum_{n=0}^\infty q^ne^{-|\bvec\xi|^2/2}L_n(|\bvec\xi|^2).\label{eq:chirhoq}
\end{align}
The generating function $\sum_nL_n(x)q^n=(1-q)^{-1}\exp[-qx/(1-q)]$, valid for $|q|<1$, gives
\begin{align}
\chi_{\rho_q}(\bvec\xi)&=\exp\left[-\frac{|\bvec\xi|^2}{2}\left(1+\frac{2q}{1-q}\right)\right]\nonumber\\
&=\exp\left(-\frac d2\bvec\xi^T\bvec\xi\right),\label{eq:chirhoq2}
\end{align}
which is the characteristic function of the Gaussian function $\mathcal N(\bvec r;0,d\iden_2)$. The mean photon number follows from $\sum_nnq^n=q/(1-q)^2$, and substituting $q=(d-1)/(d+1)$ gives $(d-1)/2$. This completes the proof.
\end{proof}

\begin{lemma}\label{lem:quasi}
The operator $\tau(\bvec\mu,\Sigma)$ of Definition~\ref{def:quasi} is Hermitian, trace class and of unit trace, and its Wigner function is $\mathcal N(\cdot\,;\bvec\mu,\Sigma)$. It does not depend on the choice of $S$ in the Williamson form. It is positive if and only if $\Sigma+i\Omega\geq0$.
\end{lemma}

\begin{proof}
Hermiticity, the trace-class property and the unit trace follow from Lemma~\ref{lem:rhoq}, because tensor products and unitary conjugation preserve them. By Lemma~\ref{lem:rhoq} and the product rule, the tensor product has the Wigner function $\mathcal N(\cdot\,;0,D)$ with $D=\mathrm{diag}(d_1,d_1,\dots,d_k,d_k)$. The Gaussian unitary turns it into
\begin{align}
\mathcal N(S^{-1}\bvec r;0,D)=\mathcal N(\bvec r;0,SDS^T)=\mathcal N(\bvec r;0,\Sigma),\label{eq:quasiW}
\end{align}
where we used $\det S=1$, and the displacement shifts the mean to $\bvec\mu$. An operator is determined by its Wigner function, so $\tau$ does not depend on the choice of $S$. If $\Sigma+i\Omega\geq0$, all $d_l\geq1$~\cite{Simon1994}, all $q_l\geq0$, and $\tau$ is positive. If some $d_l<1$, then $q_l<0$, and the Fock state with one photon in mode $l$ and none elsewhere has the negative expectation value $(1-q_l)q_l\prod_{l'\neq l}(1-q_{l'})$ in $\bigotimes_{l'}\rho_{q_{l'}}$. Unitary conjugation preserves the spectrum, so $\tau$ is not positive. This completes the proof.
\end{proof}

\begin{proof}[Proof of Theorem~\ref{thm:quasirep}]
By the partial trace rule~\eqref{eq:partialrule} and the factorization~\eqref{eq:factor},
\begin{align}
W_{\tilde\rho_{Y_j}}(\bvec y)=\mathcal N(\bvec y;0,V_{Y_j})\int E_W(\bvec\alpha)\,\mathcal N(\bvec\alpha;\Gamma_j\bvec y,\Sigma_j)\,d\bvec\alpha.\label{eq:reppf}
\end{align}
By Lemma~\ref{lem:quasi}, the Gaussian function in the integral is the Wigner function of $\tau(\Gamma_j\bvec y,\Sigma_j)$. This operator is the difference of its positive and negative parts, both trace class, and the trace rule applied to each part gives $\int E_W(\bvec\alpha)\mathcal N(\bvec\alpha;\Gamma_j\bvec y,\Sigma_j)d\bvec\alpha=\tr[\hat E\tau(\Gamma_j\bvec y,\Sigma_j)]$. For a bounded effect this trace is finite because $\tau$ is trace class. For an effect that is a polynomial in the mode operators, the matrix elements of $\hat E$ in the Fock basis of the modes that define $\tau$ grow at most polynomially in the photon numbers, while the weights $(1-q_l)q_l^{n}$ of $\tau$ decay geometrically because $|q_l|<1$, so the trace converges absolutely and the trace rule applies term by term. Inserting the trace into Eq.~\eqref{eq:reppf} gives Eq.~\eqref{eq:quasirep}, which completes the proof.
\end{proof}

%=====================================================================
\section{Proof of Theorem~\ref{thm:criterion}}\label{app:crit}
%=====================================================================

This appendix proves the criterion for simultaneous negativity. Necessity follows from the positivity of traces of products of positive operators. For sufficiency, we reduce the problem to a finite-dimensional subspace and apply the minimax theorem. Write $\Delta_m$ for the simplex of probability vectors $\bvec p=(p_1,\dots,p_m)$ and $M_{\bvec p}=\sum_jp_j\tau_j$. Each $M_{\bvec p}$ is Hermitian and trace class.

\emph{Necessity.} Suppose that $\hat E\geq0$ satisfies $\tr(\hat E\tau_j)<0$ for all $j$ and that $M_{\bvec p}\geq0$ for some $\bvec p\in\Delta_m$. The trace of a product of two positive operators is nonnegative, so
\begin{align}
0\leq\tr\left(\hat EM_{\bvec p}\right)=\sum_jp_j\tr\left(\hat E\tau_j\right)<0,\label{eq:critnec}
\end{align}
where the strict inequality holds because at least one $p_j$ is positive. This is a contradiction, so no convex combination of the $\tau_j$ can be positive semidefinite.

\emph{Sufficiency.} Assume that no $M_{\bvec p}$ is positive semidefinite. The proof has three steps.

\emph{Step 1, a finite subspace.} For every $\bvec p\in\Delta_m$ there is a unit vector $\psi_{\bvec p}$ with $\bra{\psi_{\bvec p}}M_{\bvec p}\ket{\psi_{\bvec p}}<0$. The map $\bvec p'\mapsto\bra{\psi_{\bvec p}}M_{\bvec p'}\ket{\psi_{\bvec p}}$ is affine, hence continuous, so it stays negative on an open neighborhood $U_{\bvec p}$ of $\bvec p$. The simplex is compact, so finitely many such neighborhoods, belonging to points $\bvec p_1,\dots,\bvec p_L$, cover it. Let $\mathcal F$ be the span of $\psi_{\bvec p_1},\dots,\psi_{\bvec p_L}$, let $\Pi$ be the orthogonal projector onto $\mathcal F$, and define the Hermitian matrices $A_j=\Pi\tau_j\Pi$ on $\mathcal F$ and $A_{\bvec p}=\sum_jp_jA_j$. Every $\bvec p$ lies in some $U_{\bvec p_i}$, and $\psi_{\bvec p_i}\in\mathcal F$, so
\begin{align}
\lambda_{\min}(A_{\bvec p})\leq\bra{\psi_{\bvec p_i}}M_{\bvec p}\ket{\psi_{\bvec p_i}}<0\quad\text{for all }\bvec p\in\Delta_m.\label{eq:critlmin}
\end{align}

\emph{Step 2, the minimax theorem.} Let $\mathcal D$ be the set of density matrices on $\mathcal F$, which is convex and compact, as is $\Delta_m$. The function $f(\rho,\bvec p)=\sum_jp_j\tr(\rho A_j)$ is linear in each argument, and the minimax theorem~\cite{vonNeumann1928} gives
\begin{align}
\min_{\rho\in\mathcal D}\max_{\bvec p\in\Delta_m}f(\rho,\bvec p)&=\max_{\bvec p\in\Delta_m}\min_{\rho\in\mathcal D}f(\rho,\bvec p)\nonumber\\
&=\max_{\bvec p\in\Delta_m}\lambda_{\min}(A_{\bvec p}).\label{eq:minimax}
\end{align}
The smallest eigenvalue depends continuously on the matrix, so the right side is the maximum of a continuous function on a compact set, and by Eq.~\eqref{eq:critlmin} it is a negative number, which we call $-\epsilon$.

\emph{Step 3, the effect.} Let $\rho_\star\in\mathcal D$ attain the minimum on the left of Eq.~\eqref{eq:minimax}. Taking $\bvec p$ to be the $j$th vertex of the simplex gives $\tr(\rho_\star\tau_j)=\tr(\rho_\star A_j)\leq-\epsilon<0$ for every $j$, where the first equality holds because $\rho_\star$ is supported in $\mathcal F$. The operator $\hat E=\rho_\star/\|\rho_\star\|_\infty$ satisfies $0\leq\hat E\leq\iden$, so it is an element of a measurement and can be realized as a heralded operation with Kraus operator $\hat E^{1/2}$, and it makes all neighbors negative at the chosen points. The success probability is positive, because the unnormalized heralded state of each neighbor is a positive operator whose Wigner function takes a negative value and is therefore not zero. This completes the proof of Theorem~\ref{thm:criterion}.

%=====================================================================
\section{Proof of Theorem~\ref{thm:saturation}}\label{app:sat}
%=====================================================================

This appendix proves Theorem~\ref{thm:saturation}. We first compute the conditional quasi-states of the neighbors in the tight-frame network, then derive the photon statistics of the steered modes in $\ket{\Phi_k}$, and finally combine both.

\emph{The conditional quasi-states.} In the tight-frame network, with $v=\cosh2r$, $s=\sinh2r$ and $c=\sqrt\eta\,s$, the covariance blocks that involve neighbor $Y_j$ are
\begin{align}
V_A&=v\iden_{2k},\qquad V_{Y_j}=w\iden_2,\qquad w=\eta v+1-\eta,\nonumber\\
V_{AY_j}&=\bvec u_j\otimes cZ,\label{eq:netV}
\end{align}
where $\bvec u_j\otimes cZ$ denotes the $2k\times2$ matrix whose $l$th $2\times2$ block is $u_{j,l}cZ$.

\begin{lemma}[What each neighbor steers]\label{lem:framesteer}
The conditional covariance matrix of the node given $Y_j$ is
\begin{align}
\Sigma_j=v\iden_{2k}-\frac{c^2}{w}\,\bvec u_j\bvec u_j^T\otimes\iden_2.\label{eq:frameSigma}
\end{align}
Its symplectic eigenvalues are $d=v-c^2/w$, belonging to the mode $\hat g_j$, and $v$ with multiplicity $k-1$. The quasi-state at the origin is given by Eq.~\eqref{eq:taumain} with $q_d=(d-1)/(d+1)$ and $q_v=(v-1)/(v+1)$, and
\begin{align}
q_d=-(2\eta-1)\,q_v=-\frac{q_v}{2k-1},\qquad q_v=\tanh^2r.\label{eq:qratio}
\end{align}
\end{lemma}

\begin{proof}
Equation~\eqref{eq:conditional} with $Z^2=\iden_2$ gives Eq.~\eqref{eq:frameSigma}. Complete $\bvec u_j$ to an orthogonal $k\times k$ matrix $O_j$ with first column $\bvec u_j$. The matrix $O_j\otimes\iden_2$ is orthogonal and symplectic, and $(O_j\otimes\iden_2)^T\Sigma_j(O_j\otimes\iden_2)=(v-c^2/w)\iden_2\oplus v\iden_{2(k-1)}$ is a Williamson form. Definition~\ref{def:quasi} then gives Eq.~\eqref{eq:taumain}, with $\hat U_j$ the passive unitary of $O_j\otimes\iden_2$, which maps the first mode to $\hat g_j$. With $c^2=\eta(v^2-1)$ a short calculation gives
\begin{align}
d&=\frac{(1-\eta)v+\eta}{w},\qquad1+d=\frac{v+1}{w},\nonumber\\
1-d&=\frac{(2\eta-1)(v-1)}{w},\label{eq:netd}
\end{align}
hence $q_d=(d-1)/(d+1)=-(2\eta-1)(v-1)/(v+1)$. Finally $(v-1)/(v+1)=\tanh^2r$ and $2\eta-1=1/(2k-1)$, which gives Eq.~\eqref{eq:qratio} and completes the proof.
\end{proof}

\begin{lemma}[Photon statistics]\label{lem:stats}
Let $p_n$ be the probability of finding $n$ photons in the mode $\hat g_j$ of $\ket{\Phi_k}$. Then $p_0=0$, the mean photon number in $\hat g_j$ is $(2k-1)/k=2-1/k$, and $p_1\geq1/k$. These numbers are the same for every $j$.
\end{lemma}

\begin{proof}
All creation operators commute, so $\ket{\Phi_k}\propto\hat g_j^\dagger\ket\chi$ with $\ket\chi=\prod_{l\neq j}\hat g_l^\dagger\ket0$. In a mode basis whose first mode is $\hat g_j$, the operator $\hat g_j^\dagger$ raises the photon number of that mode by one, so every Fock component of $\ket{\Phi_k}$ has at least one photon in $\hat g_j$, and $p_0=0$. For the mean, the tight-frame identity~\eqref{eq:frame} gives
\begin{align}
\sum_{j=1}^m\hat g_j^\dagger\hat g_j=\sum_{l,l'=1}^k\Big(\sum_{j=1}^mu_{j,l}u_{j,l'}\Big)\hat a_l^\dagger\hat a_{l'}=\frac mk\sum_{l=1}^k\hat a_l^\dagger\hat a_l,\label{eq:sumg}
\end{align}
and $\ket{\Phi_k}$ has $m=2k-1$ photons, so the $m$ means add up to $m^2/k$. The harmonic frame has a cyclic symmetry. The orthogonal matrix $R$ that rotates each pair of components $(\cos r\vartheta,\sin r\vartheta)$ by the angle $2\pi r/m$, and leaves the constant component of odd $k$ fixed, satisfies $R\bvec u_j=\bvec u_{j+1}$, with $\bvec u_{m+1}=\bvec u_1$. Its passive unitary maps $\hat g_j^\dagger$ to $\hat g_{j+1}^\dagger$, so it maps $\ket{\Phi_k}$ to itself, since the product in Eq.~\eqref{eq:Phi} is only reordered, and it maps the photon statistics of $\hat g_j$ to those of $\hat g_{j+1}$. All $m$ means are therefore equal to $m/k$, and the statistics are the same for every $j$. Finally, since $p_0=0$,
\begin{align}
\frac mk=\sum_{n\geq1}np_n\geq p_1+2(1-p_1)=2-p_1,\label{eq:p1bound}
\end{align}
which gives $p_1\geq2-m/k=1/k$. This completes the proof.
\end{proof}

\begin{proof}[Proof of Theorem~\ref{thm:saturation}]
By Lemma~\ref{lem:framesteer}, the quasi-state $\tau(0,\Sigma_j)$ is diagonal in the Fock basis of the modes $(\hat g_j,\dots)$ of the frame of neighbor $Y_j$, with eigenvalue $(1-q_d)(1-q_v)^{k-1}q_d^{n_1}q_v^{N-n_1}$ on a Fock state with $n_1$ photons in $\hat g_j$ and $N$ photons in total. The state $\ket{\Phi_k}$ has $N=m$ photons, so
\begin{align}
\bra{\Phi_k}\tau(0,\Sigma_j)\ket{\Phi_k}&=(1-q_d)(1-q_v)^{k-1}q_v^m\,X_k,\nonumber\\
X_k&=\sum_{n\geq1}p_n\left(\frac{q_d}{q_v}\right)^n=\sum_{n\geq1}p_n(-\varrho)^n,\label{eq:Xk}
\end{align}
with $\varrho=1/(2k-1)$ by Eq.~\eqref{eq:qratio}. The prefactor is positive. Equation~\eqref{eq:Xbound}, which uses Lemma~\ref{lem:stats}, shows $X_k\leq-\varrho^2<0$, so by Theorem~\ref{thm:quasirep} every neighbor is Wigner negative at the origin. For the normalization, the reduced state of the node is $\rho_{q_v}^{\otimes k}$, so the success probability is
\begin{align}
\tr\left(\ketbra{\Phi_k}{\Phi_k}\rho_{q_v}^{\otimes k}\right)=(1-q_v)^kq_v^m.\label{eq:psucc}
\end{align}
The normalized Wigner function at the origin is $\mathcal N(0;0,w\iden_2)=1/(2\pi w)$ times the weight~\eqref{eq:Xk}, divided by the success probability,
\begin{align}
W_{Y_j}(0)=\frac{1}{2\pi w}\,\frac{1-q_d}{1-q_v}\,X_k.\label{eq:W0pre}
\end{align}
With $1-q_v=2/(v+1)$ and $1-q_d=1+(2\eta-1)(v-1)/(v+1)=2w/(v+1)$, the ratio $(1-q_d)/(1-q_v)$ equals $w$, and $W_{Y_j}(0)=X_k/(2\pi)\leq-\varrho^2/(2\pi)$, which is Eq.~\eqref{eq:Wbound}. For $k=1$ the frame has a single vector, $\ket{\Phi_1}$ is the one-photon state, $p_1=1$, and $W_{Y_1}(0)=-1/(2\pi)$. The heralded state does not depend on the squeezing, because the projection prepares the partner modes in a state that does not depend on it, as explained in the main text. This completes the proof of Theorem~\ref{thm:saturation}.
\end{proof}

\emph{Collective steering.} For $k=2$ the neighbor at the angle $60^\circ$ or $120^\circ$ has $u_{j,1}^2=1/4$ and $u_{j,2}^2=3/4$. The conditional covariance matrix of the single node mode $\hat a_l$ given $Y_j$ is the corresponding $2\times2$ block of Eq.~\eqref{eq:frameSigma}, $(v-u_{j,l}^2c^2/w)\iden_2$, and it lies below the vacuum level if and only if
\begin{align}
u_{j,l}^2>\frac{w}{\eta(v+1)}=\frac{\eta v+1-\eta}{\eta(v+1)}.\label{eq:singlemodesteer}
\end{align}
The right side increases with $v$ from $1/(2\eta)$ at $v=1$ toward one, and $1/(2\eta)=(2k-1)/(2k)$. For $k=2$ this is $3/4$, which $u_{j,l}^2\leq3/4$ never exceeds for $v>1$, so the neighbors at $60^\circ$ and $120^\circ$ steer neither $\hat a_1$ nor $\hat a_2$. For $k\geq3$, every entry of a harmonic frame vector satisfies $u_{j,l}^2\leq2/k<(2k-1)/(2k)$, so no neighbor steers any single mode $\hat a_l$, for any squeezing.

%=====================================================================
\section{Proof of Theorem~\ref{thm:tel}}\label{app:tel}
%=====================================================================

This appendix proves the bound $F\leq1/(2-\zeta_j)$ on the fidelity of unit-gain teleportation from a neighbor into the node, in the notation of Sec.~\ref{sec:telprot}.

\begin{proof}[Proof of Theorem~\ref{thm:tel}]
The proof has three steps.

\emph{Step 1.} We reduce the fidelity to the determinant of $\Xi_j$. Since $\Xi_j\geq0$, the arithmetic-geometric mean inequality gives $\tr \Xi_j\geq2\sqrt{\det \Xi_j}$, and therefore
\begin{align}
\det(2\iden+\Xi_j)&=4+2\tr \Xi_j+\det \Xi_j\nonumber\\
&\geq4+4\sqrt{\det \Xi_j}+\det \Xi_j\nonumber\\
&=\left(2+\sqrt{\det \Xi_j}\right)^2.\label{eq:amgm}
\end{align}
By Eq.~\eqref{eq:F}, $F\leq2/(2+\sqrt{\det \Xi_j})$, and it remains to show $\sqrt{\det \Xi_j}\geq2-2\zeta_j$.

\emph{Step 2.} We show $\tr\Xi/2-1\geq1-2\zeta_j$ for the noise matrix $\Xi$ of every admissible pair. This suffices. For $S\in SL(2,\R)$, the pair $S(\hat X,\hat P)^T$ is again admissible, because the conditions $\bvec u^T\Omega_A\bvec u'=1$ and $\bvec t^TJ\bvec t'=-1$ are multiplied by $\det S=1$, and its noise matrix is $S\Xi_jS^T$. Choosing $S$ suitably, $\tr(S\Xi_jS^T)/2$ equals $\sqrt{\det \Xi_j}$, or comes arbitrarily close to it if $\Xi_j$ is singular. Applying the claimed inequality to these transformed pairs therefore gives $\sqrt{\det \Xi_j}-1\geq1-2\zeta_j$. Note that the transformed pairs only enter as admissible vectors in this inequality, so their physical implementation plays no role in the argument, although they are in fact implementable.

\emph{Step 3.} We prove the claim of Step 2. For an admissible pair define the complex vectors
\begin{align}
a=\frac{\bvec u+i\bvec u'}{\sqrt2},\qquad z=\frac{\bvec t+i\bvec t'}{\sqrt2},\qquad \upsilon=a\oplus z.\label{eq:complexvec}
\end{align}
For a real symmetric matrix $V$ and real vectors $\bvec x$ and $\bvec y$, one has $(\bvec x-i\bvec y)^TV(\bvec x+i\bvec y)=\bvec x^TV\bvec x+\bvec y^TV\bvec y$. Hence
\begin{align}
\upsilon^\dagger V^{(AY_j)}\upsilon=\frac{\langle\hat X^2\rangle+\langle\hat P^2\rangle}{2}=\frac{\tr\Xi}{2}.\label{eq:trN}
\end{align}
In the same way, $a^\dagger i\Omega_Aa=-\bvec u^T\Omega_A\bvec u'=-1$ and $z^\dagger iJz=-\bvec t^TJ\bvec t'=1$. By Proposition~\ref{prop:var} and Eq.~\eqref{eq:schur2} applied to $G=V^{(AY_j)}+i(\Omega_A\oplus0)$,
\begin{align}
1-2\zeta_j&\leq z^\dagger Q_jz\nonumber\\
&\leq \upsilon^\dagger\left[V^{(AY_j)}+i(\Omega_A\oplus0)\right]\upsilon\nonumber\\
&=\frac{\tr\Xi}{2}+a^\dagger i\Omega_Aa=\frac{\tr\Xi}{2}-1.\label{eq:telproofstep}
\end{align}
Together with Steps 1 and 2 this gives $F\leq2/(4-2\zeta_j)=1/(2-\zeta_j)$ for every protocol of the class, which completes the proof.
\end{proof}

%=====================================================================
\section{Semidefinite programs and certificates for three settings}\label{app:sdp}
%=====================================================================

This appendix states the semidefinite programs behind Sec.~\ref{sec:tight3} and explains how their solutions were turned into certificates.

\emph{Assemblage.} The two-mode squeezed vacuum with $\lambda=\tanh r$ is truncated at three photons, the second mode is sent through loss of transmission $\eta$, and the result is normalized. As explained in Sec.~\ref{sec:tight3}, this is a filtering of the node and does not affect a certificate of steering. The neighbor measures the projectors $P_{\pm|y}=(\iden\pm O_y)/2$ of qubit-block observables $O_y$ along the three Pauli directions, and the node's assemblage is $\sigma_{b|y}=\tr_{Y}[\rho(\iden\otimes P_{b|y})]$, normalized so that $\sum_b\tr\sigma_{b|y}=1$.

\emph{Primal program.} The steering robustness is
\begin{align}
R=\min\Big\{\sum_\lambda\tr\sigma_\lambda\Big\}-1,\label{eq:sdpprimal}
\end{align}
where the minimum runs over positive operators $\sigma_\lambda$, one for each of the eight assignments $\lambda=(\lambda_1,\lambda_2,\lambda_3)$ of outcomes to the three settings, subject to
\begin{align}
\sum_\lambda\delta_{b,\lambda_y}\sigma_\lambda\geq\sigma_{b|y}\quad\text{for all }b,y.\label{eq:sdpcons}
\end{align}
A local hidden-state model gives a feasible point with $\sum_\lambda\tr\sigma_\lambda=1$, so $R>0$ certifies steering.

\emph{Dual program and certificate.} The dual of Eq.~\eqref{eq:sdpprimal} maximizes $\sum_{b,y}\tr(F_{b|y}\sigma_{b|y})-1$ over operators $F_{b|y}\geq0$ with $\sum_yF_{\lambda_y|y}\leq\iden$ for every $\lambda$. For every such family, every assemblage with a local hidden-state model obeys
\begin{align}
\sum_{b,y}\tr\left(F_{b|y}\sigma_{b|y}\right)&=\sum_\lambda\tr\Big(\sum_yF_{\lambda_y|y}\,\sigma_\lambda\Big)\nonumber\\
&\leq\sum_\lambda\tr\sigma_\lambda=1,\label{eq:sdpwitness}
\end{align}
so a value above one certifies steering, independently of how the $F_{b|y}$ were found. We solved the primal program with the solver SCS through CVXPY, with tolerance $10^{-10}$, and took the $F_{b|y}$ from the dual variables of the constraints~\eqref{eq:sdpcons}. We then symmetrized them, set their negative eigenvalues to zero, added $10^{-12}\iden$ so that they are strictly positive definite, and divided all of them by the largest eigenvalue of $\sum_yF_{\lambda_y|y}$ over the eight $\lambda$, so that the second condition holds. Finally we recomputed this largest eigenvalue, the smallest eigenvalue of every $F_{b|y}$ and the left side of Eq.~\eqref{eq:sdpwitness} in $50$-digit arithmetic. In every case the smallest eigenvalue of the $F_{b|y}$ was $10^{-12}>0$, the largest eigenvalue of the local bound was one to the working precision, and the left side exceeded one by the amount listed in Table~\ref{tab:sdp}.

\begin{table*}[t]
\caption{\justifying\small Three-setting steering in the frame network with $3k-1$ neighbors. The columns give $k$, the number of neighbors, the transmission $\eta=k/(3k-1)$, the optimal $\lambda=\tanh r$ and the corresponding squeezing in decibels, the steering robustness $R$ from the primal program, and the certified violation of the steering inequality~\eqref{eq:sdpwitness}.}
\label{tab:sdp}
\centering\small\setlength{\tabcolsep}{10pt}
\begin{tabular}{rrccccc}
\hline\hline
$k$ & $3k-1$ & $\eta$ & $\lambda$ & dB & $R$ & Violation\\
\hline
1 & 2 & 0.5000 & 0.600 & 6.0 & $0.0483$ & $0.0483$\\
2 & 5 & 0.4000 & 0.544 & 5.3 & $0.0156$ & $0.0156$\\
3 & 8 & 0.3750 & 0.502 & 4.8 & $8.14\times10^{-3}$ & $8.14\times10^{-3}$\\
4 & 11 & 0.3636 & 0.468 & 4.4 & $5.06\times10^{-3}$ & $5.06\times10^{-3}$\\
5 & 14 & 0.3571 & 0.440 & 4.1 & $3.47\times10^{-3}$ & $3.47\times10^{-3}$\\
10 & 29 & 0.3448 & 0.350 & 3.2 & $1.02\times10^{-3}$ & $1.02\times10^{-3}$\\
20 & 59 & 0.3390 & 0.264 & 2.4 & $2.81\times10^{-4}$ & $2.81\times10^{-4}$\\
50 & 149 & 0.3356 & 0.175 & 1.5 & $4.79\times10^{-5}$ & $4.67\times10^{-5}$\\
\hline\hline
\end{tabular}
\end{table*}

%=====================================================================
\section{Numerical checks}\label{app:num}
%=====================================================================

This appendix describes the numerical work behind the paper, how it was carried out and what it shows.

All theorems are proven analytically, and the numerical work described here serves only as a consistency check of the derivations and of our implementation. It cannot establish the general statements. The only results that rest on numerics are those of Sec.~\ref{sec:tight3}. For the network transmissions they are certified by dual steering inequalities for each tested value of $k$, and the remaining values, near the thresholds, are reported as numerical findings.

\emph{Random states.} Random Gaussian states were generated by applying random symplectic matrices, products of exponentials $\exp(\Omega H)$ with random real symmetric $H$, to $k$ two-mode squeezed vacua with random squeezings and to vacuum modes, and by tracing out up to three environment modes. We used $k\leq3$ node modes and up to five neighbors. The criterion of Proposition~\ref{prop:hier}(a) agreed with the partial-transpose test for all $890$ neighbors tested, and the largest value of $\sum_j\zeta_j^+-k$ was $2.3\times10^{-14}$. The operator form of Proposition~\ref{prop:modeop}, maximized numerically over local symplectic matrices, agreed with the definition~\eqref{eq:share} to within $10^{-12}$. For $400$ random pairs we applied random pure-loss channels, quantum-limited amplifiers and general one-mode Gaussian channels to the neighbor, and random Gaussian channels with a thermal ancilla to the node. The largest values of the new share minus its bound in Proposition~\ref{prop:mono} were $-1.0\times10^{-4}$, $-5.5\times10^{-4}$, $0$ and $0$, respectively, so the bounds were never violated. The teleportation fidelity, maximized over the admissible pairs of Sec.~\ref{sec:telprot} by unconstrained optimization of symplectic matrices from several random starting points, never exceeded the bound~\eqref{eq:telbound} in $116$ neighbors. In $400$ random networks with up to $13$ single-mode neighbors, the number of neighbors with $\zeta_j>1/2$ never exceeded $2k-1$. For Proposition~\ref{prop:multimode} we grouped the outputs of $300$ random networks into $872$ neighbors with two modes each. For each of the $167$ neighbors that steer the node by Gaussian measurements, we followed the proof, constructed the mode of the neighbor from an eigenvector of $Q_Y$ with negative eigenvalue, and found that its share exceeds one half, by at least $1.7\times10^{-3}$.

\emph{Bell tests.} The CHSH values of Fig.~\ref{fig:bell} were computed in a Fock space truncated at fifteen photons per mode, which changes them by less than $10^{-6}$, with the four angles optimized by quasi-Newton methods from several starting points and the squeezing optimized by bounded scalar minimization. The displaced-parity values follow in closed form from the Gaussian Wigner function, with the squeezing and the four displacements optimized in the same way.

\emph{Steering robustness and $I_{3322}$.} The steering robustness curves of Sec.~\ref{sec:tight3} were computed with the conic solver SCS through CVXPY, with tolerance $10^{-9}$, in a Fock space truncated at three photons per mode, and maximized over the squeezing by bounded scalar minimization. Values below $10^{-10}$ are at solver precision and are interpreted as zero. The values reported as positive exceed $2\times10^{-5}$, except at $\eta=0.3335$, where the robustness is $3\times10^{-7}$ and we regard the result as indicative only. For the network transmissions $\eta=k/(3k-1)$ the programs were solved again with the tighter tolerance $10^{-10}$, and the positive values were certified by dual steering inequalities checked in $50$-digit arithmetic, as described in Appendix~\ref{app:sdp}. The $I_{3322}$ values were maximized with quasi-Newton methods from ten random starting points for each squeezing on a grid and each assignment of roles. For $\eta<0.45$, where the optimum moves to weak squeezing, we used fifteen starting points and nine values of $\lambda$ between $0.03$ and $0.15$.

\emph{Remote Wigner negativity.} For $k=2,\dots,6$ we constructed the harmonic frames, checked the tight-frame identity~\eqref{eq:frame}, expanded $\ket{\Phi_k}$ in the Fock basis and computed the photon statistics of every steered mode. The statistics are identical for all $2k-1$ neighbors, $p_0=0$, the mean photon number equals $2-1/k$, and the values of $p_1$ and $W_{Y_j}(0)$ quoted in Sec.~\ref{sec:saturation} follow from Eq.~\eqref{eq:Xmain}. The heralded Wigner functions of Fig.~\ref{fig:wigner}(b) were obtained by sending these statistics through the loss channel and summing Laguerre functions. The success probabilities and the values under imperfect detection in Sec.~\ref{sec:rates} were computed from Eq.~\eqref{eq:psucc} and from the analogue of Eq.~\eqref{eq:W0pre} with the modified symplectic eigenvalues. A self-contained Jupyter notebook that reproduces all checks, all numerical values quoted in the text and all figures is openly available on GitHub~\cite{Github_code}.

\bibliographystyle{unsrtnat}
\bibliography{main}
\end{document}